\documentclass[12pt]{JHEP3}

\usepackage{epsfig,amsfonts,amssymb,amscd,amsmath,mathrsfs,xspace,mathrsfs,amsthm,dsfont,bbm}

\usepackage{comment}
\usepackage{color}
\let\normalcolor\relax
\usepackage{bbold}

\newcommand{\sectiono}[1]{\section{#1}\setcounter{equation}{0}}

\newcommand{\be}{\begin{equation}}
\newcommand{\ee}{\end{equation}}
\newcommand{\ben}{\begin{eqnarray}\displaystyle}
\newcommand{\een}{\end{eqnarray}}
\newcommand{\bea}{\begin{eqnarray}}
\newcommand{\eea}{\end{eqnarray}}

\newcommand{\refb}[1]{(\ref{#1})}

\def\ZZZ{{\hbox{ Z\kern-1.6mm Z}}}
\def\RRR{{\hbox{ R\kern-2.4mm R}}}
\def\CCC{{\hbox{ C\kern-2.0mm C}}}
\def\zzz{{\hbox{z\kern-1mm z}}}

\DeclareMathOperator{\Td}{Td}
\DeclareMathOperator{\ch}{ch}

\def\ZZZ{{\mathbb Z} }
\def\RRR{{\mathbb R} }
\def\CCC{{\mathbb C} }

\newcommand{\IR}{\mathds{R}}

\newcommand{\IZ}{\mathds{Z}}

\newcommand{\p}{\partial}
\newcommand{\I}{\mathrm{i}}
\def\de{d}
\def\Tr{\,{\rm Tr}\, }

\def\Im{\,{\rm Im}\,}
\def\Re{\,{\rm Re}\,}

\def\llangl{\left\langle\!\left\langle }
\def\rrangl{\right\rangle \!\right\rangle }

\def\({\left(}
\def\){\right)}
\def\[{\left[}
\def\]{\right]}
\def\hf{{1\over 2}}

\newcommand{\non}{\nonumber}

\newcommand{\vt}{\vartheta}

\newcommand{\qeq}{{\hbox{=\kern-2.3mm ? \kern.5mm }}}
\renewcommand{\qeq}{=}

\newcommand{\bbb}{{\bar b}}

\newcommand{\eps}{\epsilon}

\newcommand{\ve}{\varepsilon}

\newcommand{\AAA}{{\cal A}}

\newcommand{\OO}{{\cal O}}

\newcommand{\XX}{{\cal X}}

\newcommand{\cC}{\mathcal{C}}

\newcommand{\cG}{\mathcal{G}}
\newcommand{\cK}{\mathcal{K}}
\newcommand{\cO}{\mathcal{O}}
\newcommand{\cA}{\mathcal{A}}

\newcommand{\cM}{\mathcal{M}}

\newcommand{\cN}{\mathcal{N}}

\newcommand{\cT}{\mathcal{T}}

\newcommand{\wt}{\widetilde}
\newcommand{\wh}{\widehat}

\newcommand{\NN}{{\cal N}}
\newcommand{\TT}{{\cal T}}

\newcommand{\ba}{\bar a}

\newcommand{\bb}{\bar b}

\def\bz{\bar z}
\def\bu{\bar u}
\def\bv{\bar v}
\def\bs{\bar s}
\def\bt{\bar t}

\def\bZ{\bar Z}

\def\btau{\bar \tau}

\newcommand{\tzeta}{\tilde\zeta}

\newcommand{\tom}{\tilde\omega}
\newcommand{\tgamma}{\tilde\gamma}
\def\tcA{\tilde\cA}
\def\tc{\tilde c}

\def\cl0{\tilde c_0}

\def\chp{\check p}
\def\chq{\check q}

\def\chgam{\check\gamma}

\def\hnabla{\hat\nabla}

\def\one{{\hbox{ 1\kern-.8mm l}}}
\def\zero{{\hbox{ 0\kern-1.5mm 0}}}

\newcommand{\dalpha}{\beta}

\def\CY{\mathfrak{Y}}

\def\NP{k}

\def\cTR{\cT^{\rm R}}
\def\ts{T}

\def\dalpha{{\dot\alpha}}
\def\dbeta{{\dot\beta}}

\def\sp{\not  \hskip -.02in p}

\def\I{i}

\def\bbb{{\bf b}}

\def\bbt{{\bf t}}

\def\bbg{{\bf g}}
\def\bbV{{\bf V}}
\def\bbG{{\bf G}}
\def\bbB{{\bf B}}
\def\bbC{{\bf C}}

\def\bbsigma{\boldsymbol{\sigma}}
\def\bbkap{\boldsymbol{\kappa}}
\def\bbom{\boldsymbol{\omega}}

\def\defparallel{P}
\def\bbgp{\bbg_\defparallel}
\def\bbBp{\bbB_\defparallel}

\def\vn{v^{(p+1)}}
\def\vi#1{v^{(#1)}}

\def\Pi#1{P^{(#1)}}

\def\bbC{{\bf C}}
\def\bbF{{\bf F}}

\def\bq{\bar q}
\def\bp{\bar p}

\def\nc{c}

\def\Lgi#1{L_\gamma^{(#1)}}
\def\chLgi#1{L_{\chgam}^{(#1)}}

\title{Euclidean D-branes in Type IIB string theory on  Calabi-Yau threefolds}

\author{Sergei Alexandrov$^1$, Ashoke Sen$^2$, Bogdan Stefa\'nski, jr.$^3$
\\
$^1$ {\it
Laboratoire Charles Coulomb (L2C), Universit\'e de Montpellier,
CNRS, F-34095, Montpellier, France}\\

$^2$ {\it Harish-Chandra Research Institute, HBNI \\ Chhatnag Road, Jhusi, Allahabad 211019, India}\\

$^3$ {\it Centre for Mathematical Science, City, University of London
\\ Northampton Square, EC1V 0HB London, UK}\\

\vspace*{2mm} {\tt e-mail:
\email{sergey.alexandrov@umontpellier.fr},
\email{sen@hri.res.in},
\email{Bogdan.Stefanski.1@city.ac.uk}
}

\vspace*{-3mm}

}

\abstract{We compute the contribution of Euclidean D-branes in type IIB string theory
on Calabi-Yau threefolds to the metric on the hypermultiplet moduli space
in the large volume, weak coupling limit. Our results are in perfect
agreement with the predictions based on S-duality, mirror symmetry and supersymmetry.

}

\begin{document}

\tableofcontents

\sectiono{Introduction} \label{s1}

Type II string theories
are known to receive non-perturbative contributions from Euclidean
D-branes wrapped on compact cycles. These contributions are expected to play an important
role in stabilizing the moduli in semi-realistic string compactifications and the determination
of Yukawa couplings in the low energy effective field theory. For this reason it is important to
to learn how to systematically compute these contributions to string amplitudes.

In a previous paper \cite{Alexandrov:2021shf}, extending the results of \cite{Sen:2021qdk,Sen:2021tpp,Sen:2021jbr},
we computed these corrections to the hypermultiplet moduli
space metric in type IIA string theory compactified
on Calabi-Yau (CY) threefolds in the weak coupling limit and compared the results with
the predictions based on S-duality, mirror symmetry and supersymmetry
\cite{Alexandrov:2008gh,Alexandrov:2009zh,Alexandrov:2014sya}.
In this paper we repeat the analysis for type IIB string theory compactified on CY threefolds.
Since the two theories are related by mirror symmetry, we could have gotten the result
just by using this symmetry. Nevertheless, since our main goal is to
learn how to compute D-instanton effects when they are not known from other considerations, it is
important to develop independent techniques for doing these computations. Indeed, as we will see,
the actual computation in type IIB string theory has a somewhat different flavor and different
subtleties than those encountered in the analysis of type IIA string theory.

The rest of the paper is organized as follows. In \S\ref{s2} we describe a few of our
conventions, but refer to \cite{Alexandrov:2021shf} for most of the world-sheet conventions.
We also describe the low energy effective
action obtained by compactifying tree level type IIB supergravity on a Calabi-Yau threefold
and the relative normalization between the variables of supergravity used in \S\ref{s2} and
\S\ref{s3} and the string theory variables that we use in later calculations.
In \S\ref{s3} we describe the
predicted result for the D-instanton contribution to the hypermultiplet metric based on
S-duality, mirror symmetry and supersymmetry. In \S\ref{s4} we describe the general
strategy for computing the instanton contribution using world-sheet techniques. This is
essentially the same as in the case of type IIA string theory. In \S\ref{s5} we compute the
normalization factor associated with the contribution from D(-1)-branes in type IIB
string theory. This case is somewhat different from the systems analyzed in type IIA string
theory in \cite{Alexandrov:2021shf}, where the Euclidean branes were assumed to be rigid,
with the only moduli being associated
with the motion along the non-compact directions and the fermion zero modes associated
with broken supersymmetry. In contrast, D(-1)-branes have additional moduli
describing motion along the CY threefold. In \S\ref{sdminus} and
\S\ref{s7} we compute the disk
amplitudes with one hypermultiplet scalar and a pair of fermion zero modes for various
scalars on various D-branes. These results are then put together following the general
algorithm of \S\ref{s4} to compute the correction to the metric on the hypermultiplet
moduli space. In \S\ref{s7} we generalize these results to multiple charge systems and
by taking into account the effect of background $B$ and RR fields. As a result, we find a perfect agreement
with the prediction for the instanton contribution to the metric presented in \S\ref{s3}.
The appendices contain various technical results that are used in the
analysis in the main text.

\section{Preliminaries}
\label{s2}

In this section we shall describe some of our conventions and the known results on the tree
level action of type IIB string theory compactified on a Calabi-Yau threefold.

\subsection{Conventions}

Since most of our world-sheet
conventions will follow those in \cite{Alexandrov:2021shf}, we shall not
repeat them here. The main difference comes in the spin fields. Unlike in type IIA
string theory, the left and right handed spin fields in type IIB string theory carry the same
ten-dimensional chirality. Therefore, now the operator product expansion of the spin fields
in the two sectors will take identical form, with the holomorphic fields / coordinates
exchanged with the anti-holomorphic fields / coordinates. These have been described in
\refb{eqb1}. The other world-sheet relations that we shall use extensively are the normalizations of
the open and closed string vacua that can be used to fix the normalization of
the disk and sphere amplitudes:
\be\label{eopennorm}
\langle k| c_{-1} c_0 c_1 \, e^{-2\phi}(0)|k'\rangle=(2\pi)^{p+1} \delta^{(p+1)}(k+k')
\ee
for the open string vacuum on a $p$-brane, and
\be\label{eclosednorm}
\langle k| c_{-1}\bar c_{-1} c_0\bar c_0 c_1 \bar c_1\, e^{-2\phi}(0) e^{-2\bar\phi}(0)|k'\rangle=
- (2\pi)^{10}\delta^{(10)}(k+k')
\ee
for the closed string.

We shall use the same gamma matrix conventions as in \cite{Alexandrov:2021shf},
but since we shall
use these results extensively, we shall review some of the relations here.
First of all, we take all the ten-dimensional gamma matrices to be $16\times 16$ symmetric
matrices. We denote
the real coordinate labels of the CY threefold
$\CY$ by lower case bold-faced letters $\bf m,n,\cdots$,
the holomorphic
coordinate labels of $\CY$ by lower case italic indices $s,t,\cdots$, those
of four-dimensional space by Greek letters $\mu,\nu,\cdots$, denote by $\wt\Gamma^{\bf m}$ for
$4\le {\bf m}\le 9$ the six-dimensional gamma matrices, by $\gamma^\mu$ for
$0\le \mu\le 3$ the four-dimensional gamma matrices and choose the ten-dimensional
gamma matrices $\Gamma^M$ as follows:
\be\label{eff1}
\wt\Gamma = \I\, \wt \Gamma^4\cdots \wt\Gamma^9,
\qquad
\Gamma^\mu = \wt\Gamma\otimes \gamma^\mu,
\qquad
\Gamma^{\bf m} = \wt\Gamma^{\bf m}\otimes I_4
\, .
\ee
We denote by $\eta$, $\bar\eta$ the covariantly constant spinors on $\CY$, for which we have the
following useful relations \cite{Alexandrov:2021shf}:
\be\label{eff2}
\bar\eta\,\eta=1,
\qquad
\bar\eta \wt\Gamma^{\bf ijklmn}\eta =-\I\, \eps^{\bf ijklmn},
\qquad
\bar\eta \wt\Gamma^{\bf ij} \eta =\I\, \bbom^{\bf ij}\, ,
\ee
where $\bbom$ is the K\"ahler form on $\CY$
related to the metric in the holomorphic coordinates by $\bbom_{s\bar t}= \I  \bbg_{s\bar t}$.

As in \cite{Alexandrov:2021shf}, the ten-dimensional spinor indices will be labelled by
$\alpha,\beta,\cdots$ and the four-dimensional spinor indices will be labelled by the
dotted indices $\dalpha,\dbeta,\cdots$ and undotted indices $\alpha,\beta,\cdots$.
Since the relevant instantons break 4 out of 8 supersymmetries, they will carry 4
fermion zero modes from the open string sector that we denote by $\tilde\chi^\alpha$ and
$\tilde\chi^\dalpha$. We can also represent them as ten-dimensional spinors
$\XX^\alpha$, $\wh\XX^\beta$ by choosing
\be \label{etendfourd}
\XX=\eta\otimes \tilde\chi^\alpha, \qquad \wh\XX = \bar\eta\otimes \tilde\chi^\dalpha\, .
\ee

In our analysis we shall use two types of Hodge dual. We denote by $*$ the Hodge dual in
ten dimensions and by $\star$ the Hodge dual in the Calabi-Yau manifold. For both of these we
use the string metric. However, we express our final result in the four-dimensional canonical
metric.

\subsection{Classical hypermultiplet moduli space}
\label{subsec-class}

Upon compactification of type II string theory on a CY threefold $\CY$,
we get a $\NN=2$ supersymmetric theory whose massless
scalars can be divided into hypermultiplet and vector multiplet fields. In particular, the
scalars labelling the hypermultiplet moduli space $\cM_H$ in type IIB formulation include the following fields:
\begin{itemize}
\item
the fields $z^a=b^a+\I t^a$ ($a=1,\dots,h^{1,1}(\CY)$) describing appropriately normalized
complexified K\"ahler moduli of $\CY$;

\item
the RR fields defined as period integrals of the RR $p$-forms $C^{(p)}$ of type IIB string theory
over a basis of cycles in $H_{\rm even}(\CY,\IZ)$
\be
\label{RRiib}
c^0= C^{(0)},
\qquad
c^a=\int_{\gamma^a} C^{(2)},
\qquad
\tc_a= \int_{\tgamma_a} C^{(4)} ,
\qquad
\tc_0=\int_{\CY} C^{(6)} ,
\ee
where $\{\gamma^a\}$ and $\{\tgamma_a\}$ are bases in $H_2(\CY,\IZ)$ and $H_4(\CY,\IZ)$,
respectively;

\item
the dilaton $\tau_2$ whose vacuum expectation value is the
inverse string coupling $1/g_s$;

\item
the axion $\sigma$ which is dual to the NSNS 2-form field in four dimensions.\footnote{Note that NS axion and the scalars defined by \eqref{RRiib}
are {\it different} from those that are usually used to express the hypermultiplet metric of type IIB string theory
and that have simple transformation properties under the S-duality group (see, e.g., \cite{Bohm:1999uk,Alexandrov:2008gh,Alexandrov:2011va}).
To make contact with the literature, we show the relations between the two sets of fields in appendix \ref{ap-fields}.}
\end{itemize}
The precise normalization of these fields will be described later.
Throughout this paper, we shall work in the large volume limit.
However, unlike
in the case of type IIA string theory, here the large volume limit is not exact,
since the K\"ahler moduli belong to the hypermultiplet.

If we denote the collection of the hypermultiplet scalars by $\{\lambda^m\}$, then the kinetic term of these
fields in the action has the form
\be \label{escalaraction}
-{1\over 2} \, \int d^4 x \, \cG^H_{mn}(\lambda) \, \p_\mu\lambda^m \, \p^\mu\lambda^n\, ,
\ee
where
in the large volume, small string coupling approximation, the metric
$\cG^H_{mn} d\lambda^m d\lambda^n$ is given by
\be
\begin{split}
\de s_{\rm cl}^2
=&\,
\frac{1}{\tau_2^2}\(2\de\tau_2+\frac{\tau_2}{2 V}\, \kappa_{ab}t^a\de t^b\)^2
+G_{ab}\(\de t^a\de t^b+\de b^a \de b^b\)
\\
&\,
+\frac{(\de c^0)^2}{\tau_2^2}
+\frac{G_{ab}}{\tau_2^2}\, \nabla c^a\nabla c^b
+\frac{G^{ab}\nabla\tc_a\nabla\tc_b}{\tau_2^2 V^2}
+\frac{\(\nabla \tc_0\)^2}{\tau_2^2 V^2}
+\frac{\(\nabla \sigma\)^2}{4\tau_2^4 V^2}\, .
\end{split}
\label{metricBtreelargeV}
\ee
Various quantities in this expression are defined as follows.
Given a basis $\{\omega_a\}$ of harmonic
(1,1)-forms, dual to the basis $\{\gamma^a\}$ of integer homology 2-cycles,
i.e. satisfying $\int_{\gamma^a}\omega_b=\delta^a_b$,
we define,
\be
\kappa_{abc}=\int_\CY \omega_a\wedge\omega_b\wedge \omega_c,
\qquad
\kappa_{ab}=\kappa_{abc}t^c,
\qquad
V=\frac16\, \kappa_{abc}t^a t^b t^c\, ,
\label{def-kapV}
\ee
and $\kappa^{ab}$ to be the matrix inverse of $\kappa_{ab}$.
Then the metrics appearing in \eqref{metricBtreelargeV} are given by
\be
\begin{split}
G_{ab}\equiv &\, -\frac{1}{V}\(\kappa_{ab} - {1\over 4 V} \,\kappa_{ac} t^c\kappa_{bd} t^d\) ,
\\
G^{ab}\equiv &\, (G^{-1})^{ab}=-V\(\kappa^{ab} - {1\over 2 V} \,t^a t^b\) ,
\end{split}
\label{defG}
\ee
while various covariant one-forms are defined as
\be
\begin{split}
&\,\nabla c^a= \de c^a -c^0 \de b^a,
\qquad
\nabla\tc_a=\de\tc_a-\kappa_{abc}c^b\de b^c,
\qquad
\nabla\tc_0=
\de\tc_0-\tc_a\de b^a,
\\
&\,\qquad\qquad\qquad
\nabla\sigma=\de\sigma+c^0\nabla \tc_0-c^a\nabla\tc_a+\tc_a\nabla c^a-\tc_0\de c^0.
\end{split}
\label{defnabla}
\ee
Note that these one-forms in $\cM_H$
can be described in a compact way using differential forms in $\CY$. Let us define
\be
B=b^a \omega_a\, ,
\qquad
C^{(\rm even)} = \sum_{k=0}^3 C^{(2k)}=c^0 + c^a\omega_a + \tc_a \tilde\omega^a +\tc_0 \omega_\CY,
\label{defBC}
\ee
where $\tilde\omega^a$ is a basis of harmonic (2,2)-forms satisfying $\int_{\tilde\gamma^a}
\tilde \omega^b=\delta_a^b$ and $\omega_\CY$ is a harmonic six-form satisfying
$\int_\CY\omega_\CY=1$.
We also introduce operator $\iota$ on even-forms which acts on $H^{2k}(\CY)$ by the sign $\iota=(-1)^{k}$.
For example, $\iota(C^{(\rm even)})$ differs from \eqref{defBC} by the sign of $c^a$ and $\tc_0$ terms.
Then we have, using \refb{basisforms} and \refb{omaomb},
\be\label{egeomcov}
\begin{split}
\nabla C^{(\rm even)} =&\,  d  C^{(\rm even)} - dB\wedge C^{(\rm even)}=
e^B d (e^{-B} C^{(\rm even)}) ,
\\
\nabla\sigma=&\, \de\sigma+\int_\CY \iota(C^{(\rm even)})\wedge \nabla C^{(\rm even)} ,
\end{split}
\ee
where all products are to be regarded as wedge products and
$d$ acts on the moduli space (not on $\CY$).

The action \refb{escalaraction} has been written down in the background in which the
four-dimensional canonical metric, normalized to have the action
\be
-\int d^4 x \, \sqrt{-\det g} \, R\, ,
\ee
has been set equal to $\eta_{\mu\nu}$. The relationship between the variables appearing
in \refb{metricBtreelargeV} and those that appear in string (field) theory (which we shall denote by boldface letters)
can be found either by comparing the kinetic terms or by comparing the disk one-point functions in the presence
of an Euclidean D-brane. This is done in appendix \ref{sd} and the results are
summarized in \refb{eccnet}.

Due to mirror symmetry, the hypermultiplets moduli spaces of type IIA and type IIB string theories
compactified on two mirror CY threefolds are identical. Therefore, the classical metric \eqref{metricBtreelargeV}
can be related by a coordinate change to the tree level metric in the type IIA formulation
evaluated for the classical prepotential
\be
F(X) = -\frac16 \kappa_{abc} \frac{X^a X^b X^c}{X^0}\, .
\label{Fcl}
\ee
This coordinate change is known as {\it classical mirror map} and in our conventions it is given by
\be
\begin{split}
&
r=\hf\,\tau_2^2 V\,,
\qquad
z^a = b^a + \I t^a,
\qquad
\zeta^0 = c^0,
\qquad
\zeta^a = -c^a + c^0 b^a,
\\ &
\tzeta_a = -\tc_a + \kappa_{abc} b^b \(c^c -\hf\, c^0 b^c\)\, ,
\qquad
\tzeta_0 = -\tc_0 +b^a\tc_a- \hf\,\kappa_{abc} b^a b^b \(c^c-\frac13\, c^0 b^c\),
\end{split}
\label{clmirmap}
\ee
where the variables on the left hand side are the same ones used in \cite{Alexandrov:2021shf}
to express the result for the metric in type IIA string theory. (The NS axion is the same on both sides.)

The metric \eqref{metricBtreelargeV} has a large isometry group.
First of all, it is invariant under $SL(2,\IR)$ transformations which include, in particular, S-duality inverting the string coupling.
As explained in appendix \ref{ap-fields}, the variables we are using are closely related (but not identical)
to the variables distinguished by having simple transformations under this symmetry.
Second, the metric is invariant under various shift symmetries.
They include the invariance under shifts of the $B$-field
\be
\label{monb}
b^a\mapsto b^a+\epsilon^a,
\ee
and the Heisenberg symmetry involving shifts of the RR fields and the NS axion.
To write them in terms of type IIB fields, it is again convenient to use the language of differential forms.
Let $\vartheta\in H^{(\rm even)}(\CY)$ be a moduli independent even form.
Then the Heisenberg symmetry is described by the following transformations
\be
C^{(\rm even)}\mapsto C^{(\rm even)}+  e^B \vartheta ,
\qquad \sigma \mapsto \sigma+\theta-\int_\CY \iota(\vartheta)\wedge C^{(\rm even)} e^{-B}\, ,
\label{Heissym}
\ee
where the operator $\iota$ was defined above \eqref{egeomcov}.
The invariance of the metric \refb{metricBtreelargeV} under these
transformations follows from the invariance of the covariant derivatives \eqref{egeomcov},
which is straightforward to check taking into account the property
\be
\iota(e^{B} \vartheta)=e^{-B}\iota(\vartheta).
\label{propiota}
\ee

\subsection{Normalization of the fields}

For our analysis we need to relate the moduli appearing in \refb{metricBtreelargeV} to
those  that appear in the string theory computation. This has been done in appendix \ref{sd}.
Here we shall summarize just the main results. We denote by $\bbg_{MN}$ and $\bbB_{MN}$
the ten-dimensional string metric and NSNS 2-form field, normalized so that the
polarizations $h_{MN}$ and $b_{MN}$ appearing in the vertex operators are
related to  these fields via
\be\label{egmnhmn}
\bbg_{MN}=\eta_{MN} + 2\kappa\, h_{MN}, \qquad \bbB_{MN} = b_{MN}\, ,
\ee
where $\kappa$ is related to the string coupling $g_s=1/\tau_2$ via
\be \label{ekappags}
\kappa = 2^3 \, \pi^{7/2}\, g_s
\, .
\ee
Upon compactification on a CY threefold $\CY$,
they give rise to the scalar fields $\bbt^a$ and $\bbb^a$ defined using the expansion in the basis of (1,1)-forms
\be \label{e626a}
\bbg_{s\bar t} = -\I \bbt^a \, \omega_{a, s\bar t},
\qquad
\bbB_{s\bar t} = {1\over 2\kappa}\, \bbb^a \, \omega_{a,s\bar t}\, .
\ee
In these variables the K\"ahler form of $\CY$ reads as $\bbom =\bbt^a \, \omega_a$.
We also denote by $\bbC^{(2k)}$ the RR $2k$-form potential whose vertex
operators are given in appendices \ref{sb} and \ref{sc}, and by $\bbsigma$ the scalar obtained by
dualizing the NSNS 2-form field $\bbB_{\mu\nu}$, whose normalization is given in \refb{e720}.
We then have the following relations between these variables and the ones
appearing in \S\ref{subsec-class}:
\be
\bbt^a = (2\pi)^2 t^a,
\qquad
\bbb^a =  (2\pi)^2 b^a,
\qquad
d\bbsigma = -{(2\pi)^6\over 2\kappa \tau_2^2} \,d\sigma,
\qquad
\bbC^{(2k)}= \frac{(2\pi)^{2k}}{2^6\pi^{7/2}}\,C^{(2k)}.
\label{eccnet}
\ee
The variable $\tau_2$, describing the ten-dimensional dilaton is the same in the
string theory variables and in those that enter the metric \refb{square-2b}.
It will also be convenient to define
\be
\bbkap_{ab}= \kappa_{abc}\bbt^c = (2\pi)^2 \, \kappa_{ab}
\label{defbbk}
\ee
and $\bbkap^{ab}$ as its inverse matrix.

Various useful quantities for Calabi-Yau threefolds and their properties may be found in
appendix \ref{ap-Kahler}.

\section{Prediction for the D-instanton amplitude} \label{s3}

In this section we shall describe the prediction for the D-instanton correction to the
hypermultiplet moduli space metric in the
weak coupling and large volume limit, where the word instanton will refer to any
Euclidean D-brane whose world-volume lies along the compact directions.
This prediction will be based on the mirror symmetry applied to the leading instanton contribution
to the metric on $\cM_H$ in type IIA string theory found in \cite{Alexandrov:2021shf}.

\subsection{D-instantons in type IIB}

The classical metric on $\cM_H$ given in \refb{metricBtreelargeV}
is not exact and receives quantum corrections,
in particular non-perturbative corrections due to D-instantons.
In contrast to type IIA string theory, where upon compactification on a CY threefold only D2-branes generated instanton corrections,
in type IIB all D$p$ branes with $p$ odd, as well as their bound states, generate non-perturbative contributions.

The charge vector  $\gamma=(q_\Lambda, p^\Lambda)$
classifying such D-instantons is given by
the expansion of the so called generalized Mukai vector in the basis
of even-dimensional cohomology \cite{Minasian:1997mm}
\be
\label{mukaimap}
\gamma \equiv \ch(E)\sqrt{\Td(\CY)} =p^0 + p^a \omega_a - q_a \tom^a + q_0 \omega_{\CY},
\ee
where $\ch(E)$ is the Chern character of $E$, $\Td(\CY)$ is the Todd class of $T\CY$,
and the normalization and other
properties of the basis forms $\omega_a$, $\tom^a$, $\omega_\CY$
are specified in appendix \ref{ap-Kahler}.
Strictly speaking, the correct mathematical description associates D-instantons to elements in
the derived category of coherent sheaves $\mathcal{D}(\CY)$ \cite{MR1403918,Douglas:2000gi}.
For our purpose however these mathematical subtleties will be unimportant\footnote{Nevertheless, one should remember that
the charges $q_0,q_a$ are not integer but belong to a shifted lattice
$$
q_a\in \IZ -\frac{p^{0}}{24}\, c_{2,a} - \frac12 \kappa_{abc} p^b p^c,
\qquad
q_0\in \IZ-\frac{1}{24}\, p^{a} c_{2,a} ,
$$
where $c_{2,a}$ are components of the second Chern class of $\CY$. For explanation how this fact is reconciled
with mirror symmetry, see \cite{Alexandrov:2010ca,Alexandrov:2011va}. In the large volume
limit that we shall be working in, these shifts in the lattice will not be important since for finite
charges the contributions from $p^0$ and $p^a$ dominate over that from $q_0$ and $q_a$.
On the other hand, if we scale the charges so that their contributions are of the same order,
then $q_0$ and $q_a$ have to be large in order to be able to compete with $p^0$ and $p^a$,
and the lattice shift becomes unimportant.\label{foot-qc}
}, and we can think about
a D-instanton of charge $\gamma=(q_\Lambda, p^\Lambda)$ as a bound state of D5-brane on $\CY$ with wrapping number $p^0$,
D3-brane wrapping the 4-cycle $-p^a\tgamma_a$, D1-brane wrapping the 2-cycle $-q_a\gamma^a$, and
D(-1)-branes with charge $-q_0$, i.e. as a homology element
\be
L_\gamma=-q_0\gamma^0 -q_a\gamma^a-p^a\tgamma_a+p^0\CY,
\label{defLg}
\ee
where $\gamma^0\in H_0(\CY)$. Note that $L_\gamma$ is Poincar\'e dual to $\iota(\gamma)$
where $\iota$ has been defined above \refb{egeomcov}.
The minus signs appearing in this definition have been fixed in appendix \ref{sd}
and are related to the choice of sign in the boundary conditions on the spin fields given in \eqref{SSbc}.
We will denote by $\Lgi{2n}$ the component of $L_\gamma$ belonging to $H_{2n}(\CY)$.
It is important to note that the BPS condition requires the wrapped cycle to be
holomorphic, i.e. to have the volume form
\be
\vi{2n}_\gamma=\frac{1}{n!}\, (P_\gamma\bbom P_\gamma)^n,
\qquad
n=\frac{p+1}{2}\, ,
\label{defvn}
\ee
where $P_\gamma$ is the projection operator along $\Lgi{2n}$. This can be defined by
using a local coordinate system in which the metric at a given point is $\delta_{\bf ij}$,
using the standard definition of the projector in flat spacetime
and then expressing it in the original coordinate system.

As in type IIA, each BPS instanton breaks 4 out 8 supercharges and is characterized by
the central charge function which can be compactly written as
\be \label{edefzgamma}
Z_{\gamma}=\int_{\CY} e^{-z^a\omega_a}\,\gamma = q_0+q_az^a +\hf\, (pz^2) -\frac{p^0}{6}\,(z^3),
\ee
where we used a convenient notation $(xyz)=\kappa_{abc}x^a y^b z^c$ and
the second equality follows from \refb{basisforms} and \refb{omaomb}.
The corresponding instanton action is given by \cite{Marino:1999af,Aspinwall:2004jr}
\be
\cT_{\gamma}=2\pi \tau_2|Z_{\gamma}|
+2\pi\I \Theta_\gamma,
\qquad
\Theta_\gamma=\int_{\CY}\gamma\wedge C^{\rm even} e^{-B}.
\label{Dbraneact}
\ee

The instanton corrections break the continuous symmetries discussed in the end of \S\ref{subsec-class}.
However, their discrete subgroups survive. In particular, the transformations \eqref{Sdualtr} with
${\scriptsize \begin{pmatrix} a & b \\ c & d  \end{pmatrix}}\in SL(2,\IZ)$ form the S-duality group
of type IIB string theory,
the shifts \eqref{Heissym} with $\vartheta\in H^{(\rm even)}(\IZ,\CY)$ and $\theta\in 2\IZ$
describe large gauge transformations of the RR fields and the $B$ field,
whereas \eqref{monb} with $\eps\in\IZ$ correspond to monodromies around the large volume point in the complexified
K\"ahler moduli space.\footnote{More precisely, to be the symmetries at non-perturbative level,
most of these transformations should be supplemented
with some additional constant terms whose origin can be traced to the non-integrality of the D-instanton charges in
the type IIB  formulation \cite{Alexandrov:2010np,Alexandrov:2014rca} (see footnote \ref{foot-qc}).
For example, \eqref{Sdualtr} requires an additional shift of $\nc_a$ by a term proportional to the second Chern class of $\CY$
and the multiplier system of the Dedekind eta function \cite{Alexandrov:2010ca}. For our purposes however all these additional
contributions to the symmetry transformations are not important and can be ignored. \label{foot-sym}}
The requirement of these symmetries provides an important constraint on the moduli space metric.

The D-instanton amplitudes are proportional to $e^{-\cT_\gamma}$, which is manifestly invariant
under the large gauge transformations generated by $\vt$ and $\theta$ for any fixed $\gamma$.\footnote{In fact, this would be
true if the charges were integer. However, their non-integrality is compensated by the constant terms mentioned
in footnote \ref{foot-sym} so that $e^{-\cT_\gamma}$ is indeed invariant.}
However, this is not the case for the monodromy transformations generated by $\eps^a$.
The fact that allows the D-instanton corrections
to be consistent with monodromy invariance \refb{monb}
is that the D-instanton action \eqref{Dbraneact} stays invariant provided
the charges also undergo a symplectic transformation to compensate
the shift of the $B$ field, namely,
\be
\gamma\mapsto \gamma[\epsilon]=e^{\eps^a\omega_a}\gamma\, .
\label{chargesymp}
\ee
This suggests to introduce a $b$-dependent version of the charges
\be
\chgam\equiv \gamma[-b]=e^{-B}\gamma,
\label{defgamB}
\ee
or, more explicitly,
\be
\begin{split}
\chp^0=p^0,
\qquad
\chp^a =&\, p^a - p^0 b^a\, ,
\qquad
\chq_a = q_a + \kappa_{abc} p^b b^c - {p^0\over 2}\, \kappa_{abc} b^b b^c,
\\
\chq_0 =&\,  q_0 + q_a b^a +{1\over 2}\, (pb^2)-{p^0\over 6}\,(b^3).
\end{split}
\label{def-qch}
\ee
These charges stay invariant under monodromies and allow us to rewrite the central charge \eqref{edefzgamma}
and the axion coupling \eqref{Dbraneact} as
\bea
Z_\gamma &=& \int_{\CY} e^{-\I t^a\omega_a}\,\chgam=\chq_0+\I \chq_a t^a-\hf\, (\chp t^2)+\I \chp^0 V ,
\label{ecentralcharge}
\\
\Theta_\gamma &=& \int_\CY \chgam\wedge C^{(\rm even)}=\chq_0 c^0-\chq_a c^a+\chp^a\tc_a+\chp^0\tc_0,
\label{hatgamTheta}
\eea
where $V$ was defined in \eqref{def-kapV}.

\subsection{The leading instanton contribution}

Our goal is to find the leading D-instanton correction to the classical metric on $\cM_H$
which can then be tested against direct calculation of string amplitudes.
Since the mirror symmetry allows to identify the metrics in type IIA and type IIB formulations,
the simplest approach to our problem is to take the leading D-instanton contribution found on
the type IIA side in \cite{Alexandrov:2021shf} and
translate it to the type IIB theory.\footnote{Note that
mirror symmetry was one of the inputs in the derivation of the D-instanton corrected metric
in \cite{Alexandrov:2008gh,Alexandrov:2009zh,Alexandrov:2014sya} (see, in particular, earlier work \cite{RoblesLlana:2007ae})
and hence can be freely used in this context.
On the other hand, when we try to reproduce the results by explicit world-sheet calculation,
we shall not make use of mirror symmetry.}
For this reason, we shall first review the results for the leading D-instanton contribution
in the type IIA variables introduced in \S\ref{s2}.

First, given a holomorphic prepotential\footnote{The prepotential is a function of $h^{1,1}+1$ variables $z^\Lambda$, homogeneous of degree 2.
The complexified K\"ahler moduli coincide with the homogeneous coordinates $z^a/z^0$. 
For simplicity, we will work in the gauge $z^0=1$ after taking the derivatives with respect to $z^\Lambda$.}
$F(z^0,\cdots, z^{h_{1,1}})$, let us define the corresponding K\"ahler potential
\be
\cK=-\log K,
\qquad
K= z^\Lambda N_{\Lambda\Sigma} \bz^\Sigma,
\qquad
N_{\Lambda\Sigma}=  -2\Im F_{\Lambda\Sigma},
\label{def-Kahlerpot}
\ee
where $F_{\Lambda\Sigma}\equiv  \p_{z^\Lambda}\p_{z^\Sigma} F$.
We also introduce
a one-form
\be
\cC_\gamma= N^{\Lambda\Sigma}\(q_\Lambda-\Re F_{\Lambda\Xi}p^\Xi\)\(\de\tzeta_\Sigma-\Re F_{\Sigma\Theta}\de\zeta^\Theta\)
+\frac14\, N_{\Lambda\Sigma}\,p^\Lambda\,\de\zeta^\Sigma\, ,
\label{connC}
\ee
where $N^{\Lambda\Sigma}$ is the matrix inverse of $N_{\Lambda\Sigma}$.
Using the form of $F$ given in \refb{Fcl} and
the change of variables \refb{clmirmap},
the instanton action $\TT_\gamma$ \refb{Dbraneact} takes the form
\be
\cT_\gamma=8\pi\sqrt{\frac{r}{K}}\, |Z_\gamma|+2\pi\I \Theta_\gamma,
\label{actD2}
\ee
where (with $z^0\equiv 1$)
\be
Z_\gamma=q_\Lambda z^\Lambda-p^\Lambda F_\Lambda\,,
\qquad
\Theta_\gamma=q_\Lambda\zeta^\Lambda-p^\Lambda\tzeta_\Lambda\, .
\label{ZThIIA}
\ee
Eqs. \refb{actD2}, \refb{ZThIIA} agree with the form of $\cT_\gamma$, $Z_\gamma$ and $\Theta_\gamma$
given in \cite{Alexandrov:2021shf}.

In terms of these quantities,
the leading instanton contribution to the metric on $\cM_H$, which follows both from
a combination of supersymmetry, mirror symmetry and S-duality as well as from a direct computation of string amplitudes,
was found to be \cite[eq.(3.5)]{Alexandrov:2021shf}:
\be
\de s_{\rm inst}^2
= \sum_\gamma \frac{\Omega_\gamma\, \Sigma_\gamma}{16 r\sqrt{2\pi \cTR_\gamma}}\(
\tcA_\gamma^2+\OO(d\TT_\gamma)\) ,
\label{square}
\ee
where the sum goes over the full charge lattice,
$\cTR_\gamma=\Re\cT_\gamma$,
$\Omega_\gamma$ is the Donaldson-Thomas (DT) invariant\footnote{Note that for pure electric charges
these invariants are known to be
$$
\Omega_\gamma=\left\{\begin{array}{ll}
-\chi, &\qquad \gamma=(0,0,0,q_0),
\\
n_{q_a}, &\qquad \gamma=(0,0,q_a,q_0), \ q_a\ne 0,
\end{array}\right.
$$
where $\chi$ is the Euler characteristic of $\CY$ and $n_{q_a}$ are its genus 0 Gopakumar-Vafa invariants.
\label{foot-DT}}
corresponding to instanton charge $\gamma$ which (roughly) counts the number of independent
supersymmetric cycles in the sector with charge $\gamma$,
and the functions $\Sigma_\gamma$ and $\tcA_\gamma$ are defined as
\be
\Sigma_\gamma=\sum_{k=1}^\infty \frac{\sigma_{k\gamma}}{\sqrt{k}}\,  e^{-k\cT_\gamma},
\label{defSumk}
\ee
\be
\tcA_\gamma=\frac{|Z_\gamma|}{\sqrt{r K}}
\(\de\sigma+\tzeta_\Lambda\de\zeta^\Lambda-\zeta^\Lambda\de\tzeta_\Lambda-\frac{4\I \sqrt{rK}}{|Z_\gamma|}\,\cC_\gamma
+8 r\Im \p\log\frac{Z_\gamma}{K} \).
\label{def-tcA}
\ee
Here $\sigma_\gamma$ is a sign factor, known as quadratic refinement, which satisfies the defining relation
$\sigma_{\gamma_1}\sigma_{\gamma_2}=(-1)^{\langle\gamma_1,\gamma_2\rangle}\sigma_{\gamma_1+\gamma_2}$,
and $\p=\de z^a\p_{z^a}$
is the Dolbeault holomorphic differential on the K\"ahler space parametrized by $z^a$.
Finally, $\OO(d\TT_\gamma)$ in \refb{square}
refers to terms proportional to $d\TT_\gamma$. As discussed in~\cite{Alexandrov:2021shf}, these terms
can be removed by field redefinition and cannot be determined by our analysis of string amplitudes.

To translate \eqref{square} to the type IIB formulation, one has to express it through the natural type IIB variables
described in the beginning of \S\ref{subsec-class}. This is the subject of the mirror map.
While its large volume, weak coupling limit is given by \refb{clmirmap}, it is known
however that it receives all possible quantum corrections {\it if we require the type IIB description to
be invariant under the S-duality transformations given in \refb{Sdualtr}} \cite{Alexandrov:2009qq,Alexandrov:2012bu}.
From the point of view of the type IIB theory, this amounts to a redefinition of the fields
appearing on the right hand side of \refb{clmirmap} and does not change the
S-matrix of the theory.
Since comparison with the string theory results are based on the S-matrix, such field redefinitions
are invisible in the string theory computation. Therefore, for our purposes we can safely use
the {\it classical} mirror map \refb{clmirmap} to express the type IIA results in terms of the variables of the type IIB theory.

Given \refb{Fcl} and \refb{clmirmap}, we find
\be
K =\frac43\, (t^3)= 8V\, ,
\qquad
\p K=-2\I \kappa_{ab}t^b \de z^a,
\qquad
\frac{\de r}{r}= \frac{2\de\tau_2}{\tau_2}+\frac{1}{2V}\, \kappa_{ab}t^a \de t^b,
\ee
\be
\begin{split}
\de \sigma + \tzeta_\Lambda \de \zeta^\Lambda - \zeta^\Lambda \de \tzeta_\Lambda
=&\,  \nabla\sigma,
\end{split}
\ee
\be
\begin{split}
\cC_\gamma=&\, \frac{\chq_0}{4V}\(\nabla\tc_0+\hf\, \kappa_{bc}t^c \nabla c^b\)
+\chq_a\(-\hf\,\kappa^{ab}\nabla\tc_b+\frac14\, t^a \de c^0\)
\\
&\, +\frac{(\chp t^2)}{8V} \(\nabla\tc_0+\hf\, \kappa_{bc}t^c \nabla c^b\)-\hf\,\chp^a \kappa_{ab}\nabla c^b
- \frac{\chp^0}{4}\( t^a\nabla\tc_a+ V \de  c^0 \) ,
\end{split}
\label{Cg-2b}
\ee
where the last relation is obtained with help of \eqref{FNN} and we used covariant derivatives and charges
defined in \refb{defnabla} and \eqref{def-qch}, respectively.
These results allow us to rewrite \eqref{square} as
\be
\de s_{\rm inst}^2= \sum_\gamma \frac{\Omega_\gamma\, \Sigma_\gamma}{8\tau_2^2 V
\sqrt{2\pi \cTR_\gamma}}\,\Bigl(\cA_\gamma^2+\cO(\de \cT_\gamma)\Bigr),
\label{square-2b}
\ee
where
\be\label{e320}
\cA_\gamma = \tcA_\gamma+\frac{\I}{\pi}\, \de\cT_\gamma
= \frac{\cTR_\gamma}{4\pi \tau_2^2 V}\,\nabla\sigma
+\frac{\I \cTR_\gamma}{\pi}\(\bar\p\log\frac{\bZ_\gamma}{K}+\frac{\de r}{2r}\) -4\I\,\cC_\gamma-2\de\Theta_\gamma\, ,
\ee
and we shifted this quantity by the irrelevant term proportional to $d\TT_\gamma$
to facilitate comparison with the amplitude calculation.
Substituting \eqref{hatgamTheta} and \eqref{Cg-2b} into \refb{e320},
one obtains the following explicit expression
\be
\begin{split}
\cA_\gamma =&\,   \frac{\cTR_\gamma}{4\pi \tau_2^2 V}\,\nabla\sigma
+\frac{\I \cTR_\gamma}{\pi}\(\frac{\de\tau_2}{\tau_2}-\frac{\I}{4V}\, \kappa_{ab}t^a \de b^b\)
+2\I\tau_2\, \frac{Z_\gamma}{|Z_\gamma|}\,\de\bZ_\gamma
\\
&\, - \(\frac{\I\chq_0}{V} + \frac{\I(\chp t^2)}{2V}+ 2 \chp^0\)\nabla\tc_0
+\( 2\I \kappa^{ab}\chq_b- 2 \chp^a+\I \chp^0t^a \)\nabla\tc_a
\\
&\,
+ \(2 \chq_a- {\I\over 4 V} \( 2 \chq_0 + (\chp t^2)\)\kappa_{ab} t^b
+  2\I\kappa_{ab}\chp^b\)\nabla c^a
- \(2 \chq_0 + \I \chq_a t^a  -\I  \chp^0 V\) \de c^0 \, .
\end{split}
\label{eagammafin}
\ee
The instanton contribution \eqref{square-2b} together with \eqref{eagammafin} is the prediction
to be reproduced by the world-sheet approach.
For future use we note that, as a consequence of \refb{ecentralcharge}, the factors $\cTR_\gamma$ and
$Z_\gamma/|Z_\gamma|$ in the presence of background $b^a$ may be obtained
using their expressions for $b^a=0$ and then replacing $\gamma$ by $\chgam$.

As discussed in \cite{Alexandrov:2021shf}, the quadratic refinement $\sigma_\gamma$
appearing in \refb{defSumk} can be changed by including constant shifts in the definition of the RR scalars,
leading to constant shift in $\Theta_\gamma$. Using this freedom, we can choose the sign factors $\sigma_\gamma$
arbitrarily for a linearly independent set of basis vectors $\gamma$.
For other charge vectors, they are fixed by the relation
$\sigma_{\gamma_1}\sigma_{\gamma_2}=(-1)^{\langle\gamma_1,\gamma_2\rangle}\sigma_{\gamma_1+\gamma_2}$
which is expected to follow from cluster property.
However, since our approximation keeps only multi-instantons wrapping the same cycle,
our analysis is not sensitive to this relation. On the other hand, in \cite{Alexandrov:2021shf}
we did show that cluster property implies the condition $\sigma_{k\gamma}=(\sigma_\gamma)^k$
and the result for the annulus amplitude given below is consistent with the choice $\sigma_{\gamma}=1$.
Therefore, in our analysis below we shall not discuss the factors $\sigma_\gamma$ any further.

We shall work in the large volume limit by scaling the fields as
\be
\tau_2, c^0\sim \lambda^0,
\qquad
t^a,b^a,c^a \sim \lambda^{1/3},
\qquad
\tc_a\sim \lambda^{2/3},
\qquad
\tc_0,\sigma\sim \lambda,
\ee
and taking $\lambda$ to be large. In this limit
all terms in the tree level metric \refb{metricBtreelargeV}
scale in the same way. If we
keep the charges fixed as we take this limit,
then we can see, using \refb{def-qch}, that the
contribution from the charge associated with brane of highest dimension dominates
in \refb{eagammafin}, {\it e.g.} when $p^0$ is non-zero then the terms proportional to
$p^0$ dominate, when $p^0=0$ but $p^a$ is non-zero then terms proportional to $p^a$ dominate and so on.
However, due to the transformation \eqref{chargesymp} of the charges under the
shift $b^a\to b^a+\eps^a$, it is more appropriate to scale the $\eps^a$'s appearing in
\eqref{chargesymp} by $\lambda^{1/3}$ as we take the large $\lambda$ limit.
This suggests the following scaling of the charges:
\be
\label{escalecharge}
\gamma=e^{-\lambda^{1/3}\bar f^a\omega_a}\,\bar\gamma
\ee
with the barred quantities kept fixed as we take the large $\lambda$ limit. We shall see in
\S\ref{s7.3x} that this limit arises naturally when we switch on gauge field strengths on the
D-branes and keep the magnitude of the field strength fixed as we take the large volume limit.
It is easy to verify that under the scaling \refb{escalecharge},
the contributions to \refb{eagammafin} (and also $\TT_\gamma$)
are still hierarchical, i.e. when $\bar p^0$ is non-zero then the terms proportional to
$\bar p^0$ dominate, when $\bar p^0=0$ but $\bar p^a$ is non-zero then terms
proportional to $\bar p^a$ dominate and so on.

\section{Strategy for computing instanton correction to the metric} \label{s4}

The general strategy that we shall follow for the computation of  D-instanton
corrections to the metric is the same as that in \cite{Alexandrov:2021shf}. For this
reason we shall now briefly recall these results. The computation involves
two parts. The first part is the computation of the exponential of the annulus amplitude.
After factoring out the integration over the collective modes, this determines the
normalization $\NN^{(0)}_{k,\gamma}$ of the $k$-instanton amplitude
with each instanton carrying charge $\gamma$. The second part of the
computation involves the disk amplitude with the insertion of a closed string vertex
operator representing a hypermultiplet scalar $\lambda^m$ and a
pair of open string vertex operators associated with the fermion zero modes
$\tilde\chi^\alpha$, $\tilde\chi^\dbeta$.  This amplitude has the form
\be \label{eterm}
\I \,a_{m,\gamma} \,
\gamma^\mu_{\dalpha\alpha} \, \tilde\chi^\alpha\tilde\chi^\dalpha \, p_\mu\lambda^m ,
\ee
where $a_{m,\gamma}$ is a computable constant, $p_\mu$ is the momentum carried by
$\lambda^m$ and we have used the convention that, while specifying an amplitude,
we also multiply it by the fields whose amplitude is being computed. In terms of these
quantities the instanton
correction to the metric takes the form \cite{Alexandrov:2021shf}:
\be\label{e75}
ds_{\rm inst}^2 = \sum_\gamma \Omega_\gamma\sum_{k=1}^\infty
\cN_{k,\gamma}\, e^{-k\TT_\gamma} \, \( \sum_m a_{m,\gamma} \, d\lambda^m\)^2\, ,
\qquad
\cN_{k,\gamma}=4\kappa^2 \, \bbV^{-1} \NN_{k,\gamma}^{(0)} .
\ee
Here $-\TT_\gamma$
is the action of a single D-brane of charge $\gamma$,
$\bbV$ is the volume of $\CY$ \eqref{defbbV} and $\kappa$ has been defined in \refb{ekappags}.
The real part $\cTR_\gamma$ of the instanton action is given by
\be
\cTR_\gamma=\ts_p |\bbV_\gamma|\, ,
\qquad
\ts_p ={1\over (2\pi)^p g_s}\, ,
\label{Realact}
\ee
where $\ts_p$ is the D$p$-brane tension and, for a D$p$-brane wrapping a cycle
$\Lgi{p+1}$, we define
\be \label{edefbbv}
\bbV_\gamma =\int_{\Lgi{p+1}} \vi{p+1}_\gamma\, ,
\ee
which can be expressed through the K\"ahler form using \eqref{defvn}.
By definition, $\bbV_\gamma$ measures the volume of the cycle $\Lgi{p+1}$,
but there are some caveats. First of all, note that \refb{edefbbv} changes sign under $\gamma\to-\gamma$
and is therefore not strictly positive. This is due to the fact that
under $\gamma\to -\gamma$, a D$p$-brane becomes an anti-D$p$-brane which has opposite intrinsic orientation,
but our definition of the volume form \eqref{defvn} remains intact.
This is why to get the volume of the cycle, we had to put the absolute value symbol in \refb{Realact}.
Second, we note that if $\gamma$ is not primitive, but has the form $l\gamma'$ for some positive integer $l$ and
some primitive vector $\gamma'$, then $\bbV_\gamma$
is defined as $l$ times the volume (with sign) of the cycle labelled by $\gamma'$. This is consistent with
\refb{Realact} since the action of an instanton is expected to be multiplied by $l$ when
the instanton charge is multiplied by $l$.

If the supersymmetric cycle associated with the charge $\gamma$ is rigid, i.e.\ has no
deformations other than the collective modes associated with the motion along
non-compact directions and the fermion zero modes associated with broken
supersymmetry, then the computation of $\NN^{(0)}_{k,\gamma}$ proceeds in the
same way as in \cite{Alexandrov:2021shf} and yields the result
\be\label{enkformula}
\NN^{(0)}_{k,\gamma}=g_{o,\gamma}\, 2^{-5}  \pi^{-13/2} \NP^{-1/2} \, ,
\ee
where $g_{o,\gamma}$ is the open string coupling on the instanton, determined from the relation
\be \label{e35xy}
\cTR_\gamma ={1\over 2\pi^2 g_{o,\gamma}^2} \quad \Rightarrow
\quad
g_{o,\gamma}^2=2(2\pi)^{p-2}\,\frac{g_s}{|\bbV_\gamma|}\, .
\ee
Therefore, we have
\be\label{e75new}
ds_{\rm inst}^2 =
\frac{\kappa^2}{ 2^{3}  \pi^{13/2}\bbV}
\sum_\gamma \Omega_\gamma \, g_{o,\gamma}\, \sum_{k=1}^\infty\, \NP^{-1/2}
\, e^{-k\TT_\gamma} \, \( \sum_m a_{m,\gamma} \, d\lambda^m\)^2  .
\ee
However, \refb{enkformula} needs to be modified for D(-1)-branes which have extra moduli
corresponding to motion along $\CY$. We shall determine this modified formula in
\S\ref{s5} where we shall also argue that this modification does not change the final formula \refb{e75new}
since it is accounted for by the DT invariant $\Omega_\gamma$ for pure D(-1)-charge.

\subsection{Normalization of the D(-1)-brane amplitudes} \label{s5}

When a D$p$-brane is wrapped on a rigid $p$-cycle, then the computation of the annulus
partition function, that determines the overall normalization of the amplitudes due to a
particular D-brane, follows from the analysis in \cite{Alexandrov:2021shf} and yields the
result \refb{enkformula}. However, this does not work for D(-1)-branes which have extra
moduli corresponding to motion inside $\CY$. In this section we shall determine this
extra contribution.

First we consider the case of a single
D(-1)-brane. This has six additional bosonic moduli describing the
location of the instanton on the Calabi-Yau space and twelve additional fermionic moduli representing
the superpartners of these bosonic modes. The twelve additional fermions can be grouped into 6
complex fermions, and furthermore, due to SU(3) holonomy of the Calabi-Yau manifold, their spinor
index can be traded for the tangent space vector index. We can represent the integral over these
additional modes as
\be \label{e3a.1}
\int\prod_{i=1}^6 \left\{{dm_i\over \sqrt{2\pi}}\, d\lambda_i \, d\lambda_i^*\right\}\,
 e^S\, ,
\ee
where $m_i$ are the coordinates of the Calabi-Yau manifold and the action is fixed by supersymmetry.
We can read this out {\it e.g.} from the dimensional reduction of the supersymmetric $\sigma$-model
in two dimensions \cite{AlvarezGaume:1983at}
to be:
\be \label{efermaction}
S = -{1\over 4} R_{ijkl} \, \lambda_i^*  \lambda_j^* \lambda_k \lambda_l\, .
\ee
The integral \refb{e3a.1} may be expressed as
\be \label{eeuler}
-\int\prod_{i=1}^6 {dm_i\over \sqrt{2\pi}}\, {1\over 3!} \, {1\over 2^6} \,
\eps^{i_1j_1i_2j_2i_3j_3} \eps^{k_1l_1k_2l_2k_3l_3} R_{i_1j_1k_1l_1}
R_{i_2j_2k_2l_2} R_{i_3j_3k_3l_3} =-\chi\, ,
\ee
where $\chi$ is the Euler number of the Calabi-Yau manifold.
Therefore, \refb{enkformula} for $k=1$ is modified to
\be\label{e1.9rep}
 \NN_{1,\gamma^0}^{(0)} = -\chi \,  g_{o,\gamma^0}\, 2^{-5} \, \pi^{-13/2}\, .
\ee
The sign in \refb{eeuler} is ambiguous since it depends on the sign of the integration measure
over the fermion zero modes. In principle the sign can be fixed using cluster property as in
appendix C of \cite{Alexandrov:2021shf}.
However, we have not done this. Instead, we have used the information
that the DT invariant associated to pure D(-1) charges is given by $-\chi$ (see footnote \ref{foot-DT}).

Next we turn to $k$-instanton amplitudes. We shall now argue that for computing the
ratio $ \NN_{k,\gamma^0}^{(0)}/ \NN_{1,\gamma^0}^{(0)}$ we can use the results in
ten-dimensional flat spacetime. The essence of the argument is as follows. Since the
D-instantons move in ten-dimensional spacetime, the extra variables that we encounter
for $k$-instantons are clearly the same whether they move in flat spacetime or the Calabi-Yau
space. Furthermore, it was shown in \cite{Alexandrov:2021shf} that the
matrix model governing the dynamics of
multiple branes restricts the relative separation between the branes to a distance scale of order
$(g_{o,\gamma^0})^{1/2}\sim g_s^{1/4}$. Therefore as long as $\CY$ has size of the order of
string scale or larger, the dynamics of the relative separation between the branes is
insensitive to the curvature of $\CY$ in the weak coupling limit
and we can use the flat spacetime results of
\cite{Sen:2021jbr}. This gives:\footnote{If we use eqs.(10) and (25) of \cite{Sen:2021jbr}, we
shall get the ratio to be $k^{9/2} \sum_{d|k}{1/ d^2}$.
However, $1/\sqrt k$ in the Chan-Paton factor of fermionic vertex
operators now produce a factor of $1/k$ in \refb{efermaction} and hence $1/k^3$ in \refb{eeuler}.
This explains the power $k^{9/2}/k^3=k^{3/2}$ in \refb{eextra11}.
}
\be \label{eextra11}
{\NN^{(0)}_k\over \NN^{(0)}_1} = \NP^{3/2}\sum_{d|k}{1\over d^2} \quad
\Rightarrow \quad \NN^{(0)}_k= -\chi \, g_{o,\gamma^0}\, 2^{-5} \, \pi^{-13/2}\,
\NP^{3/2}\sum_{d|k}{1\over  d^2}
\, .
\ee
Substituting this into \refb{e75} without the $\Omega_\gamma$ factor,
we get the following correction to the
metric due to multiple D(-1)-branes:
\be \label{enn1}
- {4\kappa^2\over \bbV}\,
\chi \, g_{o,\gamma^0}\, 2^{-5} \, \pi^{-13/2} \sum_{k=1}^\infty
\NP^{3/2}\sum_{d|k}{1\over d^2}\,
e^{-k\TT_{\gamma^0}} \( \sum_m a_{m,\gamma^0} \, d\lambda^m\)^2\, .
\ee

We now claim that, taking into account that $\Omega_{q_0\gamma^0} = -\chi$,
this result agrees with the contribution to \refb{e75new} from charges of the form $\gamma=\ell\gamma^0$ with $\ell\in\IZ^+$.
Indeed, using the relabelling $d=n$, $k=n\ell$, one finds
\be
\sum_{k=1}^\infty\NP^{3/2}\sum_{d|k}{1\over d^2}\,e^{-k\TT_{\gamma^0}} =
\sum_{\ell=1}^\infty \ell^{3/2}\sum_{n=1}^\infty n^{-1/2} e^{-n\ell\TT_{\gamma^0}}.
\ee
Hence, one can rewrite \eqref{enn1} as
\be \label{enn2}
\frac{\kappa^2}{ 2^{3}  \pi^{13/2}\bbV}
\sum_{\ell=1}^\infty\Omega_{\ell\gamma^0} \,\ell^{-1/2}g_{o,\gamma^0}\sum_{k=1}^\infty k^{-1/2} e^{-k\ell\TT_{\gamma^0}}
\( \sum_m \ell\, a_{m,\gamma^0} \, d\lambda^m\)^2\, .
\ee
It remains to take into account that due to $\TT_{\ell\gamma^0}= \ell \, \TT_{\gamma^0}$ and \refb{e35xy}, one has
$g_{o,\ell\gamma_o}=\ell^{-1/2} g_{o,\gamma_o}$, while the disk one-point function satisfies
$a_{m,\ell\gamma^0}=\ell \, a_{m,\gamma^0}$ because it
has an overall multiplicative factor of $\cTR_\gamma$.
Therefore, we can express \refb{enn2} as
\be \label{enn2new}
\frac{\kappa^2}{ 2^{3}  \pi^{13/2}\bbV}
\sum_{\ell=1}^\infty\Omega_{\ell\gamma^0} \, g_{o,\ell\gamma^0}\sum_{k=1}^\infty k^{-1/2} e^{-k\TT_{\ell\gamma^0}}
\( \sum_m a_{m,\ell\gamma^0} \, d\lambda^m\)^2\, .
\ee
This can now be interpreted as part of the contribution to \refb{e75new}, with $\gamma$ taking
values $\ell\gamma^0$.

In our analysis we have assumed that the 2 and 4-cycles of the CY are rigid so that
D1 and D3-branes wrapped on these cycles have no extra moduli. When these conditions
fail, there will be extra integrals in these cases as well.
Our result for D(-1)-branes suggests that such extra contributions
will also arise in the counting of BPS
index of D1 and D3-branes wrapped on these cycles and will be included in the definition
of the DT invariants $\Omega_{\gamma}$ so that \refb{e75new} will continue
to hold in these cases as well.

\section{Explicit computation of D$p$-brane corrections} \label{sdminus}

Our main goal is to reproduce the prediction \refb{square-2b} for the instanton contribution to the metric on $\cM_H$
by explicit world-sheet computation. In this section we will reproduce the `linear terms' in
$\AAA_\gamma$, namely those that survive when we set the background NSNS 2-form
and RR fields to zero. In this case the $b$-dependent charges $\chq_0$, $\chq_a$ and $\chp^a$
defined in \eqref{def-qch} reduce to
$q_0$, $q_a$ and $p^a$ and all covariant derivatives \eqref{defnabla} are replaced by ordinary
ones. We shall further assume that only one type of charge is present,
corresponding to D$p$-brane wrapped on a $(p+1)$-cycle $\Lgi{p+1}$ for fixed $p$.
In \S\ref{s7} we shall relax these assumptions.

The following equations will be used for the computation of the disk
amplitudes \cite{Alexandrov:2021shf}:
\be\label{edisk1}
\{ V_c\}= {1\over 2} \, \kappa \, T_p \, \langle c_0^- V_c \rangle\, ,
\ee
and
\be\label{edisk2}
\left\{V_c  \prod_{k=1}^n V_o^{(k)} \right\} =
\I\, \pi\, \kappa\, T_p \,  \int \left\langle V_c  \prod_{k=1}^n V_o^{(k)} \right\rangle\, .
\ee
Here $V_c$ and $V_o^{(k)}$ denote closed and open string vertex operators, respectively,
$k$ labels different operators, and $\langle~\rangle$ denotes correlation
function on the upper half plane.
Note that the formula takes a different form when no open string
vertex operators are present --- this can be  traced to the presence of a conformal Killing
vector on the disk with just one closed string insertion. On the right hand side of
\refb{edisk2} the integration is performed over the locations of the $(n-1)$ open string
vertex operators along the real axis.

The main tools in our analysis will be the operator product expansion \refb{eqb1} and the
boundary conditions \refb{boundarycond},
\refb{SSbc} that allow us to replace the anti-holomorphic fields in the
upper half plane by the holomorphic fields at the complex conjugate points.
The final correlation function will be evaluated using the normalization condition
\refb{eopennorm}.

\subsection{NSNS axion contribution} \label{s7.1}

The analysis of this contribution
is almost identical to the D2-brane contribution computed in \cite{Alexandrov:2021shf}
once we replace the tension $\ts_{2}$ by $\ts_{p}$.

The disk amplitude with one 2-form vertex operator and a pair of open string zero modes
in the presence of a flat  Euclidean D$p$-brane is given by:
\be \label{e2.1}
\I \pi\kappa \, \ts_{p} \, 2\, b_{\mu\nu} \, \XX^\alpha\wh\XX^\beta
\int_{-\infty}^\infty dz \left\langle
c\, \bar c \(\p X^{\mu} + \I\, p_\rho\, \psi^\rho\psi^\mu\)
e^{\I p.X}
e^{-\bar\phi}\bar\psi^\nu(\I) \, c e^{-\phi/2} S_\alpha(0) \, e^{-\phi/2} S_\beta(z)
\right\rangle .
\ee
Here we are using the same world-sheet notations as in \cite{Alexandrov:2021shf}.
Dropping $\partial X^\mu$ and $e^{ip.X}$ terms whose contributions
vanish due to the physical state condition,
and using the doubling trick to replace the anti-holomorphic
fields by holomorphic ones at the complex conjugate point, this becomes
\be\label{ecal1}
-2 \I \pi\, \kappa\, \ts_{p} \,  b_{\mu\nu} \,\XX^\alpha\wh\XX^\beta
\int_{-\infty}^\infty dz \, \Bigl\langle  \I \,
c \, p_\rho\psi^\rho\psi^\mu (\I)\,  c\, e^{-\phi}\psi^\nu(-\I)
\, c\, e^{-\phi/2} S_\alpha(0) \, e^{-\phi/2} S_\beta(z) \Bigr\rangle\, .
\ee
The minus sign arises due to Dirichlet boundary conditions on $\bar\psi^\nu$.
As in \cite{Alexandrov:2021shf}, we shall take the $z$ integration
contour to pass above the origin.
We now deform it to pick up the residue from the pole at $\I$.
The effect of this is to drop the integration over $z$ together with the factor
$e^{-\phi/2} S_\beta(z)$ inside the correlator and replace the factor
$\psi^\rho\psi^\mu(\I)$
by $(\I\pi/2) \left(\Gamma^{\rho\mu}\right)_\beta^{\ \gamma} e^{-\phi/2} S_\gamma (\I)$.
The resulting correlation function can be easily evaluated and gives:
\be\label{e610}
{1\over 2}\, \pi^2\kappa \, \ts_{p} \,   p_\rho\, b_{\mu\nu} \,  \XX^\alpha\wh\XX^\beta\,
(\Gamma^{\rho\mu}\Gamma^{\nu})_{\beta\alpha} \, (2\pi)^{p+1} \delta^{(p+1)}(0)\ \rightarrow\
{1\over 2}\, \pi^2\kappa \, \cTR_\gamma \,   p_\rho\, b_{\mu\nu} \,  \XX^\alpha\wh\XX^\beta\,
(\Gamma^{\rho\mu\nu})_{\beta\alpha}\, ,
\ee
where in the second step we have interpreted the $(2\pi)^{p+1}  \delta^{(p+1)}(0)$ as
integration over the D$p$-brane world-volume yielding a factor of $|\bbV_\gamma|$
and used \refb{Realact} to express
$\ts_p |\bbV_\gamma|$ as $\cTR_\gamma$.
In terms of four-dimensional spinors the above expression becomes
\be
{1\over 2}\, \pi^2\kappa \, \cTR_\gamma \,   p_\rho\, b_{\mu\nu} \,  \tilde\chi^\alpha\tilde\chi^\dalpha\,
(\gamma^{\rho\mu\nu})_{\dalpha\alpha}\, .
\label{ampl-diskB}
\ee
Following the same steps as in \cite{Alexandrov:2021shf},
we can express this in terms of a dual scalar $\bbsigma$
(called $\tilde\sigma$ in \cite{Alexandrov:2021shf}) using the relation
\be \label{ehpsi}
p_\mu b_{\nu\rho} + p_\nu b_{\rho\mu}+p_\rho b_{\mu\nu}
={1\over 2 \bbV}\, \eps_{\mu\nu\rho\tau}\, p^\tau\bbsigma\, ,
\ee
where $\bbV$ is the volume of the CY manifold given in \eqref{defbbV}.
In terms of $\bbsigma$, the expression \refb{ampl-diskB} takes the form
\be\label{eq:d-1-dual-s-amp}
\begin{split}
{1\over 2}\,  \pi^2 \kappa \,\cTR_\gamma \,   {1\over 6\bbV}\, {\eps_{\rho\mu\nu}}^{\tau} \, p_\tau\bbsigma\,
(\gamma^{\rho\mu\nu})_{\dalpha\alpha}\, \tilde\chi^\alpha \tilde\chi^\dalpha
=&\,
-{\I\pi^2 \over 2\bbV}\,  \kappa \, \cTR_\gamma\, p_\mu \bbsigma\, (\gamma^\mu)_{\dalpha\alpha}
\, \tilde\chi^\alpha \tilde\chi^\dalpha
\\
=&\, {\I\pi^2 \over 4V\tau_2^2}\,  \cTR_\gamma \, p_\mu \sigma\, (\gamma^\mu)_{\dalpha\alpha}
\, \tilde\chi^\alpha \tilde\chi^\dalpha\, ,
\end{split}
\ee
where in the last step we have used relations \eqref{eccnet}.
Comparing this with \refb{eterm}, we get
\be
a_{\sigma,\gamma}d\sigma
= \frac{\pi^2}{4V\tau_2^2}\, \cTR_\gamma \de\sigma \, .
\ee

\subsection{Dilaton contribution} \label{sdilaton}

We shall now compute the coefficient $a_{m,\gamma}$
associated with the dilaton field $\tau_2$.
As in \cite{Alexandrov:2021shf}, we start with the dilaton vertex operator in the
$(-1,-1)$ picture which, at non-zero
momentum, is given by
\be \label{ebrstoriginal}
\wt V_{-1,-1}= f_{\mu\nu}\,  c\, \bar c\, e^{-\phi}\, \psi^\mu \, e^{-\bar\phi} \bar\psi^\nu \, e^{\I p.X}
\, ,
\ee
where
\be \label{emunuform}
f_{\mu\nu} \propto \(\eta_{\mu\nu} - (n.p)^{-1} \, \(n_\mu p_\nu + n_\nu p_\mu\)\) ,
\ee
and $n$ is any four-vector for which $n.p\ne 0$. Our first step will be to determine the
normalization of $f_{\mu\nu}$. For this we compute the disk one-point
function of the dilaton on a D(-1)-brane. This is given by:
\be
{\kappa\, \ts_{-1}\over 2} \, {1\over 2} \,\bigl\langle
\(\p c(\I)-\bar\p\bar c(\I)\) \wt V_{-1,-1}(\I)\bigr\rangle
= -{\kappa\, \ts_{-1}\over 4}\, f_\mu^{~\mu} \, .
\ee
Let $\tilde \tau_2=\tau_2 \sqrt \bbV$ be the inverse of the four-dimensional string coupling.
Then $f_\mu^{~\mu}$ must be proportional to $\delta\tilde\tau_2/\tilde\tau_2$.
On the other hand, since $\ts_{-1}=2\pi \tau_2$ we see that the expected
one-point coupling of $\delta \tau_2$ is $-\ts_{-1}\delta \tau_2/\tau_2$.
Therefore, we choose
\be \label{edilnorm}
f_\mu^{~\mu} = {4\over \kappa} {\delta \tilde\tau_2\over \tilde\tau_2}
= {4\over \kappa} \({\delta \tau_2\over \tau_2} +{\delta \bbV\over 2\bbV}\).
\ee
Once the normalization has been fixed this way, we can use it to compute the dilaton amplitude on
any D$p$-brane.

We shall now
analyze the disk amplitude with one dilaton and a pair of open string
fermion zero modes on a D$p$-brane. It can be found using the dilaton vertex operator in the $(-1,0)$ picture
\be
V_{-1,0} = - f_{\mu\nu} \,
c\, \bar c\, \left\{\p X^{\mu} + i\, p_\rho\, \psi^\rho\psi^\mu\right\}
e^{\I p.X}
e^{-\bar\phi}\bar\psi^\nu(\I) +\cdots\, .
\ee
This operator has the same form as the NSNS axion vertex operator.
Therefore, the analysis of the amplitude follows exactly the same route as in
\S\ref{s7.1} and
leads us to the left hand side of \refb{e610} with $b_{\mu\nu}$ replaced by $-f_{\mu\nu}/2$:
\be
-{1\over 4}\, \pi^2\kappa \, \ts_{p} \,   p_\sigma\, f_{\mu\nu} \,  \XX^\alpha\wh\XX^\beta\,
(\Gamma^{\sigma\mu}\Gamma^{\nu})_{\beta\alpha} \, (2\pi)^{p+1}\delta^{(p+1)}(0)
\ \rightarrow\
-{1\over 4}\, \pi^2\kappa \, \cTR_\gamma\,   p_\sigma\, f_{\mu\nu} \,  \tilde\chi^\alpha\tilde\chi^\dalpha\,
(\gamma^{\sigma\mu}\gamma^{\nu})_{\dalpha\alpha}\, .
\ee
Expressing
$(\gamma^{\sigma\mu}\gamma^{\nu})$ as $\gamma^{\sigma\mu\nu}+\gamma^\sigma
\eta^{\mu\nu}- \gamma^\mu\eta^{\sigma\nu}$, and using the form of $f_{\mu\nu}$
given in \refb{emunuform}, we see that only the $\gamma^\sigma
\eta^{\mu\nu}$ term contributes. Finally, using \refb{edilnorm},
the resulting expression may be written as:
\be
- \pi^2 \, \cTR_\gamma \,   p_\sigma  \({\delta \tau_2\over \tau_2} +{\delta \bbV \over 2\bbV }\)
(\gamma^{\sigma})_{\dalpha\alpha}\, \tilde\chi^\alpha\tilde\chi^\dalpha\,.
\label{edilfin}
\ee
Comparing this with \refb{eterm} and using  \refb{eccnet}, we obtain
\be \label{earfindp}
(a_{m,\gamma} \de\lambda^m)_{\rm dilaton} =   {\I\pi^2\over \tau_2} \, \cTR_\gamma
\(d\tau_2 + {\tau_2\over 2 V} \,dV\) .
\ee

\subsection{K\"ahler and B-field moduli contribution}

A general closed string  vertex operator in the $(0,-1)$ picture, polarised in the internal CY directions, has the form
\be
V_e = 2\, e_{\bf ij} \,
c\, \bar c \(\p X^{\bf i} + i\, p_\rho\, \psi^\rho\psi^{\bf i}\)
e^{\I p.X}
e^{-\bar\phi}\bar\psi^{\bf j}(\I) +\cdots\, ,
\ee
so that the amplitude is given by
\be\label{ens1}
2 \I \pi \kappa\, \ts_{p}\, e_{\bf ij} \, \XX^\alpha \wh\XX^\beta
\int dz \left\langle c\bar c \(\p X^{\bf i} + \I\, p_\rho\, \psi^\rho\psi^{\bf i}\)
e^{\I p.X}
e^{-\bar\phi}\bar\psi^{\bf j}(\I)\,  c e^{-\phi/2} S_\alpha(0)\,
e^{-\phi/2}S_\beta(z)\right\rangle .
\ee
We can drop the $\p X^{\bf i}$ and $e^{\I p.X}$ terms since they do not contribute to the
correlation function. Also, using the doubling trick, we can replace the
$\bar c\, e^{-\bar\phi}\bar\psi^{\bf j}(\I)$ term by  $(P^{\bf j}_{\bf k}-Q^{\bf j}_{\bf k})\,
c \, e^{-\phi} \psi^{\bf k}(-\I)$, where $P$ and $Q$ are respectively projection operators
tangential and transverse to the brane. This leads to,
\be
2 \I \pi \kappa\, \ts_{p}\, \I\, p_\rho\, \XX^\alpha  \wh\XX^\beta\,
e_{\bf ij} (P^{\bf j}_{\bf k}-Q^{\bf j}_{\bf k})
\int dz \,\Bigl\langle c \, \psi^\rho\psi^{\bf i} (\I)\,
c\, e^{-\phi}\psi^{\bf k}(-\I)\,  c e^{-\phi/2} S_\alpha(0)\,
e^{-\phi/2}S_\beta(z)\Bigr\rangle\, .
\ee
Picking up the residue at $z=\I$, evaluating the
remaining correlator and expressing the result in terms of four-dimensional spinors, we get
\be \label{ee3.7}
-{1\over 2}\, \pi^2 \kappa\, \ts_{p}\,
p_\rho\,  \tilde\chi^\alpha \tilde\chi^\dbeta\,\gamma^\rho_{\dbeta\alpha}\, e_{\bf ij}  \,
(P^{\bf j}_{\bf k}-Q^{\bf j}_{\bf k})
\,\bar \eta\tilde \Gamma^{\bf i} \tilde \Gamma^{\bf k}\eta\, (2\pi)^{p+1}\delta^{(p+1)}(0) \, .
\ee
Since $\bar \eta\tilde \Gamma^{\bf i} \tilde \Gamma^{\bf k}\eta= \bbg^{{\bf i}{\bf k}}+\I\bbom^{{\bf i}{\bf k}}$
and $Q^{\bf j}_{\bf k} =\delta^{\bf j}_{\bf k} - P^{\bf j}_{\bf k}$,
we have
\be
e_{\bf ij}(P-Q)^{\bf j}_{\bf k}\, \bar \eta\tilde \Gamma^{\bf i} \tilde \Gamma^{\bf k}\eta
=2e_{\bf ij}P^{\bf j}_{\bf k}\,\(\bbg^{{\bf i}{\bf k}}+\I\bbom^{{\bf i}{\bf k}}\)
-\(e^{\bf i}_{\bf i}+\I e_{\bf ij}\bbom^{{\bf i}{\bf j}}\).
\label{comb-ePQ}
\ee

The second contribution can be easily evaluated.
Indeed, using \refb{egmnhmn} which gives
\be\label{ebijdef}
e_{\bf ij}=\frac{\delta \bbg_{\bf ij}}{2\kappa} + \delta \bbB_{\bf ij},
\ee
this term can be written as
\be \label{eginter}
-\frac{1}{2\kappa} \delta \bbg^{\bf i}_{\bf i}
-\I \delta \bbB_{\bf ij}\,\bbom^{{\bf i}{\bf j}}
=-\frac{\I}{\kappa}\delta \bbg_{s\bt}\, \bbom^{s\bt} - 2\I\, \delta \bbB_{s\bt} \,\bbom^{s\bt}\, ,
\ee
where we have used $\bbom^{s\bar t}= - \I  \bbg^{s\bar t}$ in the holomorphic indices $s,t$.
Using \refb{e626a}, \refb{eb7} and \refb{eccnet}, we can rewrite \refb{eginter} as,
\be
-\frac{\I}{2\kappa \bbV}\, \kappa_{abc} \bbt^b \bbt^c\(\delta \bbb^a-\I\delta \bbt^a\)=
-{\I \over 2\kappa V}\, \delta b^a \kappa_{abc} t^b t^c - {1\over \kappa} \, {\delta V\over V}\, .
\ee
This result should still be integrated over the wrapped cycle producing the factor
$|\bbV_\gamma|$. Using \refb{Realact}, this gives the following contribution to \refb{ee3.7}:
\be \label{eonecon}
-{1\over 2}\, \pi^2 \kappa \,\cTR_\gamma p_\rho\,  \tilde\chi^\alpha \tilde\chi^\dbeta\,
\gamma^\rho_{\dbeta\alpha} \(-{\I \over 2\kappa V}\, \delta b^a \kappa_{ab} t^b  - {1\over \kappa} \, {\delta V\over V}\) .
\ee

To evaluate the first term on the r.h.s. of \eqref{comb-ePQ}, we use the BPS condition \eqref{defvn}
which implies that the projector $P_\gamma$
on the wrapped cycle (replacing $P$ in the above formul\ae) satisfies
$P_\gamma^{st}=P_\gamma^{\bs\bt}=0$. Thus, we get
\be
\begin{split} \label{eomegaproject}
2e_{\bf ij}P^{\bf j}_{\gamma,\bf k}\,\(\bbg^{{\bf i}{\bf k}}+\I\bbom^{{\bf i}{\bf k}}\)
=&\, 2 e_{\bf ij} P_\gamma^{\bf ij}+2(e_{s\bt}-e_{\bt s})P_\gamma^{s\bt}
= \frac{2}{\kappa}\(\delta \bbg_{s\bt}+2\kappa\,\delta \bbB_{s\bt} \)P_\gamma^{s\bt}
\\
=&\, \frac{2}{\kappa}\(\delta \bbb^a-\I \delta \bbt^a \)\omega_{a,s\bt} P_\gamma^{s\bt}=
{8\pi^2\over \kappa} \, \delta \bz^a \, \omega_{a,s\bt} P_\gamma^{s\bt},
\end{split}
\ee
where we expressed the result in terms of complexified K\"ahler moduli
$\bz^a\equiv b^a-\I t^a= {1\over 4\pi^2} (\bbb^a-\I\bbt^a)$.
For $p=-1$, $\omega_a$ has no component along the directions tangential to the brane
and \refb{eomegaproject} vanishes.
For $p\ge 1$, the contribution of \refb{eomegaproject} to \refb{ee3.7} takes the form:
\be \label{efirstterm}
- 4\pi^4 \,\ts_{p}\,p_\rho\,  \tilde\chi^\alpha \tilde\chi^\dbeta\,\gamma^\rho_{\dbeta\alpha}\, I_p \, ,
\ee
where
\be\label{eorient}
\begin{split}
I_p=&\, \delta\bz^a\,  {\bbV_\gamma\over |\bbV_\gamma|}
\int_{\Lgi{p+1}}\vn_\gamma\, \omega_{a,s\bt} P_\gamma^{s\bt}
=
\frac{\I \delta\bz^a}{\(\frac{p-1}{2}\)!}\, {\bbV_\gamma\over |\bbV_\gamma|}
\int_{\Lgi{p+1}}\omega_a\wedge \bbom^{\frac{p-1}{2}},
\end{split}
\ee
and we used \refb{eyintegral} and \eqref{relform}.
Note the factor
\be
{\bbV_\gamma\over |\bbV_\gamma|}={ \int_{\Lgi{p+1}}\bbom^{\frac{p+1}{2}}\over |\int_{\Lgi{p+1}} \bbom^{\frac{p+1}{2}}|}
\ee
that is needed to ensure that we choose the
correct sign of the volume form.
For D1-brane wrapped on $\Lgi{2}$, one obtains from \refb{eorient}
\be\label{intD1}
I_1=\I\delta\bz^c\, {(-q_a \bbt^a) \over |q_b \bbt^b|}\int_{\Lgi{2}}\omega_c
=\I (\delta\bz^c q_c)\,
{q_a \bbt^a \over |q_b \bbt^b|}\, .
\ee
For D3-brane wrapped on $\Lgi{4}$, one obtains
\be
\begin{split}
I_3=\I\delta\bz^a\, {-(p\bbt^2)\over|(p\bbt^2)|} \int_{\Lgi{4}}\omega_a\wedge \bbom
=&\,-\I\, {(p\bbt^2)\over|(p\bbt^2)|} \,
\delta\bz^a\int_{\CY}\omega_a \wedge \bbom\wedge (-p^{b} \omega_{b})
\\
=&\, {\I(p\bbt^2)\over |(p\bbt^2)|}\, \kappa_{abc}\delta\bz^a p^b \bbt^c \, ,
\end{split}
\label{intD3}
\ee
where $-p^a\omega_a$ is the Poincar\'e dual form to $\Lgi{4}$.
Finally, for D5-brane wrapped on $p^0\CY$, one finds
\be\label{intD5}
I_5= \frac{\I p^0}{2}\,\delta\bz^a\, {p^0\over |p^0|}\int_{\CY}\omega_a\wedge\bbom\wedge \bbom
=\frac{\I |p^0|}{2}\,\kappa_{abc}\delta\bz^a \bbt^b \bbt^c.
\ee
We shall be considering D-branes carrying only one type of charge so that
only one of the terms is present.
Using \eqref{edefzgamma} and \refb{eccnet},
all three contributions \refb{intD1}, \refb{intD3} and \refb{intD5}, as well
as the vanishing result for $p=-1$, can be written
as
\be\label{eq:4pisq}
I_p= (2\pi)^{p-1} \, {Z_\gamma\over |Z_\gamma|} \, \delta\bZ_\gamma\, ,
\ee
for vanishing background $b^a$.
Substituting this into \refb{efirstterm}, we get
the contribution from the first term on the r.h.s.\ of \refb{comb-ePQ} to \refb{ee3.7}:
\be \label{etwocon}
-\pi^2(2\pi)^{p+1} \, \ts_{p}\,p_\rho\,  \tilde\chi^\alpha \tilde\chi^\dbeta\,
\gamma^\rho_{\dbeta\alpha}\, {Z_\gamma\over |Z_\gamma|}
\, \delta\bZ_\gamma\, .
\ee
Adding \refb{eonecon} and \refb{etwocon} and using \refb{eterm}, \refb{Realact}, we get
\be\label{eq:fin-a-kb}
\begin{split}
a_{t^a,\gamma} dt^a+a_{b^a,\gamma} db^a
=&\, \pi^2 \cTR_\gamma\(\frac{1}{4V}\, \kappa_{ab} t^b \de b^a-\frac{\I\de V}{2V} \)
+2\I \pi^3 \tau_2\, {Z_\gamma\over |Z_\gamma|} \,\de\bZ_\gamma.
\end{split}
\ee

\subsection{RR contributions} \label{srrcont}

Using the $(-1/2,-1/2)$ picture vertex operator for RR fields given in appendix \ref{sc},
the disk amplitude
with one RR closed string field with polarization
$F^{\gamma\delta}$ and two open string zero mode fields $\XX^\alpha$, $\wh\XX^\beta$
for a general D$p$-brane can be expressed as
\be\label{eq:rr-2zm-disc-start}
\I \pi \kappa\, \ts_{p} \, F^{\gamma\delta} \, \XX^\alpha\, \wh\XX^\beta\,
\int dz \, \Bigl\langle c\bar c e^{-\phi/2} S_\gamma e^{-\bar\phi/2} \bar S_\delta(\I)\, c \,e^{-\phi/2} S_\alpha(0)\,
e^{-\phi/2}S_\beta(z)\Bigr\rangle\, .
\ee
We shall take the $z$ contour to lie above the origin as before.
Using the boundary condition on the real line described in \refb{SSbc}
and the doubling trick, we represent the closed string vertex operator as
\be\label{eq:D1-RR-bc}
{\I\over (p+1)!}\, c\, e^{-\phi/2} \, S_\gamma(\I) \, c\, e^{-\phi/2} \,
v_{I_1\cdots I_{p+1}}(\Gamma^{I_1\cdots I_{p+1}})^{\delta\delta'}
S_{\delta'}(-\I)\, ,
\ee
where $v$ is the volume form along the brane.
We can now calculate the correlation function by first deforming the $z$ contour to pick up the residue
at $\I$ using the following operator product expansion
\be
e^{-\phi/2} S_\gamma(w)\, e^{-\phi/2} S_\beta(z) = \I\, (w-z)^{-1} \, (\Gamma^M)_{\gamma\beta}
e^{-\phi}\psi_M(w) +\dots\, .
\ee
The operator at $\I$ then becomes $2\pi\, c\, e^{-\phi} \psi_M (\Gamma^M)_{\beta\gamma}$.
We are now left with a $\psi$-$S$-$S$ correlator that gives a term proportional to
$(\Gamma_M)_{\delta'\alpha}$. The result is
\be\label{eq:ge-rr-disc-2zmD1}
{\I\over (p+1)!}\, \pi^2 \kappa\, \ts_{p} \, F^{\gamma\delta}\,
\XX^\alpha \wh\XX^\beta\, (\Gamma^M)_{\beta\gamma}  \, v_{I_1\cdots I_{p+1}}
(\Gamma^{I_1\cdots I_{p+1}}\Gamma_M)_{\delta\alpha}\,  (2\pi)^{p+1}\delta^{(p+1)}(0) \, .
\ee

Our interest will be in the couplings proportional to $p_\mu\bbC^{(2k)}_{\bf m_1\cdots m_{2k}}$.
For this we pick the following term in the expansion of $F^{\gamma\delta}$ given in \refb{ebb7}, taking into account
the doubling of the multiplicative factor due to the self-duality constraint \refb{edualcon}:
\be \label{e541a}
{\I\over (2k)!} \bbF^{(2k+1)}_{\mu\bf m_1\cdots m_{2k}} (\Gamma^{\mu\bf
m_1\cdots m_{2k}})^{\gamma\delta}
= {\I\over (2k)!} \, \I  p_\mu \bbC^{(2k)}_{\bf m_1\cdots m_{2k}} (\Gamma^\mu
\Gamma^{\bf
m_1\cdots m_{2k}})^{\gamma\delta}.
\ee
Also since the brane lies along a subspace of $\CY$, $v$ is replaced by the volume form
$v^{(p+1)}_\gamma$ on the $(p+1)$-cycle $\Lgi{p+1}$ wrapped by the D$p$-brane with
only internal components. Finally, the $(2\pi)^{p+1}\delta^{(p+1)}(0)$ term has to be interpreted
as integration along this cycle.
Substituting \refb{e541a} into \refb{eq:ge-rr-disc-2zmD1} and using the expression for the
tension from \eqref{Realact}
and the last relation in \eqref{eccnet},
we get the relevant term to be
\be\label{emasterd1pre}
-  {\pi^2 (2\pi)^{2k-p}\over 8(p+1)!}\, {\I p_\mu\over (2k)!}\,
\XX^\alpha \wh\XX^\beta \int_{\Lgi{p+1}} v^{(p+1)}_{\gamma}\,
 (v^{(p+1)}_{\gamma})_{\bf i_1\cdots i_{p+1}}
\, C^{(2k)}_{\bf m_1\cdots m_{2k}}
(\Gamma^M   \Gamma^\mu \Gamma^{\bf m_1\cdots m_{2k}}
\Gamma^{\bf i_1\cdots i_{p+1}}\Gamma_M)_{\beta\alpha} \, .
\ee
Due to the presence of two factors of $v^{(p+1)}_{\gamma}$, we do not need to include
the compensating sign that accompanied \refb{eorient}.
By splitting the sum over $M$ into non-compact and compact directions, and
using \refb{eff1}, we can write
\be
\Gamma^M   \Gamma^\mu \Gamma^{\bf m_1\cdots m_{2k}}
\Gamma^{\bf i_1\cdots i_{p+1}}\Gamma_M
= \gamma^\mu \Gamma^{\bf m}   \wt \Gamma \Gamma^{\bf m_1\cdots m_{2k}}
\Gamma^{\bf i_1\cdots i_{p+1}}\Gamma_{\bf m}
- 2 \gamma^\mu  \wt \Gamma \Gamma^{\bf m_1\cdots m_{2k}}
\Gamma^{\bf i_1\cdots i_{p+1}}\, .
\ee
Substituting this into \refb{emasterd1pre}, using four-dimensional spinor notation
\refb{etendfourd} and $\wt\Gamma\eta=1$,
we can express the amplitude as
\be\label{emasterd1kr}
\begin{split}
&  {\pi^2 (2\pi)^{2k-p}\over 8(p+1)!}\, {\I p_\mu\over (2k)!}\,
\tilde\chi^\alpha \tilde\chi^\dbeta  \gamma^\mu_{\dbeta\alpha}
\int_{\Lgi{p+1}} v^{(p+1)}_{\gamma}\,
 (v^{(p+1)}_{\gamma})_{\bf i_1\cdots i_{p+1}}
\, C^{(2k)}_{\bf m_1\cdots m_{2k}} \\
&\hskip 1in \times
\bar\eta\[\wt\Gamma^{\bf m}   \wt\Gamma^{\bf m_1\cdots m_{2k}}
\wt\Gamma^{\bf i_1\cdots i_{p+1}}\wt\Gamma_{\bf m}
+2  \,\wt\Gamma^{\bf m_1\cdots m_{2k}}
\wt \Gamma^{\bf i_1\cdots i_{p+1}}\]\eta  \, .
\end{split}
\ee
Comparison with \refb{eterm} turns this into the following contribution to $a_{m,\gamma} \de\lambda^m$:
\be\label{emasterd1}
\begin{split}
&{\pi^2 (2\pi)^{2k-p}\over 8(p+1)!}\, {1\over (2k)!}\,
\int_{\Lgi{p+1}} v^{(p+1)}_{\gamma}\,
 (v^{(p+1)}_{\gamma})_{\bf i_1\cdots i_{p+1}}
\, \de C^{(2k)}_{\bf m_1\cdots m_{2k}} \\
&\hskip 1in \times
\bar\eta\[\wt\Gamma^{\bf m}   \wt\Gamma^{\bf m_1\cdots m_{2k}}
\wt\Gamma^{\bf i_1\cdots i_{p+1}}\wt\Gamma_{\bf m}
+2  \,\wt\Gamma^{\bf m_1\cdots m_{2k}}
\wt \Gamma^{\bf i_1\cdots i_{p+1}}\]\eta  \, ,
\end{split}
\ee
with the understanding that the differential $\de$ acting on $C^{(2k)}$ is in the moduli space and not
in spacetime.

The general procedure that we shall follow for analyzing the integrand of \refb{emasterd1}
is as follows. We first express the product $\wt\Gamma^{\bf m_1\cdots m_{2k}}\wt\Gamma^{\bf i_1\cdots i_{p+1}}$
using the identity
\be\label{e544}
\begin{split}
\wt \Gamma^{\bf i_1\cdots i_{\it 2a}}\, \wt \Gamma^{\bf j_1\cdots i_{\it 2 b}}
&\, = \sum_{\ell={\rm max}(0, (a+b-3)) }^{\rm min(\it 2a,2b)}  (-1)^{\ell (\ell+1)/2}\,
\delta^{\bf i_1\cdots i_\ell\, ,\,  j_1\cdots j_\ell}
\wt \Gamma^{\bf i_{\ell+1}\cdots i_{\it 2 a} j_{\ell+1}\cdots j_{\it 2 b}} \\ &\, \hskip 1.2in +
\hbox{inequivalent perm. with sign},
\end{split}
\ee
where
\be \label{edefdelta}
\delta^{\bf i_1\cdots i_n\, ,\,  j_1\cdots j_n}
\equiv g^{\bf i_1 j_1}\cdots g^{\bf i_n j_n} + \hbox{$(-1)^P$ weighted
perm. of {$\bf j_1,\cdots, j_n$}}\, ,
\ee
and inequivalent permutation in \refb{e544}
means all ${2a\choose \ell}$ inequivalent permutations
of $\bf i_1,\cdots, i_{\it 2a}$ and independently, all ${2b\choose \ell}$ inequivalent permutations
of $\bf j_1,\cdots, j_{\it 2b}$.
Then we can manipulate the resulting expression with the help of the identity
\be\label{egammaidentity}
\bar\eta \[\wt\Gamma^{\bf m}   \wt\Gamma^{\bf j_1\cdots j_{2\ell}}\wt\Gamma_{\bf m}
+ 2 \, \wt\Gamma^{\bf j_1\cdots j_{2\ell}}\] \eta
=
(8-4\ell)\, \bar\eta\,  \wt\Gamma^{\bf j_1\cdots j_{2\ell}}\, \eta\, ,
\ee
and then use \refb{eff2} to evaluate the right hand side.
This allows us to
express the integrand in terms of geometric quantities on $\CY$.
Note that for $\ell=2$ the right hand side
vanishes and so we do not need the expression for $\bar\eta \wt\Gamma^{\bf ijkl}\eta$.

We shall now evaluate \refb{emasterd1} for different values of $k$ and $p$.
Note that $p$ is always odd.
Even though $\bbF^{(3)}$ and $\bbF^{(7)}$ are related by the duality relation \refb{edualcon},
and therefore the corresponding potentials $\bbC^{(2)}$ and $\bbC^{(6)}$ are related by
\refb{ekk4}, we shall treat $\bbC^{(2)}_{\bf mn}$ and $\bbC^{(6)}_{\bf m_1\cdots m_6}$
as independent fields in our analysis. This is due to the fact that
the internal components of $\bbC^{(6)}$ are equivalent to the ones obtained by dualizing the field
$\bbC^{(2)}_{\mu\nu}$, but are not related to the internal components
of $\bbC^{(2)}$. Since much of our analysis will be repetitive, we shall give most of
the details at the beginning, but will be progressively less explicit as we proceed.

\subsubsection{D(-1)-brane}

For $p=-1$,  we can use \refb{egammaidentity} to simplify \refb{emasterd1} to
\be\label{emaster}
- \pi^3 (2\pi)^{2k}\,  {(4k-8)\over 4(2k)!}\,
\de C^{(2k)}_{\bf m_1
\cdots m_{2k}}
\,  \bar \eta\, \wt \Gamma^{\bf m_1\cdots m_{2k}}\,\eta.
\ee
It remains to make this result explicit for different $k$.

\subsubsection*{0-form potential}

For $k=0$, using $\bar\eta\eta=1$ and taking into account that $C^{(0)}= c^0$,
\refb{emaster} gives
\be
a_{ c^0,\gamma^0} d c^0=2\pi^3  d c^0 \, ,
\ee
where $\gamma^0$ refers to the charge vector corresponding to a single D(-1)-brane.

\subsubsection*{2-form potential}

For $k=1$, using the relation $ \bar \eta \wt \Gamma^{\bf ij}\eta=\I \, \bbom^{\bf ij}$
and expressing the 2-form field through the moduli $c^a$ (see \eqref{defBC}), one finds that
\refb{emaster} reduces to
\be
a_{c^a,\gamma^0}\de c^a=2 \I \pi^5   \de C^{(2)}_{\bf ij} \bbom^{\bf ij}=
4 \I \pi^5(\omega_a,\bbom)\,\de c^a
=4 \I  \pi^5 \,  \frac{1}{2\bbV}\, \kappa_{abc} \bbt^{b} \bbt^{c}\de c^a
= {\I\pi^3\over 2 V}\,  \kappa_{ab} t^b \de c^a  \, ,
\ee
where we have used the identity \refb{eb7} in the third step and the first relation in \refb{eccnet} in the last one.

\subsubsection*{4-form potential}

Since for $k=2$ \refb{emaster} vanishes, there is no coupling to $C^{(4)}$, or equivalently
its components $\tc_a$ along $\CY$.

\subsubsection*{6-form potential}

Using the relation
$\bar \eta \wt \Gamma^{\bf m_1\cdots m_{6}}
\eta=-\I \eps^{\bf m_1\cdots m_{6}}$, we can express \eqref{emaster} for $k=3$ as
\be
a_{\tilde c_0,\gamma^0} d\tilde c_0 =\I\pi^3(2\pi)^6 \, (\star\, \de C^{(6)})\, = {\I\pi^3\over V}\, \de\tilde c_0\, ,
\ee
where in the second step we have used \eqref{star1} and \eqref{eccnet}.

\smallskip

The results derived above are for a single D$(-1)$-brane, which, according to \refb{defLg}, carries
charge $q_0=-1$. For D$(-1)$-brane charge $q_0$, the above results should be multiplied
by $-q_0$.

\subsubsection{D1-brane}
\subsubsection*{0-form potential}

For $p=1$, $k=0$, after using \refb{egammaidentity}, \refb{eff2} and \refb{eyintegral},
we can express \refb{emasterd1} as
\be
\begin{split}
&\,
\frac{\pi\I}{8}\, \de  C^{(0)}
\int_{\Lgi{2}} \vi{2}_\gamma
(\vi{2}_\gamma)_{\bf ij}\,\bbom^{\bf ij}
= \frac{\pi\I}{4}\, \de  c^0
\int_{-q_a\gamma^a}\bbom
=
-\frac{\pi\I}{4}\, q_a \bbt^a \de c^0
\, .
\end{split}
\ee
Using again \eqref{eccnet}, we find
\be \label{eq:tau1-metric}
a_{ c^0,q_b\gamma^b} \de  c^0=- \I \pi^3  q_a t^a \,\de  c^0 \, .
\ee

\subsubsection*{2-form potential}

For $p=1$, $k=1$, using  \refb{e544}, \refb{egammaidentity} and \refb{eff2},
we can express  \refb{emasterd1} as
\be
\pi^3\, \int_{\Lgi{2}} \vi{2}_\gamma \(\I\, \de C^{(2)}_{\bf ij}(\vi{2}_\gamma)^{\bf j}_{~\bf k}\, \bbom^{\bf ik}
+\de C^{(2)}_{\bf ij}(\vi{2}_\gamma)^{\bf ji}\).
\ee
The first term in the integrand vanishes which can be seen by expressing
it in terms of holomorphic and anti-holomorphic components.
Thus, taking into account that $C^{(2)}=c^a\omega_a$, one remains with
\be
a_{c^a,q_b\gamma^b}dc^a=-2 \pi^3\, \de c^{a}
\int_{-q_a\gamma^a}\omega_{a}
=
2\pi^3\, q_a \, \de c^{a}\, .
\ee

\subsubsection*{4-form potential}

For $p=1$, $k=2$, the contribution \refb{emasterd1} gives the following expression
\be
\begin{aligned}
&\,
{2\over  4!}\,\pi^5 \int_{\Lgi{2}} \vi{2}_\gamma\,(\vi{2}_\gamma)_{\bf ij}\de C^{(4)}_{\bf klmn}
\(\I\,  \epsilon^{\bf ijklmn}-12\I\, \delta^{\bf im}\delta^{\bf jn}\bbom^{\bf kl}\)
\\
=&\, -\I  \pi^5
\int_{\Lgi{2}}\(-4\star \de C^{(4)}+
\vi{2}_\gamma\, \de C^{(4)}_{\bf ijkl}\,(\vi{2}_\gamma)^{\bf ij}\,\bbom^{\bf kl}\).
\end{aligned}
\label{D1tca-intermed}
\ee
The second term in the integrand can be rewritten using \eqref{Hodgeom} as follows
\be \label{e555}
\begin{split}
\de C^{(4)}_{\bf ijkl}\,(\vi{2}_\gamma)^{\bf ij}\,\bbom^{\bf kl}=&\,
\frac{1}{8}\, \de
C^{(4)}_{\bf ijkl}\,(\vi{2}_\gamma)^{\bf ij}\,\eps^{\bf klmnpq}\bbom_{\bf mn}\bbom_{\bf pq}
\\
=&\, \frac{1}{16}\, \eps_{\bf ijklrs}(\star\, \de C^{(4)})^{\bf rs}\,(\vi{2}_\gamma)^{\bf ij}
\,\eps^{\bf klmnpq}\bbom_{\bf mn}\,\bbom_{\bf pq}
\\
=&\, (\vi{2}_\gamma)^{\bf ij}(\star\, \de C^{(4)})^{\bf kl}
\(\bbom_{\bf ij}\,\bbom_{\bf kl}+2\, \bbom_{\bf ik}\,\bbom_{\bf lj}\)
\\
=&\, 4(\vi{2}_\gamma, \bbom)(\star\, \de C^{(4)},\bbom)
+4(\vi{2}_\gamma)^{s\bt}\bbom_{s\bu}\,(\star\,\de C^{(4)})^{\bu v}\,\bbom_{v\bt}\, .
\end{split}
\ee
Using \refb{defBC}, \refb{innerwedge} and \refb{basisforms}, one can show that
$(\star\, \de C^{(4)},\bbom)=\bbt^a d\tc_a/\bbV$. Using also $\bbom_{s\bar t}=\I\bbg_{s\bar t}$
in the second term on the r.h.s. of \refb{e555}, one obtains
\be
\de C^{(4)}_{\bf ijkl}\,(\vi{2}_\gamma)^{\bf ij}\,\bbom^{\bf kl}=
 \frac{4}{\bbV}\,(\vi{2}_\gamma, \bbom)\, \bbt^a\de\tc_a-4(\vi{2}_\gamma, \star \,\de C^{(4)}).
 \ee
Thus, \eqref{D1tca-intermed} becomes
\be
\I\pi^5 \int_{-q_a\gamma^a}
\(8\star \de C^{(4)}- \frac{4}{\bbV}\,\bbt^a\de\tc_a\, \bbom\)
=8\I \pi^5 \bbkap^{ab} q_a (\de \tc_b)\, ,
\ee
where
we used \eqref{Hodge-om2}. Due to  \eqref{defbbk},
this gives
\be
a_{\tc_a,q_b\gamma^b}\de\tc_a= 2\I \pi^3\kappa^{ab}q_a \de\tc_b.
\label{res-D1tca}
\ee

\subsubsection*{6-form potential}

For $k=3$, \refb{emasterd1} vanishes due to \refb{egammaidentity}.
This shows that
\be
a_{\tc_0,\gamma} \, d\tc_0=0\, .
\ee

\subsubsection{D3-brane}
\subsubsection*{0-form potential}

For $p=3$, $k=0$,
the integrand in \refb{emasterd1} vanishes due to \refb{egammaidentity} and we have
\be
a_{ c^0,p^b\gamma_b}  \de  c^0=0\, .
\ee

\subsubsection*{2-form potential}

In this case we have $p=3$, $k=1$ and can express \refb{emasterd1} as
\be
{ \pi\I\over 4!\cdot 8}
\int_{\Lgi{4}} v^{(4)}_{\gamma}\, \de C^{(2)}_{\bf ij} \, (\vi{4}_\gamma)_{\bf klmn}
\(\eps^{\bf ijklmn} - 12\, \delta^{\bf im}\, \delta^{\bf jn}\, \bbom^{\bf kl}\).
\ee
This can further be rewritten as
\be
\begin{split}
\frac{\pi\I}{4}\,
\int_{\Lgi{4}} \(\star\, \de C^{(2)}-\de C^{(2)}\wedge \bbom\)
=&\, - \frac{\pi\I}{4}\,
\int_{-p^a\tgamma_a} \(2 \de C^{(2)}\wedge \bbom-\frac{(\bbt^2\de c )}{4\bbV}\, \bbom^2\)
\\
=&\,\frac{\pi\I}{4} \,
\(2(p\bbt\de c)-\frac{(\bbt^2\de c)}{4\bbV}\,(p\bbt^2) \) ,
\end{split}
\ee
where we used \eqref{basisforms} and \eqref{Hodge-om}.
Therefore, taking into account the first relation in \eqref{eccnet}, we have
\be
a_{c^a,p^b\gamma_b} \de c^a
=2\I\pi^3\kappa_{ab}p^a\de c^b-\frac{\I\pi^3}{4V}\, (pt^2)\kappa_{ab}t^a \de c^b  .
\ee

\subsubsection*{4-form potential}

In this case we have $p=3$, $k=2$ and
can express \refb{emasterd1} as
\be\label{e3.62a}
\begin{split}
&  {2\pi^3\over 4!}\,
\int_{\Lgi{4}}
v^{(4)}_{\gamma}\, (v^{(4)}_{\gamma})_{\bf ijkl} \Big[  \de C^{(4)\bf ijkl}
+2\I  \, {\de C^{(4)\bf ijk}}_{\bf m} \, \bbom^{\bf ml}
-{2\I\over 3!} \,{\de C^{(4)\bf i}}_{\bf mnp}\,
\eps^{\bf  mnpjkl}\Big]\, .
\end{split}
\ee
The last two terms actually vanish: the vanishing of the second term becomes
evident after writing it in the holomorphic
and anti-holomorphic components, whereas the vanishing of the third one follows just from the anti-symmetrization of indices
as the index $\bf i$ can equal neither $\bf jkl$ nor $\bf mnp$, and these six indices must all be different.
Thus, one remains with
\be
a_{\tc_a,p^b\gamma_b}\de \tc_a= 2 \pi^3
\int_{-p^a\tgamma_a} \de C^{(4)}
=-2 \pi^3\, p^a \, \de \tc_a\, .
\ee

\subsubsection*{6-form potential}

In this case we have $p=3$, $k=3$ and can express \refb{emasterd1} as
\be
\begin{split}
{2\I\pi^5 \over 4! }\,
(\star\, \de C^{(6)})
\int_{\Lgi{4}} v^{(4)}_{\gamma}\, (v^{(4)}_{\gamma})_{\bf ijkl} \,
\eps^{\bf ijklmn}\, \, \bbom_{\bf mn}
= &\, 4\I \pi^5(\star\, \de C^{(6)})\,
\int_{-p^a\tgamma_a} \star \,\bbom
\\
=&\, -2\I \pi^5 {(p\bbt^2)\over \bbV}\, \de \tc_0\, ,
\end{split}
\ee
where we used \eqref{Hodgeom} and \eqref{basisforms}.
Therefore, due to \eqref{eccnet}, we have
\be
a_{\tc_0,p^b\gamma_b}  \de \tc_0=-\frac{\I\pi^3}{2V}\,(pt^2)\de\tc_0\, .
\ee

\subsubsection{D5-brane}
\subsubsection*{0-form potential}

In this case we can express \refb{emasterd1} as
\be \label{e570}
a_{ c^0,\tgamma_0} \de  c^0 = \frac{\I\,\de c^0}{8(2\pi)^3}\,
\int_{\CY} v^{(6)}_\gamma
= \frac{\I\,\bbV }{8(2\pi)^3}\,  \, \de c^0
=\I \pi^3 V \de  c^0\, ,
\ee
where in the last step we have used \refb{eccnet}.
Here $\tgamma_0$ corresponds to the charge vector describing a D5-brane wrapped on $\CY$.

\subsubsection*{2-form potential}

In this case the integrand in \refb{emasterd1} vanishes and we have
\be
a_{c^a,\tgamma_0} d c^a=0\, .
\ee

\subsubsection*{4-form potential}

In this case we can express the contribution \refb{emasterd1} as
\be
\begin{split}
& {\pi^2 \over 8 \cdot 6! \cdot 4!} {1\over 2\pi} {6\choose 2} \times 4! \times 4\I
\int_\CY v^{(6)}_\gamma\, \(v^{(6)}_\gamma\)^{\bf i_1\cdots i_6}
dC^{(4)}_{\bf i_1\cdots i_4} \,
\bbom_{\bf i_5i_6}
\\
= &\, {\I\pi \over 4 }
\int _\CY v^{(6)}_\gamma\, \( v^{(6)}_\gamma, dC^{(4)}\wedge \bbom\)
=  {\I\pi \over 4} \int_{\CY} dC^{(4)}\wedge \bbom =  {\I \pi\over 4}\, \bbt^a d\tc_a\, .
\end{split}
\ee
Using the first relation in \eqref{eccnet}, one finds
\be
a_{\tc_a,\tgamma_0}\de\tc_a
=\I\pi^3  t^a  \de\tc_a.
\ee

\subsubsection*{6-form potential}

For $p=5$, $k=3$, we can express \refb{emasterd1} as
\be \label{e574}
a_{\tilde c_0,\tgamma_0} \de\tilde c_0 = -  2 \pi^3 \int_{\CY}\de C^{(6)}
= -2 \pi^3\,\de \tc_0\, .
\ee

Eqs. \refb{e570}-\refb{e574} hold for a single D5-brane carrying charge $p^0=1$. For general $p^0$, these
expressions should be multiplied by $p^0$.

\subsection{Final result}

Adding the various contributions to $a_{m,\gamma} d\lambda^m$
obtained in this section, we get
\ben
\sum_m a_{m,\gamma} d\lambda^m  &=& \frac{\pi^2}{4V\tau_2^2}\, \cTR_\gamma \de\sigma
+ {\I\pi^2\over \tau_2} \, \cTR_\gamma d\tau_2 +
\frac{\pi^2}{4V}\, \cTR_\gamma\,\kappa_{ab} t^b \de b^a
+ 2\I \pi^3 \tau_2\, {Z_\gamma\over |Z_\gamma|} \,\de\bZ_\gamma
\non\\
&&
- q_0 \[{\I \pi^3\over V} \, d\tilde c_0 + {\I\pi^3\over 2 V}\, \kappa_{ab} t^b \de c^a
+ 2\pi^3  d c^0\]
\non\\
&&+  q_a  \biggl[2\I \pi^3\kappa^{ab}\, \de\tc_b + 2\pi^3 \de c^{a}- \I \pi^3  t^a \de  c^0\biggr]
\label{efinalam}\\
&&+ \[-\frac{\I\pi^3}{2V}\,(pt^2)\de\tc_0-2\pi^3  p^a \de\tc_a
+ 2\I\pi^3\kappa_{ab}p^a\de c^b-\frac{\I\pi^3}{4V}\, (pt^2)\kappa_{ab}t^a \de c^b\]
\non\\
&&+  p^0 \biggl[-2\pi^3 \de \tilde c_0+\I \pi^3t^a \de\tc_a + \I \pi^3  V\, \de  c^0\biggr] ,
\non
\een
with the understanding that this result is valid when only one type of charge is
present. Comparing this with
\eqref{eagammafin} for vanishing background values of $b^a$ and RR fields, we conclude that
\be\label{e696}
\sum_m a_m \, d\lambda^m=\pi^3\cA_\gamma\, .
\ee

Now, using \refb{e35xy} and \refb{e75new}, we can express the instanton correction to the
metric as:
\be\label{e75newer}
ds_{\rm inst}^2 = \frac{\kappa^2}{ 2^{3}  \pi^{13/2}\bbV}
\sum_\gamma \frac{\Omega_\gamma }{\(2\pi^2 \cTR_\gamma\)^{1/2}}
\, \sum_{k=1}^\infty\, \NP^{-1/2}
\, e^{-k\TT_\gamma} \, \( \sum_m a_{m,\gamma} \, d\lambda^m\)^2\,  .
\ee
Using \refb{ekappags},  \refb{eccnet} and \refb{e696} this reduces to
\be
ds_{\rm inst}^2 =
2^{-7/2} \,  \pi^{-1/2} (\tau_2^2 V)^{-1}
\sum_\gamma \frac{\Omega_\gamma}{( \cTR_\gamma)^{1/2}}
\sum_{k=1}^\infty\, \NP^{-1/2}
\, e^{-k\TT_\gamma} \, \AAA_\gamma^2\,  .
\ee
This is in perfect agreement with the prediction \refb{square-2b}.

\section{Non-linear terms} \label{s7}

The results of explicit instanton calculation given in \refb{efinalam},
\refb{e75newer} match the prediction \refb{square-2b} when
the background values of $B$ and RR fields are set to zero
in the expression for $\AAA_\gamma$.
Furthermore, in \refb{efinalam} we had to restrict to instantons carrying only
one type of charge, while the result \refb{square-2b} includes sum over
instantons that can carry multiple types of charges corresponding to D$p$-branes for different
values of $p$. Our goal in this section will be to rectify these defects.

\subsection{Disk amplitude with one RR field and a pair of zero modes}
\label{s7.1x}

First, we shall study the effect of switching on the $B$ field along $\CY$ on the
amplitudes involving the  RR
fields. In the presence of such a background, the tension of a D$p$-brane
is modified to:
\be \label{etsnew}
 \qquad
\tilde\ts_{p}\equiv \ts_{p} \,  \sqrt{\mbox{det}(\bbg_\parallel+2\kappa \bbB_\parallel)} \bigg/
\sqrt{\mbox{det}\bbg_\parallel} \, ,
\ee
where the subscript $\parallel$ denotes the pullback along the world-volume directions of the
D-brane. The boundary condition on
the spin fields given in \refb{SSbc} get modified to \cite{DiVecchia:1999uf}
\be\label{eaa11new}
\begin{split}
e^{-\bar\phi/2}\bar S_\alpha(\bar z) =&\,  - {\I\over (2k)!} {\sqrt{\mbox{det}\bbg_\parallel}
\over \sqrt{\mbox{det}(\bbg_\parallel+
2\kappa \bbB_\parallel)}}
\, v_{\bf m_1\cdots m_{2k}}
(\Gamma^{\bf m_1\cdots m_{2k}}\,\hat{e}^{-\kappa(\bbB_P)_{\bf ij}
\Gamma^{\bf ij}})_\alpha^{~\beta} \, e^{-\phi/2} S_\beta(z)\, ,
\\
e^{-3\bar\phi/2}\bar S^\alpha(\bar z) = &\,  {\I\over (2k)!} {\sqrt{\mbox{det}\bbg_\parallel}
\over \sqrt{\mbox{det}(\bbg_\parallel+
2\kappa \bbB_\parallel)}}
\, v_{\bf m_1\cdots m_{2k}}
(\Gamma^{\bf m_1\cdots m_{2k}}\,\hat{e}^{-\kappa(\bbB_P)_{\bf ij}
\Gamma^{\bf ij}})^\alpha_{~\beta} \, e^{-3\phi/2} S^\beta(z)\, ,
\end{split}
\ee
where $\bbB_P=P\bbB P$ denotes the projection of $\bbB$ along the brane.\footnote{While the
pullback and projection are closely related objects, they are not the same. The former has its
indices labelled by the intrinsic coordinates on the brane that run over $(p+1)$ values while the latter
has its indices labelled by the coordinates of the full spacetime even though it will be a matrix
of rank $\le (p+1)$.}
The factor of $2\kappa$ comes from
the particular normalization of the 2-form field we have chosen, reflected in
\refb{egmnhmn}. $\hat{e}$  should be interpreted as an
exponential on forms: after Taylor expanding,
one has to anti-symmetrize the tangent-space indices of the $\Gamma$-matrices.
Due to this, the series terminates after a finite number of terms; {\it e.g.}
in the case of the D1-brane, only the first two terms survive.

The  modifications \refb{etsnew}, \refb{eaa11new} due to the
presence of background $B$ field can affect the computation of the $a_m$'s in \S\ref{sdminus} by changing the result
of the disk amplitude with one closed string vertex operator and two fermion zero modes from
the open string sector.
In the absence of $B$-field background,
the integral in \refb{emasterd1} has the general form
\be \label{e7.3ab}
{1\over (p+1)!} \, \int_{\Lgi{p+1}} v^{(p+1)}_\gamma \, v^{(p+1)}_{\gamma, \bf i_1\cdots i_{p+1}}
Y^{(p+1)\bf i_1\cdots i_{p+1}}
=\int_{\Lgi{p+1}} \, Y^{(p+1)}
\ee
for some $(p+1)$-form $Y^{(p+1)}$.
Using the Poincar\'e duality  between $(p+1)$-cycles and $(5-p)$-forms on $\CY$, we can associate
a $(5-p)$-form $\omega_\gamma^{(5-p)}$ with
$\Lgi{p+1}$ and express the integral \eqref{e7.3ab} as
\be\label{e3.84}
\int_\CY \, \omega_\gamma^{(5-p)} \wedge Y^{(p+1)}\, .
\ee
In particular, we have
\be
\omega_{\gamma}^{(0)}=p^0,
\qquad
\omega_{\gamma}^{(2)}=-p^a\omega_a,
\qquad
\omega_\gamma^{(4)}=-q_a \tom^a,
\qquad
\omega_\gamma^{(6)}=-q_0 \omega_\CY,
\label{list-volume}
\ee
while their sum, which we denote by $\omega_\gamma$, can be seen as Poincar\'e dual to the homology element \eqref{defLg}.
It follows from the remark below \refb{defLg} that
$\omega_\gamma=\iota(\gamma)$.

Now let us consider the effect of switching on the $B$-field.
The net effect is to make the following replacement in the expression for the integrand in \refb{emasterd1}:
\be \label{e66}
\begin{split}
& v^{(p+1)}_{\gamma, \bf i_1\cdots i_{p+1}} \Gamma^{\bf i_1\cdots i_{p+1}}
\qquad \Rightarrow \qquad
v^{(p+1)}_{\gamma, \bf i_1\cdots i_{p+1}} \Gamma^{\bf i_1\cdots i_{p+1}}
\hat e^{- \kappa (\bbB_P)_{\bf ij} \Gamma^{\bf ij} }\\
= &\ v^{(p+1)}_{\gamma, \bf i_1\cdots i_{p+1}} \bigg[\Gamma^{\bf i_1\cdots i_{p+1}}
+ 2\kappa \bbB^{\bf i_1i_2} \Gamma^{\bf i_3\cdots i_{p+1}}
+{(2\kappa)^2\over 2}  \bbB^{\bf i_1i_2} \bbB^{\bf i_3i_4}\Gamma^{\bf i_5\cdots i_{p+1}}
\\ & \hskip 2in + \ {(2\kappa)^3\over 6}  \bbB^{\bf i_1i_2} \bbB^{\bf i_3i_4} \bbB^{\bf i_5i_6}\Gamma^{\bf i_7\cdots i_{p+1}}
\bigg]\, ,
\end{split}
\ee
where we have terminated the sum using the fact that $p$ is at most 5. The rest of the factors
remain the same. Note that on the r.h.s.\ of \refb{e66} we have used the full $\bbB_{\bf ij}$
since contraction with $v_\gamma^{(p+1)}$ given in \refb{defvn}
automatically ensures projection
along the brane. Since the factors $\sqrt{\mbox{det}(\bbg_\parallel+2\kappa \bbB_\parallel)}
\bigg/ \sqrt{\mbox{det}\bbg_\parallel}$
cancel between \refb{etsnew} and \refb{eaa11new}, we effectively make the replacement
\be
Y^{(p+1)} \ \Rightarrow \
Y^{(p+1)} + {2\kappa\over (2\pi)^2}\, \bbB \wedge Y^{(p-1)}
+{1\over 2}\, {(2\kappa)^2\over (2\pi)^4}\,\bbB\wedge \bbB\wedge Y^{(p-3)}
+{1\over 6}\, {(2\kappa)^3\over (2\pi)^6} \, \bbB\wedge \bbB\wedge \bbB \wedge Y^{(p-5)}\, .
\ee
In writing the above expression we have used
the fact that when we go from $p$-brane contribution
to the $(p-2)$-brane contribution, we have to replace the $\Gamma^{\bf i_1\cdots i_{p+1}}$
contracted with the volume form on the brane by $\Gamma^{\bf i_3\cdots i_{p+1}}$ and
also the tension that multiplies the amplitudes gets multiplied by a factor of $(2\pi)^2$.
Substituting this into \refb{e3.84} and using the relation ${2\kappa\over (2\pi)^2}\, \bbB=
\sum_a b^a\omega_a=B$
following from \eqref{e626a} and \eqref{eccnet}, we obtain
\be
\int_\CY \omega_\gamma^{(5-p)} \wedge Y^{(\rm even)}e^B \, ,
\qquad
Y^{(\rm even)}=\sum_{k=0}^3 Y^{(2k)}.
\label{contrBY}
\ee

Let us combine all charge vectors together by considering the formal sum of contributions \eqref{contrBY} for different $p$
keeping in mind that at this stage, a physical instanton will carry only one of the four types of charges corresponding to a
given value of $p$.
With this understanding we can express the net contribution from the formal sum as
\ben \label{e3.88}
\int_\CY \omega_\gamma \wedge e^{B}\, Y^{(\rm even)}
=\int_\CY \iota(\chgam)\wedge Y^{(\rm even)}=\int_\CY \omega_{\chgam} \wedge Y^{(\rm even)},
\een
where we used $\omega_\gamma=\iota(\gamma)$, \eqref{propiota} and \eqref{defgamB}.
Since we are allowed to have only a fixed value of $p$, \refb{e3.88}
should be interpreted as follows. The contribution due to a given D$p$-brane of fixed charge
in the presence of a background $B$ field is given by the sum of the contributions from
different types of branes, carrying
charges $\chp^\Lambda,\chq_\Lambda$, with no background $B$ field.
However, these charges are not independent, since they are all determined
by the charge labelling the original D$p$-brane which could be $p^0$, $p^a$, $q_a$ or $q_0$.
We shall partially overcome this restriction in \S\ref{s7.3x}.

So far the analysis of this section has been done for constant $B$ field and
vanishing RR background. Therefore, we could not determine the
terms proportional to $db^a$ inside the covariant derivatives in \refb{defnabla}.
However, the following argument can be used to
show that what appear in the expression for $\sum_m a_m d\lambda^m$ are the full
covariant derivatives. The point is that perturbative amplitudes in type IIB string theory as well as all
instanton amplitudes, with the sole exception of a disk with one closed string vertex operator,
are computed using the $(-1/2,-1/2)$ picture vertex operators of the RR fields, which involve
the field strengths $\bbF^{(2k+1)}$ instead of the potentials $\bbC^{(2k)}$. This leads to a
symmetry under constant shift of the potentials $\bbC^{(2k)}$
which translates to the shift symmetry \eqref{Heissym}.
The disk one-point functions that are responsible for the breaking of the shift symmetry can
be summed to produce the factor $e^{-k\TT_\gamma}$ that accompanies the instanton
amplitude --- this has been checked explicitly in appendix \S\ref{sg}.
Therefore, $\sum_m a_m d\lambda^m$ must be invariant under the continuous
shift symmetry. This is precisely achieved by the covariant derivatives whose invariance under transformations \eqref{Heissym}
was verified in the end of \S\ref{subsec-class}.\footnote{Note that while the shift symmetry in
string field theory shows the existence of a shift symmetry of $\sum_m a_m d\lambda^m$, the
actual transformation laws may be different due to the possibility of non-trivial transformations
relating the fields and the transformation parameters in string field theory and effective field
theory description. What we are using here to fix the form of $a_m$ is the knowledge that the
missing terms must be proportional to $db^a$ together with the shift symmetry transformation laws
\refb{Heissym} that leave the tree level action \refb{metricBtreelargeV}, \refb{defnabla} invariant.}

To summarize, we found that the RR contribution to $a_m d\lambda^m$
in the presence of the $B$ field is given by the RR contribution to \refb{efinalam} with the charges replaced by
$\chp^\Lambda,\chq_\Lambda$ and the differentials of the RR scalars replaced by their
covariant versions \eqref{defnabla}.
This gives
\ben \label{efinalammod}
\frac{1}{\pi^3}\(\sum_m a_{m,\gamma} d\lambda^m\)_{RR}
&=&  -\chq_0 \biggl[ {\I\over V} \, \nabla \tc_0 + {\I \over 2 V}\,\kappa_{ab} t^b\nabla c^a + 2 \de  c^0\biggr]
 +  \chq_a  \biggl[ 2\I \kappa^{ab}\, \nabla\tc_b + 2\nabla c^{a}- \I  t^a \de  c^0\biggr]
\non\\
&& + \[-\frac{\I}{2V}\,(\chp t^2)\nabla\tc_0 - 2\chp^a \nabla\tc_a
+ 2\I\kappa_{ab}\chp^a\nabla c^b-\frac{\I}{4V}\, (\chp t^2)\kappa_{ab}t^a \nabla c^b\]
\non\\
&& + p^0 \biggl[-2\nabla \tc_0 +\I t^a \nabla\tc_a+ \I V \de  c^0\biggr]\, .
\een
This agrees with the RR part of \refb{eagammafin}.

\subsection{Disk amplitude with one NSNS field and a pair of zero modes} \label{s7.4x}

Turning on a constant $\bbB$ field in the internal directions along the D-brane modifies the
boundary condition on $\psi^M$,
with the doubling trick now leading to the replacement~\cite{Abouelsaood:1986gd}
\be
\bar c\, e^{-\bar\phi}\bar\psi^{M}(\I)\rightarrow
O^{M}_{~ N}\,
 c \, e^{-\phi} \psi^{N}(-\I)\,,
\ee
where $O$ is the orthogonal matrix
\be
O=P (\bbgp+2\kappa\,  \bbBp)^{-1} (\bbgp-2\kappa\,\bbBp)  - Q \,,
\qquad
\bbBp=P\bbB P, \qquad \bbgp=P\bbg P,
\ee
and $P$ and $Q$ are the projection operators tangential and transverse to the brane, respectively.
Note that the inverse of $(\bbgp+2\kappa\,  \bbBp)$ exists since it is taken inside the subspace
projected by $P$.
In addition, the D-brane tension is modified as in~\eqref{etsnew}.
Since the action of $O$ on directions transverse to the brane does not depend on $\bbB$,
the dilaton and NSNS axion amplitudes are formally the same as those found in \S\ref{s7.1} and \S\ref{sdilaton}
before with the D-brane tension modified as above.

In the case of  K\"ahler and B-field moduli,  \refb{ee3.7} generalizes to
\be\label{eq:amp-bkdb-kb}
-\hf\, \pi^2\kappa\, \tilde\ts_{p}\, p_\rho\, \gamma^\rho_{\dbeta\alpha}\,\tilde\chi^\alpha \tilde\chi^\dbeta\,
e_{\bf ij}   O^{\bf j}_{~\bf k} \(\bbg^{\bf ik}+\I\bbom^{\bf ik}\)
(2\pi)^{p+1}\delta^{(p+1)}(0)  \, .
\ee
Since $Q=1-P$, we can write as in \eqref{comb-ePQ}
\be
e_{\bf ij}   O^{\bf j}_{~\bf k} \(\bbg^{\bf ik}+\I\bbom^{\bf ik}\)
=2 e_{\bf ij} \(P \(\bbgp+2\kappa\, \bbBp\)^{-1} \bbgp \)^{\bf j}_{~\bf k}  \(\bbg^{\bf ik}+\I\bbom^{\bf ik}\)
-\(e^{\bf i}_{\bf i}+\I e_{\bf ij}\bbom^{{\bf i}{\bf j}}\).
\label{comb-ePQ-B}
\ee
The second term gives the same contribution as \eqref{eonecon} with $\TT_\gamma^R$ including
the effect of the background $B$ field as shown in \refb{eg.7},
whereas the first can be rewritten as (cf. \eqref{eomegaproject})
\be\label{e6.15}
 \frac{2}{\kappa}\(\delta \bbb^a-\I \delta \bbt^a \) \omega_{a,s\bt}
\(P \(\bbgp+2\kappa\, \bbBp\)^{-1} P\)^{\bt s}
=  {2\over \kappa} \,  (\delta \bbgp + 2\kappa \delta \bbBp)_{s \bar t}
\(\(\bbgp+2\kappa\,  \bbBp\)^{-1}\)^{\bt s},
\ee
where $\delta\bbgp$ and $\delta\bbBp$
are defined as the projection of $\delta\bbg$ and $\delta\bbB$ on the brane, i.e.\ we do not
take into account the possible variation of $P$ in defining these quantities.\footnote{Of course,
the cycle itself, determined from the requirement of holomorphy,
does not depend on the choice of $\bbg$ and $\bbB$, but  $P$, expressed in a fixed coordinate
system, could still depend on the choice of the metric.}

If we work within the subspace projected by $P$, then we can regard
$(\delta \bbgp + 2\kappa \delta \bbBp)$ as a $(p+1)/2 \times (p+1)/2$ matrix obtained by
this projection and $\(\bbgp+2\kappa\,  \bbBp\)^{-1}$ as the inverse of this matrix.
There is a closely related object where instead of using projection
we use the pullback
operation for covariant tensors. The projection depends on the choice of metric but
the pullback does not. For a rank (1,1) tensor $M$, the pullback $M_{\rm \parallel}$ and the
projection $M_\defparallel$ are related by $M_{\rm\parallel} = W M_\defparallel \bar W^{T}$
for some matrix
$W$. However, in the combination appearing in \refb{e6.15} the dependence on the matrix $W$
cancels, showing that we can replace the matrices by their pullback.
This allows us to express it as:
\be\label{e6.16}
{2\over \kappa} \,  \mbox{Tr$_{\rm h}$}\[(\delta \bbg_\parallel + 2\kappa \delta \bbB_\parallel)
\(\bbg_\parallel+2\kappa\,  \bbB_\parallel \)^{-1}\]
=\frac{2}{\kappa}\, \delta\log\mbox{det$_{\rm h}$}\left(\bbg_\parallel+2\kappa\bbB_\parallel\right),
\ee
where the subscript `h' indicates
that we are considering the trace and determinant using the holomorphic
coordinates on the brane as in \refb{e6.15}. Since pullback of
a covariant tensor
does not depend on the choice of metric, in this form it does not matter if in computing
$\delta\bbg$ and $\delta\bbB$ we first compute the variation and then take the pullback or take the
variation of the pullback, and therefore we could pull the $\delta$ outside the trace and the determinant.
Using \refb{e6.16}, the contribution of the first term in \refb{comb-ePQ-B} to
\refb{eq:amp-bkdb-kb} is given by:
\be \label{e6.18}
- \pi^2\, \tilde\ts_{p}\, p_\rho\, \gamma^\rho_{\dbeta\alpha}\,\tilde\chi^\alpha
\tilde\chi^\dbeta\,
 {\bbV_\gamma\over |\bbV_\gamma|}
\int_{\Lgi{p+1}}\vn_\gamma\,  \delta\log\mbox{det$_{\rm h}$}
\left(\bbg_\parallel+2\kappa\bbB_\parallel\right) .
\ee
The factor $\bbV_\gamma/|\bbV_\gamma|$ has the same origin as in \refb{eorient}.
Now using \refb{evaldet} with $M=\bbg+2\kappa\bbB$ and
$N$ some matrix having the expansion $N_{s\bt}=N^a\omega_{a,s\bt}$ with coefficients independent of $t^a$ and $b^a$, we get
\be
\log\mbox{det$_{\rm h}$}\left(\bbg_\parallel+2\kappa\bbB_\parallel\right)
=\log \bar Z_\gamma + \log\mbox{det$_{\rm h}$} N_\parallel- \log \bar Z_\gamma(N)\, ,
\ee
where $\bar Z_\gamma(N)$ is $\bar Z_\gamma$ evaluated at $z^a=N^a/(2\pi)^2$.
Since the last two terms are independent of the moduli,
their variation vanishes and we obtain
\be
\delta \log\mbox{det$_{\rm h}$}\left(\bbg_\parallel+2\kappa\bbB_\parallel\right)
= {\delta \bar Z_\gamma\over \bar Z_\gamma}\, .
\ee
Since this is a constant in $\CY$, it comes out of the integral in \refb{e6.18}. The integration now yields
a factor of $\bbV_\gamma$.
Using the relation $\bar\ts_p |\bbV_\gamma|= \cTR_\gamma=2\pi \tau_2|Z_\gamma|$ in the
presence of a background $B$ field, as verified in
\refb{eg.7}, we can express \refb{e6.18} as
\be\label{etwoconnew}
- \pi^2\, p_\rho\, \gamma^\rho_{\dbeta\alpha}\,\tilde\chi^\alpha
\tilde\chi^\dbeta\,
 {2\pi \tau_2} \, |Z_\gamma|\, {\delta \bar Z_\gamma\over \bar Z_\gamma}\, .
\ee
Adding \refb{eonecon} and \refb{etwoconnew} and using \refb{eterm}, we get
\be\label{e6fin}
\begin{split}
a_{t^a,\gamma} dt^a+a_{b^a,\gamma} db^a
=&\, \pi^2 \cTR_\gamma\(\frac{1}{4V}\, \kappa_{ab} t^b \de b^a-\frac{\I\de V}{2V} \)
+2\I \pi^3 \tau_2\, {Z_\gamma\over |Z_\gamma|} \,\de\bZ_\gamma.
\end{split}
\ee
This has the same form as equation~\eqref{eq:fin-a-kb}, except that it is now valid even in the
presence of background $B$ field.

Finally, we shall discuss the replacement of $d\sigma$ in \refb{efinalam} by the full covariant
derivative $\nabla\sigma$ that
appears in the expression for $\AAA_\gamma$ in \refb{eagammafin}.
This follows essentially from the same argument as in \cite{Alexandrov:2021shf}.
Since the tree level
kinetic term \refb{metricBtreelargeV} also involves $\nabla\sigma$,
such a term must be present already in the dualization
rule \refb{ehpsi} due to the presence of an additional term in the action
involving the NSNS 3-form field strength before dualization. In other words, the Fourier
transform of $d\sigma$ is replaced by the Fourier transform of $\nabla\sigma$ on the
r.h.s.\ of \refb{ehpsi}. Therefore, $d\sigma$ in \refb{efinalam}
should be replaced by $\nabla\sigma$.

In summary, we have shown that the contribution to $\sum_m a_{m,\gamma}
d\lambda^m/\pi^3$ from the NSNS sector fields $\lambda^m$
reproduces the result given in the first line of
\refb{eagammafin} even in the presence of
background $B$ field.

\subsection{Multiple-charge system} \label{s7.3x}

Notwithstanding our use of
the formal sum  in \refb{e3.88}, our analysis up to now applies to the case where
the instanton carries a single type of D$p$-brane charge.
We shall now rectify this defect by using the fact that the original D$p$-brane could carry
world-volume gauge field with field strength $F_{\bf ij}$ and this could induce D$q$-brane
charges with $q<p$. It is known that in the world-volume theory the combination
$F+B$ appear together. Therefore, if we decompose $F$
as\footnote{Since $F$ must be along the world-volume of the brane, not all $f^a$'s are
allowed. But since the final formula is sensitive only to the components of $F$ along
the brane, the extra variables are automatically eliminated.}
\be
F = f^a \omega_a\, ,
\ee
then in \refb{def-qch} the combination $f^a+b^a$ will appear together. For reasons that will
become clear soon, let us denote the charge carried by the original system by
$\bar\gamma=(\bp^\Lambda, \bq_\Lambda)$ --- as before only one of these charges will be taken
to be non-zero.
Then, using the form notation, \refb{defgamB} will now be replaced by
\be
\chgam= e^{-B-F}\bar\gamma=e^{-B}\gamma,
\label{def-qindependent}
\ee
where in the last step we defined
\be
\gamma=e^{-F}\bar\gamma.
\label{def-qbar}
\ee
As a result, \refb{def-qindependent} has the form identical to \refb{defgamB}.
The analysis of the earlier sections
will now proceed as before with $\gamma$ replaced by $\bar\gamma$ and $B$ replaced by
$B+F$. But by regarding $\gamma$ as the instanton charge and
by choosing the $f^a$'s appropriately, we can now make all of the charges $(p^\Lambda,q_\Lambda)$
(or a subset of them) non-zero. We still have the constraint that
in terms of the barred variables, only one charge can be non-zero at a time.
Since $f^a$'s, representing flux of gauge field strengths through 2-cycles, are quantized, the charges
$(p^\Lambda,q_\Lambda)$ remain quantized since $(\bp^\Lambda,\bq_\Lambda)$ are quantized.
In contrast, $b^a$'s are continuous variables
and as a result $\chp^a,\chq_a,\chq_0$ are not quantized. Eq.\eqref{def-qindependent} shows
that the final result, expressed in terms of $\gamma$ and $B$, takes the same form as before.

We can now make contact with the large volume limit described in \refb{escalecharge} if we
identify $f^a$ with $\lambda^{1/3}\bar f^a$. Therefore, $f^a$ scales in the same way as $b^a$.
This is natural since $f^a+b^a$ appear in this combination in all formul\ae, but also from the
point of view that with this choice $F_{\bf ij}F^{\bf ij}$ on the brane
remains finite as we take the large volume limit.
Hence, if we keep the magnitude of the gauge field strength on the brane fixed
as we take the large volume limit then we naturally arrive at the scaling described in
\refb{escalecharge}. Put another way, even though according to \refb{escalecharge}
the charges associated with lower dimensional branes are scaled up in the large volume limit,
the charge is spread out over the higher dimensional brane so that the charge density remains
finite.
The comments below \refb{escalecharge} also show that in terms of
the barred variables, only the one associated with the highest dimensional brane contributes
in the large volume limit. Therefore, the surviving contribution can
be obtained by starting with a single D-brane and switching on gauge fields on the brane, keeping
$F_{\bf ij} F^{\bf ij}$ on the brane finite as we take the large  volume limit.

Even though by varying the $f^a$'s we can vary the charges carried by a system, we do not have
enough variables to vary all charges independently. For example, if we begin with a D5-brane along
$\CY$, we have $h^{1,1}$ $f^a$'s and $\bar p^0$ which cannot be used to vary all
$2h^{1,1}+2$ charges independently.  This can be rectified by allowing the gauge fields to take
more general form than the restriction of $f^a \omega_a$ to the brane, but we shall now argue that
in the natural large volume limit these additional charges will not scale up appropriately so as to
be able to contribute to the D-instanton amplitude. For this let us again consider the case where
the parent brane is a D5-brane ($\bar p^0=1$). In the presence of a background 2-form field,
the U(1) gauge field on the brane is known to have a non-commutative instanton
solution \cite{Nekrasov:1998ss} that
carries D1-brane charge and is localized with finite width on a codimension 4 subspace. In this
case we expect supersymmetry to require this codimension 4 subspace to be a supersymmetric
cycle. In principle, the D1-brane charge induced this way can be varied independently by varying the
number of non-commutative instantons. However, since the D1-brane charge is
localized on the D5-brane, if we want to scale it up by
$\lambda^{2/3}$ as will be required to be able to contribute to the instanton amplitude, we need
to have large localized D1-brane
charge density on the D5-brane. This is not a natural large volume limit. For
this reason, we stick to the limit \refb{escalecharge} even though in this limit not all components of
charge contribute to the amplitude.

\bigskip

\noindent {\bf Acknowledgement:}
The work of A.S. was supported by the Infosys chair professorship and the
J. C. Bose fellowship of the Department of Science and Technology, India.
B.S. acknowledges funding support
from an STFC Consolidated Grant `Theoretical Particle Physics at City, University of London' ST/T000716/1.

\appendix

\section{Manifestly S-duality invariant metric}
\label{ap-fields}

In this paper we expressed the moduli space metric in terms of the NSNS axion $\sigma$,
identical to the one used in the type IIA formulation,
and the RR scalar fields \eqref{RRiib} that are  period integrals of the RR $p$-form potentials.
This choice is particularly convenient for comparison with  calculations based on string amplitudes
since these are the potentials $C^{(p)}$ that enter the expressions for the RR vertex operators, see appendix \ref{sc}.
However,  under the S-duality group
the fields $\tc_a$, $\tc_0$ and $\sigma$ transform  in a very complicated way,
making it hard to see the S-duality invariance of the metric \eqref{metricBtreelargeV}.
For this reason,
usually the metric is expressed in terms of a somewhat different set of fields.
If one introduces \cite{Louis:2002ny,Alexandrov:2008gh}
\be
\begin{split}
\nc_a=&\, -\tc_a+\hf\,\kappa_{abc} b^b c^c,
\qquad
\nc_0= -\tc_0+b^a\tc_a-\frac13\, (b^2c),
\\
&\quad
\psi= -\hf\,\sigma+\hf\, c^0\tc_0-\hf\, c^a\tc_a+\frac16\, (bc^2),
\end{split}
\ee
and defines $\tau=c^0+\I \tau_2 $, then the classical metric \eqref{metricBtreelargeV} takes the form
(see, e.g., \cite{Bohm:1999uk,Alexandrov:2017mgi})
\be
\begin{split}
\de s_{\rm cl}^2
=&\,
\frac{1}{\tau_2^2}\(2\de\tau_2+\frac{\tau_2}{2 V}\, \kappa_{ab}t^a\de t^b\)^2+\frac{(\de c^0)^2}{\tau_2^2}
+G_{ab}\de t^a\de t^b
\\
&\,
+\frac{G_{ab}}{\tau_2^2}\, (\de c^a-\tau \de b^a)(\de c^b-\bar\tau \de b^a)
+\frac{G^{ab}}{\tau_2^2 V^2}\, \nabla \nc_a\nabla \nc_b
\\
& +\frac{1}{\tau_2^4 V^2} \,\Bigl[\left| \de\psi+\tau\de\nc_0\right|^2
+ \left|(c^a-\btau b^a) \hnabla \nc_a\right|^2 \Bigr]
\\
&
- \frac{1}{\tau_2^4 V^2}\Bigl[(\de\psi+\tau\de\nc_0)(c^a-\btau b^a) + (\de\psi+\btau\de\nc_0)(c^a-\tau b^a)  \Bigr]
\hnabla \nc_a,
\end{split}
\label{classmetric}
\ee
where
\be
\begin{split}
\nabla \nc_a=&\, \de \nc_a +\frac12 \kappa_{abc} (c^b\de b^c-b^b\de c^c),
\\
\hnabla \nc_a=&\, \de \nc_a +\frac16 \kappa_{abc} (c^b\de b^c-b^b\de c^c).
\end{split}
\ee
Except the first line, \refb{classmetric}
is manifestly invariant under the following $SL(2,\IR)$ transformations
\be
\begin{split}
\tau\to\frac{a\tau+b}{c\tau+d},
&\qquad
t^a\to |c\tau+d| \, t^a,
\\
\(\begin{array}{c} c^a \\ b^a \end{array}\)\to \(\begin{array}{cc} a & b \\ c & d \end{array}\)\(\begin{array}{c} c^a \\ b^a \end{array}\),
&\qquad
\(\begin{array}{c} \nc_0 \\ \psi \end{array}\)\to \(\begin{array}{cc} d & -c \\ -b & a \end{array}\)\(\begin{array}{c} \nc_0 \\ \psi \end{array}\).
\end{split}
\label{Sdualtr}
\ee
In particular, the variable $\tau$ is the usual complexified coupling constant on which  $SL(2,\IR)$ acts by the familiar fractional-linear transformations. The first line in \eqref{classmetric} can also be rewritten in manifestly invariant way as follows:
\be
\frac{|\de\tau|^2}{\tau_2^2}
-\frac{1}{\tau_2 V}\, \kappa_{ab}\de (\sqrt{\tau_2} t^a) \de (\sqrt{\tau_2} t^b)
+\frac{1}{2 \tau_2 V^2 }\, (\kappa_{ab} t^a\de (\sqrt{\tau_2} t^b))^2.
\ee
Thus, this set of fields is adapted for studying S-duality of the hypermultiplet moduli
space in the type IIB formulation
and has been widely used in this context. However, in this paper S-duality is not used and therefore we prefer to work
in terms of $\tc_a$, $\tc_0$ and $\sigma$.

\section{K\"ahler moduli space and holomorphic cycles}
\label{ap-Kahler}

In this appendix we collect various useful relations that hold on the K\"ahler moduli space of a CY threefold.

First of all, note that for the classical prepotential \eqref{Fcl},
one finds the following expressions for the matrix of its second derivatives
\be
\begin{split}
\Re F_{\Lambda\Sigma} =&\,
\begin{pmatrix}
(t^2b)-\frac13\,(b^3)  & \hf\, \kappa_{abc}(b^b b^c -t^b t^c)\\ \hf\, \kappa_{abc}(b^b b^c -t^b t^c) & -\kappa_{abc}b^c
\end{pmatrix},
\\
N_{\Lambda\Sigma} =&\,
-2\Im F_{\Lambda\Sigma}=
\begin{pmatrix}
2\kappa_{cd}b^c b^d - 4V & \, -2\kappa_{ac} b^c \\ -2\kappa_{bc} b^c & 2\kappa_{ab}
\end{pmatrix},
\\
N^{\Lambda\Sigma} =&\,  -\frac{1}{4V}
\begin{pmatrix}
1 & b^a \\ b^b & -2V \kappa^{ab} + b^a b^b
\end{pmatrix} .
\end{split}
\label{FNN}
\ee

Next, let $\omega_a\in H^{1,1}(\IZ,\CY)$ be a basis of harmonic (1,1)-forms and
$\tom^a\in H^{2,2}(\IZ,\CY)$ be a basis of harmonic (2,2)-forms
such that they satisfy the following relations \cite{Candelas:1990pi}
\be
\omega_a\wedge \omega_b =\kappa_{abc}\, \tom^c\, ,
\qquad
\omega_a\wedge \tom^b=\delta_a^b \,\omega_{\CY}\, ,
\label{basisforms}
\ee
where $\omega_\CY$ is a 6-form normalized as $\int_\CY\omega_\CY=1$ and $\kappa_{abc}$ are intersection numbers
introduced in \eqref{def-kapV}.
This implies
\be
\omega_a\wedge \omega_b\wedge \omega_c =\kappa_{abc} \,\omega_\CY\, .
\label{omaomb}
\ee
The K\"ahler form $\bbom$ expanded in this basis gives rise to the moduli $\bbt^a$: $\bbom =\bbt^a\omega_a$,
and measures the volume of the CY threefold
\be
\bbV=\frac16\int_\CY \bbom \wedge \bbom \wedge \bbom =\frac16\, \kappa_{abc}\bbt^a\bbt^b \bbt^c.
\label{defbbV}
\ee

The action of the Hodge operator on this basis is~\cite{Strominger:1985ks}
\bea
\star\,\omega_a&=& -\bbom\wedge \omega_a+\frac{\bbkap_{ab}\bbt^b}{4\bbV}\,\bbom\wedge \bbom
=\bbV\bbG_{ab}\,\tom^b,
\label{Hodge-om}
\\
\star\,\tom^a&=&-\bbkap^{ab} \omega_b+\frac{\bbt^a}{2\bbV}\,  \bbom=\bbV^{-1}\bbG^{ab}\,\omega_b,
\label{Hodge-om2}
\\
\star\,\omega_\CY&=&\bbV^{-1},
\label{star1}
\eea
where $\bbkap_{ab}=\kappa_{abc}\bbt^c$, the metric $\bbG_{ab}$ is given by
\be
\bbG_{ab}\equiv -\frac{1}{\bbV}\(\bbkap_{ab} - {1\over 4\bbV}\, \bbkap_{ac}\bbt^c \bbkap_{bd} \bbt^d\) ,
\label{defbbG}
\ee
and $\bbG^{ab}$ is its inverse.
A particular case of these relations is
\be
\star\,\bbom=\hf\, \bbom^2,
\qquad
\star\,\bbom^2=2\,\bbom.
\label{Hodgeom}
\ee
Since for any two $p$-forms one has
\be
\alpha\wedge \star \beta=(\alpha, \beta)\, \bbV\omega_{\CY},
\label{innerwedge}
\ee
where we defined the inner product on $p$-forms
\be
(\alpha,\beta)=\frac{1}{p!}\, \alpha_{\bf i_1\cdots i_p}\beta^{\bf i_1\cdots i_p},
\label{definner}
\ee
the relations~\eqref{Hodgeom} and~\eqref{omaomb} imply
\be
(\omega_a,\bbom)=\frac{\omega_a\wedge \bbom^2}{2\bbV\omega_\CY} =\frac{1}{2\bbV}\,
\kappa_{abc}\bbt^b \bbt^c.
\label{eb7}
\ee
Note that this relation implies that the quantity on the l.h.s. is constant on $\CY$.

\medskip

Let us consider a $2n$-dimensional holomorphic cycle $\Lgi{2n}$ with the volume form
$\vi{2n}_\gamma$ given in \eqref{defvn} and a Poincar\'e dual form $\omega_\gamma^{(6-2n)}$
given in \refb{list-volume}. For any $2n$-form $\alpha^{(2n)}$, one has the obvious identity
\be\label{eyintegral}
\int_{\Lgi{2n}} \vi{2n}_\gamma  \(\vi{2n}_\gamma, \alpha^{(2n)}\)
= \int_{\Lgi{2n}}\alpha^{(2n)}\, .
\ee
Then we also have the relation
\be\label{eextraidentity}
\bbom^{n}\wedge \omega_\gamma^{(6-2n)}=n! \bbV_\gamma\omega_\CY\, ,
\ee
where $\bbV_\gamma$ is the volume of $\Lgi{2n}$ defined in \refb{edefbbv}.
To prove this, we first note that
it follows from \refb{list-volume}, \refb{basisforms} and \refb{omaomb} that the left hand side of
\refb{eextraidentity} is proportional to $\omega_\CY$ with the proportionality factor
being a constant on $\CY$. We can then integrate both sides over $\CY$. The left hand side
takes the form $\int_{\Lgi{2n}} \bbom^n$ which evaluates to $n!\, \bbV_\gamma$ due to
\refb{defvn}. On the right hand side $\int_\CY\omega_{\CY}$ gives 1. This fixes the constant
of proportionality to be $n!\bbV_\gamma$.

We now claim that for any $2n$-form $\alpha^{(2n)}$ on $\CY$ one has
\be
\(\alpha^{(2n)},\vi{2n}_\gamma\)=\frac{\bbV}{\bbV_\gamma}\(\alpha^{(2n)},\star\, \omega_\gamma^{(6-2n)}\)
\quad \hbox{on $\Lgi{2n}$}.
\label{equal-scpr}
\ee
To prove this relation, note that we can write
\be
\begin{split}
\int_{\Lgi{2n}}\alpha^{(2n)}=&\,
\int_\CY \alpha^{(2n)}\wedge \omega_\gamma^{(6-2n)}=
\bbV \int_\CY\omega_\CY\(\alpha^{(2n)},\star\, \omega_\gamma^{(6-2n)}\)
\\
=&\, \frac{\bbV}{n!\bbV_\gamma}\int_\CY \bbom^{n}\wedge \omega_\gamma^{(6-2n)}\(\alpha^{(2n)},\star\, \omega_\gamma^{(6-2n)}\),
\\
=&\,  \frac{\bbV}{\bbV_\gamma}\int_{\Lgi{2n}}\vi{2n}_\gamma\(\alpha^{(2n)},\star\, \omega_\gamma^{(6-2n)}\),
\end{split}
\label{int1}
\ee
where we used \eqref{innerwedge} and  \refb{eextraidentity}.
One the other hand, the original integral can also be rewritten as
\be
\int_{\Lgi{2n}}\vi{2n}_\gamma\(\alpha^{(2n)},\vi{2n}_\gamma\).
\label{int2}
\ee
Hence, the integrals \eqref{int1} and \eqref{int2} must be equal.
But in fact, since $\alpha^{(2n)}$ is arbitrary and, in particular, can be multiplied by any function on $\CY$,
the equality holds for the integrands as well. This proves \eqref{equal-scpr}.

Next, let $\bbom_\gamma$ denote the K\"ahler form projected along the cycle.
It should not be confused with $\omega_\gamma^{(6-2n)}$ appearing above.
It turns out that the mixed components of $\bbom_\gamma$ are proportional to those of
the projector $P_\gamma$ on the cycle $L_\gamma^{(2n)}$. Indeed,
\be
\bbom_\gamma^{s\bt}=(P_\gamma)^s_{s'}(P_\gamma)^{\bt}_{\bt'}\bbom^{s'\bt'}=-\I P_\gamma^{s\bt},
\label{prom}
\ee
where we used $\bbom^{s\bt}=-\I \bbg^{s\bt}$.

Now we claim that for any (1,1)-form $\alpha$ one has
\be
\alpha_{s\bt}P_\gamma^{s\bt}=\frac{\I}{(n-1)!} (\alpha\wedge \bbom^{n-1},\vi{2n}_\gamma).
\label{relform}
\ee
Indeed, for $n=1$ it is a direct consequence of \eqref{prom}:
\be
\I (\alpha,\vi{2}_\gamma)=\I\alpha_{s\bt}\bbom_\gamma^{s\bt}
=\alpha_{s\bt}P_\gamma^{s\bt}.
\ee
For $n=2$, similarly we have
\be
\begin{split}
\I (\alpha\wedge\bbom,\vi{4}_\gamma)=&\,
\frac{\I\, 6^2}{2\cdot 3\cdot 4!}\, \alpha_{\bf ij}\,\bbom_{\bf kl}\(\bbom_\gamma^{\bf ij}\,\bbom_\gamma^{\bf kl}
+2\,\bbom_\gamma^{\bf ik}\,\bbom_\gamma^{\bf lj}\)
\\
=&\, -\frac{\I}{2}\, \alpha_{s\bt}\(2\,P_\gamma^{s\bt}\,(\bbom_{u\bv}\,P_\gamma^{u\bv})
+2\,P_\gamma^{s\bu}\,\bbom_{\bu v}\,P_\gamma^{v\bt}\)
=\alpha_{s\bt}P_\gamma^{s\bt},
\end{split}
\ee
where we again used \eqref{prom} and that for $2n$-dimensional cycle, $(P_\gamma)^s_s=n$.
Finally, for $n=3$ there is no projection and we find
\be
P_\gamma^{s\bt}= \bbg^{s\bt},
\label{projhol}
\ee
so that
\be
\begin{split}
\frac{\I}{2}\, (\alpha\wedge\bbom^2,\vi{6}_\gamma)=&\,
\frac{\I}{12\cdot 15\cdot 6!}\(\frac{6!}{8}\)^2 \alpha_{\bf ij}\,\bbom_{\bf kl}\,\bbom_{\bf mn}
\(\bbom^{\bf ij}\,\bbom^{\bf kl}\,\bbom^{\bf mn}
+4\,\bbom^{\bf ik}\,\bbom^{\bf lj}\,\bbom^{\bf mn}
\right.
\\
&\, \left. \qquad
+2\, \bbom^{\bf ij}\, \bbom^{\bf km}\, \bbom^{\bf nl}
+8\, \bbom^{\bf ik}\bbom^{\bf mj}\, \bbom^{\bf nl}\)
\\
=&\, \frac{1}{16}\, \alpha_{s\bt}\bbg^{s\bt} \, 2\(6^2-24-12+8\)
=\alpha_{s\bt}P_\gamma^{s\bt}.
\end{split}
\ee

\section{Boundary condition on spin fields} \label{espinboundary}

In this appendix we shall discuss the choice of boundary conditions on the spin fields.
We shall work in background with vanishing NSNS 2-form field. When this field is present
in the background, \refb{eqb1} remains unchanged but \refb{boundarycond}
and subsequent equations
are modified in a way described in \S\ref{s7}.

We begin with the operator product expansions:
\begin{subequations} \label{eqb1}
\bea\label{eaa1}
e^{-\phi}\psi^M(z) \ e^{-\phi/2} S_\alpha(w) &=& {\I\over 2}\, (z-w)^{-1}\,
(\Gamma^M)_{\alpha\beta} e^{-3\phi/2}\, S^\beta(w)
+\cdots ,
\\
\label{eaa2}
e^{-3\phi/2} S^\alpha(z)\ e^{-\phi/2} S_\beta(w)   &=& (z-w)^{-2} \, \delta^\alpha_\beta \,
e^{-2\phi}(w)+\cdots ,
\\
\label{eaa3}
e^{-\bar\phi}\bar\psi^M(\bar z) \ e^{-\bar \phi/2} \bar S_\alpha(\bar w)  &=& {\I\over 2}\,
(\bar z-\bar w)^{-1}\,
(\Gamma^M)_{\alpha\beta} e^{-3\bar \phi/2}\, \bar S^\beta(\bar w)
+\cdots ,
\\
\label{eaa4}
e^{-3\bar \phi/2} \bar S^\alpha(\bar z)\ e^{-\bar \phi/2} \bar S_\beta(\bar w)   &=&
(\bar z-\bar w)^{-2} \, \delta^\alpha_\beta \,
e^{-2\bar \phi}(\bar w)+\cdots ,
\eea
\end{subequations}
where $\cdots$ denote less singular terms. We also consider the possible choice of boundary
conditions on the real  axis:
\begin{subequations}\label{boundarycond}
\bea
\label{eaa7}
e^{-\bar\phi}\bar\psi^{M_\perp}(\bar z)
&=&
-e^{-\phi}\psi^{M_\perp}(z),
\quad
e^{-\bar\phi}\bar\psi^{M_\parallel}(\bar z)
=e^{-\phi}\psi^{M_\parallel}(z)
\vphantom{A\over A}
\\
\label{eaa5}
e^{-\bar\phi/2}\bar S_\alpha(\bar z)  &=& {\ve\over (2k)!} \, v_{M_1\cdots M_{2k}}
(\Gamma^{M_1\cdots M_{2k}})_\alpha^{~\beta} \, e^{-\phi/2} S_\beta(z)\, ,
\\
\label{eaa6}
e^{-3\bar\phi/2}\bar S^\alpha(\bar z)  &=&  {\ve'\over (2k)!} \,  v_{M_1\cdots M_{2k}}
(\Gamma^{M_1\cdots M_{2k}})^\alpha_{~\beta} \, e^{-3\phi/2} S^\beta(z)\, ,
\eea
\be \label{eaa8}
e^{-2\bar\phi}(\bar z) = e^{-2\phi}(z)\, ,
\ee
\end{subequations}
where $\ve$ and $\ve'$ are possible phases that we want to determine and
$v_{M_1\cdots M_{2k}}$ is the volume form along the Euclidean D$(2k-1)$-brane.
Even though we have displayed the arguments of holomorphic fields as $z$ and the
anti-holomorphic fields as $\bar z$, we must remember that on the real axis $z=\bar z$.
In \refb{eaa7}, $M_\perp$ and $M_\parallel$ denote directions transverse and tangential
to the D-brane, respectively.

Using \refb{boundarycond}, we can express \refb{eaa3} as
\be
\begin{split}
&\, - e^{-\phi}\psi^{M_\perp}(z) \, {\ve\over (2k)!} \,  v_{M_1\cdots M_{2k}}
(\Gamma^{M_1\cdots M_{2k}})_\alpha^{~\beta} \, e^{-\phi/2} S_\beta(z)
\\
 =& \, {\I\over 2}\, (z-w)^{-1}\,
(\Gamma^{M_\perp})_{\alpha\beta} \, {\ve'\over (2k)!} \, v_{M_1\cdots M_{2k}}
(\Gamma^{M_1\cdots M_{2k}})^\beta_{~\gamma} \, e^{-3\phi/2} S^\gamma(z)\, .
\end{split}
\ee
Manipulating the left hand side using \refb{eaa1}, we get
\be \label{eaa9}
- \ve \, v_{M_1\cdots M_{2k}} \Gamma^{M_1\cdots M_{2k}} \Gamma^{M_\perp} =
\ve' \,v_{M_1\cdots M_{2k}} \Gamma^{M_\perp} \Gamma^{M_1\cdots M_{2k}}  \quad
\Rightarrow \quad \ve = -\ve'\, .
\ee
Note that in type IIA string theory $2k$ is replaced by $2k+1$ and we would get an
extra minus from the gamma matrix commutators, giving $\ve=\ve'$.

Next we use the boundary conditions \refb{boundarycond} in \refb{eaa4} to get
\be
\begin{split}
&\,{\ve'\over (2k)!} \,  v_{M_1\cdots M_{2k}}
(\Gamma^{M_1\cdots M_{2k}})^\alpha_{~\gamma} \, e^{-3\phi/2} S^\gamma(z)\,
{\ve\over (2k)!} \, v_{N_1\cdots N_{2k}}
(\Gamma^{N_1\cdots N_{2k}})_\beta^{~\delta} \, e^{-\phi/2} S_\delta(w)
\\
= &\, (z-w)^{-2} \, \delta^\alpha_\beta \,
e^{-2\phi}(w).
\end{split}
\ee
Using \refb{eaa2}, this gives
\be\label{eaa10}
{\ve'\over (2k)!} \, v_{M_1\cdots M_{2k}} \, {\ve\over (2k)!} \, v_{N_1\cdots N_{2k}}
\, \[ (\Gamma^{M_1\cdots M_{2k}}) (\Gamma^{N_1\cdots N_{2k}})^T\]^\alpha_{~\beta}
= \delta^\alpha_{~\beta} \quad \Rightarrow \quad
\ve\, \ve' = 1\, .
\ee
Using \refb{eaa9} and \refb{eaa10}, we see that we have $\ve=\pm \I$, $\ve'=\mp \I$. We shall
choose $\ve=-\I$ and $\ve'=\I$ so that the boundary conditions \refb{eaa5} and \refb{eaa6} take the form:
\be
\begin{split}
e^{-\bar\phi/2}\bar S_\alpha(\bar z) =&\,  - {\I\over (2k)!} \, v_{M_1\cdots M_{2k}}
(\Gamma^{M_1\cdots M_{2k}})_\alpha^{~\beta} \, e^{-\phi/2} S_\beta(z)\, ,
\\
e^{-3\bar\phi/2}\bar S^\alpha(\bar z) = &\, {\I\over (2k)!} \, v_{M_1\cdots M_{2k}}
(\Gamma^{M_1\cdots M_{2k}})^\alpha_{~\beta} \, e^{-3\phi/2} S^\beta(z)\, .
\end{split}
\label{SSbc}
\ee

\section{Normalization of the RR fields} \label{sb}

In this appendix we shall fix the normalization of the RR fields by analyzing the two-point
function of the gauge invariant field strength. The results
of this appendix will be insensitive to the presence of a constant background NSNS 2-form field, since this
does not affect the closed string sector.

We introduce the off-shell level zero RR field in Siegel gauge in the  $(-1/2,-1/2)$ picture:
\be\label{ephiRrep}
|\phi_R\rangle = \int{d^{10}p\over (2\pi)^{10}}\,
F^{\alpha\beta}(p) \, c\, \bar c\, e^{-\phi/2} S_\alpha \,e^{-\bar\phi/2}
\bar S_\beta \, e^{\I p.X}(0)|0\rangle\, ,
\ee
and in the $(-3/2,-3/2)$ picture
\be \label{efg1rep}
|\wt\phi_R\rangle = \int{d^{10}p\over (2\pi)^{10}}\,
A_{\alpha\beta}\, c\, \bar c \,e^{-3\phi/2}\, S^\alpha \, e^{-3\bar\phi/2}\, \bar S^\beta (0)
 e^{\I p.X}(0)|0\rangle\, .
\ee
We have chosen to work in  Siegel gauge for convenience; the two-point function of the
gauge invariant field strength will be independent of the choice of gauge.
The quadratic part of the string field theory action is given by \cite{deLacroix:2017lif}
\be\label{erraction}
S_{RR}=4\, \[ -{1\over 2} \langle \tilde \phi_R | c_0^- \, \XX_0\bar\XX_0\,
(Q_B+\bar Q_B)\, |\tilde \phi_R\rangle
+ \langle \tilde \phi_R | c_0^- \, (Q_B+\bar Q_B)\, |\phi_R\rangle \],
\ee
where $Q_B,\bar Q_B$ are the holomorphic and anti-holomorphic BRST operators and
$\XX_0,\bar \XX_0$ are the zero modes of the holomorphic and anti-holomorphic
picture changing operators. All these operators have been reviewed in \cite{Alexandrov:2021shf}.
The overall normalization factor of 4 has been chosen to ensure that the kinetic term in the NSNS
sector has conventional normalization \cite{Sen:2021tpp},
but once this has been fixed, we must use the same
normalization in the RR sector unless we change the rules for the interaction terms in string
field theory. Using the operator
product expansions of the world-sheet fields which can also be found in \cite{Alexandrov:2021shf},
the action \refb{erraction} evaluates to
\be
S_{RR}= \int{d^{10}p\over (2\pi)^{10}}\, \[-{1\over 8}\,  A_{\alpha\beta}(-p)
A_{\gamma\delta}(p) \, p^2 \!\sp^{\alpha\gamma}\!\!\sp^{\beta\delta}
+ p^2 A_{\alpha\beta}(-p) F^{\alpha\beta}(p)\] .
\ee
We can eliminate $A_{\alpha\beta}$ using its equation of motion and write the action as
\be\label{errcompact}
S_{RR} = 2\, \int{d^{10}p\over (2\pi)^{10}}\,  {1\over p^2} \, F^{\alpha\beta}(-p)
\!\sp_{\alpha\gamma}\!\sp_{\beta\delta} \, F^{\gamma\delta}(p)\, .
\ee
From this action we can calculate the propagator
of $F^{\alpha\beta}$:
\be\label{ebb6}
\llangl F^{\alpha\beta}(p') \,F^{\gamma\delta}(p) \rrangl = -
(2\pi)^{10} \delta^{(10)}(p+p') \, {1\over 4p^2}\! \sp^{\alpha\gamma}
\!\!\sp^{\beta\delta} ,
\ee
where $\llangl\,\cdot\, \rrangl$ denotes expectation value in string field theory.

We now expand $F^{\alpha\beta}$ as
\be\label{ebb7}
F^{\alpha\beta}={\I\over 2}\, \sum_{k=0}^4 {1\over (2k+1)!} \, \bbF^{(2k+1)}_{M_1\cdots M_{2k+1}}
\, (\Gamma^{M_1\cdots M_{2k+1}})^{\alpha\beta} \, ,
\ee
where due to chirality of the spinors $S_\alpha$,
$\bbF^{(2k+1)}$ satisfies a self-duality constraint
\be\label{edualcon}
* \bbF^{(2k+1)} = (-1)^k\, \bbF^{(9-2k)}\, .
\ee
Here we have used the following conventions
\be
\eps^{01\cdots 9}=1,
\qquad
(\Gamma^{01\cdots 9})_\alpha^{~\beta}=\delta_\alpha^{~\beta}.
\ee
The reason for including the factor of $\I$ in \refb{ebb7} will become clear soon.
The factor of $2$ in the denominator has been included to compensate for the fact that
the contribution from  $\bbF^{(2k+1)}$ and $\bbF^{(9-2k)}$ in \refb{ebb7} are
identical due to the self-duality constraint \refb{edualcon}. With this understanding,
we have
\be\label{ebb8}
\bbF^{(2k+1)}_{M_1\cdots M_{2k+1}} = -{\I\over 16}\, \Tr(\Gamma_{M_{2k+1}\cdots M_{1}}
F)  \, .
\ee
For $k=2$ there is a self-duality constraint on the field strength $\bbF^{(5)}$. So even though
$\bbF^{(5)}$ appears only once in the sum in  \refb{ebb7},  the same component
$\bbF^{(5)}_{M_1\cdots M_{5}}$ appears
twice in the sum: once as a coefficient of $(\Gamma^{M_1\cdots M_{5}})^{\alpha\beta}$
and once more as a coefficient of $\Gamma$ with complementary indices.
Therefore, in \refb{ebb8} we still get a factor of 2 that cancels the factor of 2 in the denominator
of \refb{ebb7}.
Using \refb{ebb6} and \refb{ebb8} we get
\be \label{ebb9}
\begin{split}
& \llangl \bbF^{(2k+1)}_{M_1\cdots M_{2k+1}} (p')\,  \bbF^{(2k+1)}_{N_1\cdots N_{2k+1}} (p)\rrangl
\\
=&\, - {1\over 256} \(\Gamma^{M_{2k+1}\cdots M_{1}}\)_{\alpha\beta}
\(\Gamma^{N_{2k+1}\cdots N_{1}}\)_{\gamma\delta} \llangl F^{\alpha\beta}(p')\,F^{\gamma\delta}(p) \rrangl
\\
=&\, (2\pi)^{10} \delta^{(10)}(p+p') \, {1\over 4p^2} \,  {1\over 256}
\Tr ( \Gamma^{M_1\cdots M_{2k+1}} \!\sp \, \Gamma^{N_{2k+1}\cdots N_{1}} \! \sp)
\\
=&\, (2\pi)^{10} \delta^{(10)}(p+p') \, {1\over 64 p^2} \, \bigg[
- p^2 \delta^{M_1\cdots M_{2k+1},N_1\cdots N_{2k+1}}
\\ &\,
+ 2 \( p^{M_1} p^{N_1} \,
\delta^{M_2\cdots M_{2k+1},N_2\cdots N_{2k+1}} +\hbox{cyclic perm.}\)
\bigg],
\end{split}
\ee
where the $\delta$ symbol can be defined as in \refb{edefdelta} after suitable generalization to ten
dimensions and `cyclic perm.' include cyclic permutations of two sets, $M_1,\cdots, M_{2k+1}$ and
$N_1,\cdots, N_{2k+1}$, producing $(2k+1)^2$ terms. Note that only the
second term produces a pole in the propagator.

We shall now compare this with the two-point function constructed from the action
\be\label{ebb10}
S = -{a_k \over 2 \cdot (2k+1)!} \int d^{10}x\, \bbF^{(2k+1)}_{M_1\cdots M_{2k+1}}
\bbF^{(2k+1)M_1\cdots M_{2k+1}},
\ee
where $a_k$ is a constant and
\be\label{ebb11}
\bbF^{(2k+1)}_{M_1\cdots M_{2k+1}}=\p_{M_1} \bbC^{(2k)}_{M_2\cdots M_{2k+1}} +
\hbox{cyclic perm. with sign}\, .
\ee
Our goal will be to compute the two-point function of $\bbF^{(2k+1)}$ in this theory.
Since it is gauge invariant, we can use the harmonic gauge:
\be
\p^{M_1} \bbC^{(2k)}_{M_1\cdots M_{2k}}=0\, .
\ee
In this gauge the action takes the form
\be\label{eactiontrial}
\begin{split}
S =&\, {a_k \over 2 \cdot (2k)!} \int d^{10}x\, \bbC^{(2k)}_{M_1\cdots M_{2k}} \p_M\p^M
\bbC^{(2k)M_1\cdots M_{2k}}
\\
=&\, - {a_k \over 2 \cdot (2k)!}
\int{d^{10}p\over (2\pi)^{10}}\, \bbC^{(2k)}_{M_1\cdots M_{2k}}(-p)\,
p^2\, \bbC^{(2k)M_1\cdots M_{2k}}(p).
\end{split}
\ee
From this we get
\be
\llangl \bbC^{(2k)}_{M_1\cdots M_{2k}} (p')  \, \bbC^{(2k)}_{N_1\cdots N_{2k}} (p)\rrangl
=  (2\pi)^{10} \delta^{(10)}(p+p') \, {1\over a_k p^2}\,
\delta_{M_1\cdots M_{2k},N_1\cdots N_{2k}}\, ,
\ee
and hence
\be\label{ebb15}
\begin{split}
&\llangl \bbF^{(2k+1)}_{M_1\cdots M_{2k+1}} (p') \, \bbF^{(2k+1)}_{N_1\cdots N_{2k+1}} (p)\rrangl
\\
=&\,  (2\pi)^{10} \delta^{(10)}(p+p') \, {1\over a_k p^2}
\( p^{M_1} p^{N_1} \, \delta^{M_2\cdots M_{2k+1},N_2\cdots N_{2k+1}} +\hbox{cyclic perm.}\) .
\end{split}
\ee

Let us now compare the pole terms in \refb{ebb9} and \refb{ebb15}. The reason that we can only
compare the pole terms is that off-shell, the string field theory action is different from the
one given in \refb{ebb10}, \refb{ebb11}. This will become clear in appendix \ref{sc}, where we shall
see that in string field theory the field strength $\bbF^{(2k+1)}$ is given
both in terms of a $2k$-form potential and a dual $(8-2k)$-form potential, and therefore
the propagator of the field strength off-shell is not expected to agree with the one
computed from the action \refb{ebb10}, \refb{ebb11}
where we use only the $2k$-form potential. Comparison of
the pole terms in \refb{ebb9} and \refb{ebb15} gives
\be
a_k=32\, .
\ee
Therefore, the action for $\bbC^{(2k)}$ is
\be\label{ebb17}
S = -{16 \over (2k+1)!} \int d^{10}x \,
\bbF^{(2k+1)}_{M_1\cdots M_{2k+1}} \bbF^{(2k+1)M_1\cdots M_{2k+1}}
=-16\int \bbF^{(2k+1)} \wedge * \bbF^{(2k+1)}\, .
\ee

Note that we could also formulate the theory in terms of the dual fields,
{\it e.g.} $\bbC^{(2k)}$ could have been traded for $\bbC^{(8-2k)}$ by replacing $k$ by
$4-k$ in \refb{ebb17}. Either action could be used to compute the
propagator of the field strength as well as its dual, and we shall get the same result for
the residue at the pole even though the finite parts will in general differ.
However, we need to use only one of the two actions and not add them.

Finally, note that for $k=2$ the action \refb{ebb17} vanishes if we impose the self-duality
constraint from the beginning. However, we can still use \refb{ebb17} with the
understanding that we first use this to calculate two-point function of
$\bbF^{(5)}$ and then use it only for self-dual $\bbF^{(5)}$. If on the other hand we want
to couple the action to gravity and compute the energy-momentum tensor, then we
get a double counting since the energy momentum tensor will receive a particular
contribution twice  --- once from the field and once more from the dual, while in actual
practice we should have this contribution only once. In this case we need to include
an extra factor of 1/2 compared to \refb{ebb17}.
Therefore such an action has the form:
\be\label{ebb18}
S = -8\int \bbF^{(5)} \wedge * \bbF^{(5)}\, .
\ee

\section{Vertex operator for  the RR fields and string field theory action} \label{sc}

In this appendix we shall
describe the construction of the on-shell
vertex operators of the RR fields in different pictures
in ten-dimensional type IIB string theory.
The results
of this appendix will also be insensitive to the presence of background NSNS 2-form field.
Most of them follow from the results of appendix D in \cite{Alexandrov:2021shf} after exchange
of chiral and anti-chiral spinor indices in the anti-holomorphic sector.

In the $(-1/2,-1/2)$ picture, the off-shell RR field at level 0, describing massless fields, is given
by a state $|\phi_R\rangle$ satisfying the conditions
$b_0^-|\phi_R\rangle=0$, $L_0^-|\phi_R\rangle=0$.
The general form of $|\phi_R\rangle$ is:
\be\label{ephiR}
|\phi_R\rangle = \int{d^{10}p\over (2\pi)^{10}}\,
F^{\alpha\beta}(p) \, c\, \bar c\, e^{-\phi/2} S_\alpha \,e^{-\bar\phi/2}
\bar S_\beta \, e^{\I p.X}(0)|0\rangle\, ,
\ee
for some set of functions $F^{\alpha\beta}(p)$.
This has the same form as \refb{ephiRrep} since there are no additional states at level
zero even after relaxing the Siegel gauge condition.
Acting with the BRST operator, we find
\be
\begin{split}
& (Q_B+\bar Q_B) |\phi_R\rangle =
\int{d^{10}p\over (2\pi)^{10}}\,
F^{\alpha\beta}(p) \, \Bigg[ {p^2\over 4} (\p c+\bar \p\bar c) \,
c\, \bar c\, e^{-\phi/2} S_\alpha \,e^{-\bar\phi/2}
\bar S_\beta \, e^{\I p.X}(0)|0\rangle
\\
& \quad
- {1\over 4} (\sp)_{\alpha\gamma} c\, \bar c\, \eta\, e^{\phi/2} S^\gamma e^{-\bar\phi/2}
\bar S_\beta \, e^{\I p.X}(0)|0\rangle   + {1\over 4} (\sp)_{\beta\delta} \,
c\, \bar c\, e^{-\phi/2} S_\alpha \bar\eta\, e^{\bar\phi/2}
\bar S^\delta \, e^{\I p.X}(0)|0\rangle\Bigg]\, .
\end{split}
\ee
Therefore, BRST invariance of the vertex operator will require
\be\label{errbrst}
p^2\, F^{\alpha\beta}(p) =0,
\qquad
F^{\alpha\beta}(p) (\sp)_{\alpha\gamma}=0,
\qquad
F^{\alpha\beta}(p) (\sp)_{\beta\delta} =0\, ,
\ee
where $\sp\equiv p_M\Gamma^M$.
We now expand $F^{\alpha\beta}$ as in \refb{ebb7}.
Substituting this into \refb{errbrst} we get:
\begin{align}
p^2 \bbF^{(2k+1)}_{M_1\cdots M_{2k+1}}(p)&=0,
\qquad
p^{M_1}  \bbF^{(2k+1)}_{M_1\cdots M_{2k+1}}=0,
\qquad
p_{[M_0} \bbF^{(2k+1)}_{M_1\cdots M_{2k+1}]}=0\, .
\end{align}
These are the usual on-shell conditions for a $(2k+1)$-form
field strength. $\bbF^{(5)}$ also satisfies a self-duality constraint.

We will also need the BRST invariant vertex operator for these states in the $(-1/2, -3/2)$ picture.
To find it, we begin with the general form of level zero string field in the $(-3/2,-3/2)$ picture:
\bea \label{efg1}
|\wt\phi_R\rangle &=& \int{d^{10}p\over (2\pi)^{10}}\, \bigg[
A_{\alpha\beta}\, c\, \bar c \,e^{-3\phi/2}\, S^\alpha \, e^{-3\bar\phi/2}\, \bar S^\beta (0)
+ E_{\alpha}^{~\beta} \, (\p\, c+\bar\p\, \bar c)\, c\, \bar c\, e^{-3\phi/2}\, S^\alpha \, \bar\p \bar\xi \,
e^{-5\bar\phi/2}\bar S_\beta(0)
\non\\
&&
+ D^\alpha_{~\beta} \, (\p\, c+\bar\p\, \bar c)\,  c\, \bar c\, \p \xi \,
e^{-5\phi/2} S_\alpha \, e^{-3\bar\phi/2}\, \bar S^\beta(0)
\bigg] e^{\I p.X}(0)|0\rangle\, .
\eea
From this we get
\be\label{efg2}
\begin{split}
(Q_B+\bar Q_B) |\wt\phi_R\rangle =&\, \int{d^{10}p\over (2\pi)^{10}} \[ {p^2\over 4} \,
A_{\alpha\beta}
- {1\over 4} \, E_{\alpha}^{~\gamma}  (\sp)_{\gamma\beta}
+ {1\over 4}\, D^{\gamma}_{~\beta} (\sp)_{\gamma\alpha} \]
\\ & \hskip 1in
\,\times(\p c+\bar\p\bar c) \,
c\, \bar c \,e^{-3\phi/2} S^\alpha \, e^{-3\bar\phi/2}\, \bar S^\beta\,
e^{\I p.X}(0)|0\rangle\, , \end{split}
\ee
\be\label{efg3}
\begin{split}
|\wt\phi_{-1/2,-3/2}\rangle \equiv\ & \bar\XX_0 \, |\wt\phi_R\rangle =
\int{d^{10}p\over (2\pi)^{10}}\, \bigg[{1\over 2}\,
A_{\alpha\beta}\, (\sp)^{\beta\gamma} \, c\, \bar c \,e^{-3\phi/2}\, S^\alpha \,
e^{-\bar\phi/2}\, \bar S_\gamma (0) \\ &
- {1\over 2} \, E_{\alpha}^{~\beta} \, c\, \bar c\, e^{-3\phi/2} S^\alpha \,  e^{-\bar\phi/2}\,
\bar S_\beta (0) \\ &
+ {1\over 2} \, D^{\alpha}_{~\beta} \, (\sp)^{\beta\gamma}
 (\p\, c+\bar\p\, \bar c)\,  c\, \bar c\, \p \xi \,
e^{-5\phi/2} S_\alpha \, e^{-\bar\phi/2}\, \bar S_\gamma(0) \\ &
- {1\over 2} \, D^{\alpha}_{~\beta} \, c\, \bar c\, \p \xi \,
e^{-5\phi/2} S_\alpha \, \bar\eta\, e^{\bar\phi/2}\, \bar S^\beta(0)
\bigg] e^{\I p.X}(0)|0\rangle\, ,
\end{split}
\ee
and the $(-1/2,-1/2)$ picture string field
\be \label{efg4}
\begin{split}
|\phi_{R}\rangle = \XX_0 \, |\wt\phi_{-1/2,-3/2}\rangle = &
\int{d^{10}p\over (2\pi)^{10}}\, \[{1\over 4}\,
A_{\alpha\beta}\, (\sp)^{\beta\gamma} (\sp)^{\alpha\delta} - {1\over 4} \, E_{\alpha}^{~\gamma} \, (\sp)^{\alpha\delta}
+ {1\over 4} \, D^{\delta}_{~\beta} \, (\sp)^{\beta\gamma}\]
\\
& \hskip 1in
\,\times c\, \bar c \,e^{-\phi/2}\, S_\delta \,
e^{-\bar\phi/2}\, \bar S_\gamma \, e^{\I p.X}(0)|0\rangle\, .
\end{split}
\ee
Comparison of \refb{ephiR} and \refb{efg4} leads to the identification
\be \label{efg5}
F^{\alpha\beta} = {1\over 4}\( \sp \, A \!\sp \, - \!\sp\, E + D\! \!\sp\)^{\alpha\beta}\, .
\ee
Also \refb{efg2} leads to the on-shell condition
\be
p^2  A - E\!\! \sp\, + \!\sp \, D =0\, ,
\ee
which implies  \refb{errbrst}.

From \refb{efg5} we see that the description in terms of the matrices $A$, $E$ and $D$
provides a redundant description of the field strength $F^{\alpha\beta}$.  Indeed,
$F^{\alpha\beta}$ is invariant under the transformation
\be
A\to A+\Lambda,
\qquad
E\to E + \Lambda'\! \!\sp\, ,
\qquad
D\to D+\!\sp\, (\Lambda' -\Lambda)\, .
\ee
The redundant degrees of freedom are related to gauge transformation parameters in the $(-3/2,-3/2)$ picture.
We shall choose a particular gauge in which $A=0$. Therefore, we have on-shell
\be\label{efg6}
F = {1\over 4} \(D \!\sp\, -\!\! \sp\, E\),
\qquad
\sp\, D = E \!\!\sp\, .
\ee
We can decompose $E_{\alpha}^{~\beta}$ and $D^{\alpha}_{~\beta}$ as:
\be\label{efg7}
E_{\alpha}^{~\beta} = {1\over 2}\,
\sum_{k=0}^5 {1\over (2k)!}\,
E^{(2k)}_{M_1\cdots M_{2k}} (\Gamma^{M_1\cdots M_{2k}})_{\alpha}^{~\beta}\, ,
\qquad
D^{\alpha}_{~\beta}= {1\over 2}\sum_{k=0}^5 {1\over (2k)!}\,
D^{(2k)}_{M_1\cdots M_{2k}} (\Gamma^{M_1\cdots M_{2k}})^{\alpha}_{~\beta}\, ,
\ee
with the coefficients satisfying the self-duality constraints
\be\label{edualbc}
* E^{(2k)} = (-1)^{k+1} \, E^{(10-2k)}, \qquad * D^{(2k)} = (-1)^{k} \, D^{(10-2k)}\, .
\ee
Using \refb{ebb8}, \refb{efg6}, \refb{efg7} and \refb{edualbc}, we get
\bea\label{efg8}
\bbF^{(2k+1)}_{M_1M_2M_{2k+1}} &=& -{\I\over 16}\, \Tr(\Gamma_{M_{2k+1}M_{2k}\cdots
M_{1}} F) = {\I\over 4}
p^M \(D^{(2k+2)}+E^{(2k+2)}\)_{MM_1 \cdots M_{2k+1}}
\\ &&
-{\I \over 4} \[
p_{M_1} \(D^{(2k)}-E^{(2k)}\)_{M_2\cdots M_{2k+1}} +\hbox{cyclic perm. of $M_1,
\cdots, M_{2k+1}$
with sign}\].
\non
\eea
Also, multiplying the second equation in \refb{efg6} by $\Gamma_{M_1\cdots M_{2k+1}}$
and taking the trace, we get
\be\label{efg9}
\begin{split}
&\, p^M \(D^{(2k+2)}+E^{(2k+2)}\)_{MM_1 \cdots M_{2k+1}}
\\
=-&\,  \[
p_{M_1} \(D^{(2k)}-E^{(2k)}\)_{M_2\cdots M_{2k+1}} +\hbox{cyclic perm. of $M_1,
\cdots, M_{2k+1}$
with sign}\].
\end{split}
\ee
Using this, we can express \refb{efg8} as
\be\label{efg10}
\bbF^{(2k+1)}_{M_1\cdots M_{2k+1}} = -{\I\over 2} \[
p_{M_1} \(D^{(2k)}-E^{(2k)}\)_{M_2\cdots M_{2k+1}} +\hbox{cyclic perm. of $M_1,
\cdots, M_{2k+1}$
with sign}\].
\ee
One can check that \refb{efg8}-\refb{efg10} are consistent with the duality relations given in
\refb{edualcon} and \refb{edualbc}.

Let us now define
\be \label{eccdef}
\bbC^{(2k)} = {1\over 2} \( E^{(2k)}-D^{(2k)}\)  .
\ee
Then the fields $\bbC^{(2k)}$ have no self-duality constraint and
we have, using \refb{edualbc},
\be
*\bbC^{(10-2k)} = {1\over 2} \, (-1)^k\, \( E^{(2k)}+D^{(2k)}\) ,
\ee
which gives
\be
E^{(2k)}= \bbC^{(2k)} + (-1)^k *\bbC^{(10-2k)}, \qquad D^{(2k)}= -\bbC^{(2k)} + (-1)^k *\bbC^{(10-2k)}
\, .
\ee
Also \refb{efg7} may be expressed as,
\be\label{eEBC}
E_{\alpha}^{~\beta} =
\sum_{k=0}^5 {1\over (2k)!}\,
\bbC^{(2k)}_{M_1\cdots M_{2k}} (\Gamma^{M_1\cdots M_{2k}})_{\alpha}^{~\beta}\, ,
\qquad
D^{\alpha}_{~\beta}= - \sum_{k=0}^5 {1\over (2k)!}\,
\bbC^{(2k)}_{M_1\cdots M_{2k}} (\Gamma^{M_1\cdots M_{2k}})^{\alpha}_{~\beta}\, .
\ee
Due to the absence of self-duality constraint on $\bbC^{(2k)}$,
the two  expressions in \refb{eEBC} are
not simply related despite the similarity in their appearance, since the $\Gamma$ matrices
in the two expressions are different.
Eq. \refb{efg10} now gives
\be \label{efcrelation}
\bbF^{(2k+1)}_{M_1\cdots M_{2k+1}} = {\I} \[
p_{M_1} \(\bbC^{(2k)}\)_{M_2\cdots M_{2k+1}} +\hbox{cyclic perm. of $M_1,
\cdots, M_{2k+1}$
with sign}\] .
\ee

Note that while the zero modes of the fields $\bbC^{(2k)}$ are all independent, their field
strengths are related due to the
self-duality constraint \refb{edualcon}
on $\bbF^{(2k+1)}$. This is reflected in \refb{efg9}, which can be written in terms
of $\bbC^{(2k)}$ using \refb{eEBC}.
In our analysis we shall need to consider the
fields $\bbC^{(2k)}_{\bf m_1\cdots m_{2k}}$ along the internal directions of $\CY$ with
the only non-vanishing components of fields strengths proportional to
$\bbF^{(2k+1)}_{\mu \bf m_1\cdots m_{2k+1}}=\I p_\mu \bbC^{(2k)}_{\bf m_1\cdots m_{2k}}$.
Self-duality constraint \refb{edualcon} on $\bbF^{(2k+1)}$ relates these
components to different components  $\bbF^{(9-2k)}_{\nu\rho\sigma\bf n_1\cdots n_{6-2k}}$.
Therefore, we can treat the $\bbF^{(2k+1)}_{\mu \bf m_1\cdots m_{2k+1}}$ given above as
independent and replace $\bbF^{(9-2k)}_{\nu\rho\sigma\bf n_1\cdots n_{6-2k}}$ in terms of
these components to express
\refb{ebb7} as
\be
F^{\alpha\beta} = -\sum_{k=0}^3 {1\over (2k)!} \, p_{\mu} \bbC^{(2k)}_{\bf m_1\cdots m_{2k}}
\(\Gamma^{\mu \bf m_1\cdots m_{2k}}\)^{\alpha\beta} .
\label{ecnorm}
\ee
Eqs. \refb{efg3} with $A_{\alpha\beta}=0$ and \refb{eEBC} with the components along $\CY$
can be used to read out the vertex
operator of the $(2k)$-form field $\bbC^{(2k)}_{\bf m_1\cdots m_{2k}}$
in the $(-1/2,-3/2)$ picture, while \refb{ephiR} and \refb{ecnorm} may be
used to read out the same vertex operator in the $(-1/2,-1/2)$ picture.

It is instructive to express \refb{efg8} and \refb{efg9} in position space using the language
of differential forms. We have
\be
\begin{split}
\bbF^{(2k+1)} & = {1\over 4} * d * \(D^{(2k+2)}+E^{(2k+2)}\) -{1\over 4} \, d\(D^{(2k)}-E^{(2k)}\),
\\
& * d * \(D^{(2k+2)}+E^{(2k+2)}\) = -d\(D^{(2k)}-E^{(2k)}\) .
\end{split}
\ee
Using the self-duality constraints  \refb{edualbc}, we can
rewrite these relations as
\be\label{ekk1}
\begin{split}
\bbF^{(2k+1)}&
= (-1)^{k+1} \, {1\over 4} * d  \(D^{(8-2k)}-E^{(8-2k)}\) -{1\over 4}\, d\(D^{(2k)}-E^{(2k)}\),
\\
&
(-1)^{k+1} * d  \(D^{(8-2k)}-E^{(8-2k)}\) = -d\(D^{(2k)}-E^{(2k)}\)\, .
\end{split}
\ee
Using \refb{eccdef}, we can express \refb{ekk1} as
\be\label{ekk3}
\bbF^{(2k+1)}  ={1\over 2}\, d\bbC^{(2k)} + {(-1)^k\over 2} * d \bbC^{(8-2k)}
\ee
and
\be\label{ekk4}
d\bbC^{(2k)} = (-1)^k *d\bbC^{(8-2k)}\, .
\ee

Let us compare these with the equations of motion derived from the
string field theory action \refb{erraction} in the $A_{\alpha\beta}=0$ gauge. The
action takes the form:
\be
\sum_{k=0}^4 \int \( \bbF^{(2k+1)} - {1\over 4}\, d\bbC^{(2k)} - { (-1)^k\over 4} * d \bbC^{(8-2k)}\)
\wedge * \Bigl(d\bbC^{(2k)} -(-1)^k  * d \bbC^{(8-2k)}\Bigr)\, ,
\ee
up to a constant of proportionality. Here $\bbF^{(2k+1)}$ is subject to the algebraic
constraint \refb{edualcon}.
The equation of motion of $\bbF^{(2k+1)}$ leads to \refb{ekk4} and the equation of
motion of $\bbC^{(2k)}$ leads to the exterior derivative of \refb{ekk3}. Therefore, the solutions
to string field theory equations of motion contain additional degrees of freedom. As has been
discussed in detail in \cite{deLacroix:2017lif}, the extra modes describe free fields and
decouple from the theory even
after adding interaction terms to the action, which involve $\bbF^{(2k+1)}$ but not
$\bbC^{(2k)}$. Note that $\bbC^{(10)}$ does not appear in the action since $d \bbC^{(10)}$
vanishes, but we include this as a field in the theory. This demands that in a consistent theory
the one-point function of $\bbC^{(10)}$ on the disk must vanish. This is part of the
tadpole cancellation constraint in the theory.

Note that while $\bbC^{(2k)}$'s for different $k$ are independent fields to begin with,
\refb{ekk4} puts a constraint on their field strengths so that on-shell we only have
half the number of degrees of freedom. In perturbation theory this can be made explicit by
restricting $k$ to be in the range
0 to 2, but this will allow us to couple these fields only to D$p$-branes
for $p\le 5$. When we compactify the theory
so that the ten-dimensional Lorentz invariance is broken,
we could also use $\bbC^{(2k)}$ for certain components of the fields and $\bbC^{(8-2k)}$
for the other components. We make use of this explicitly in our analysis where instead of
dualizing $\bbC^{(2)}_{\mu\nu}$ to a scalar field in four dimensions, we directly use the
components of $\bbC^{(6)}$ along the Calabi-Yau threefold to describe the same scalar.
In this way we can ensure that we have all the components of
$\bbC^{(2k)}$ that couple to at least those D-branes that
do not extend along the non-compact spacetime directions. This is captured in \refb{ecnorm}.

\section{Normalization of the hypermultiplet moduli}
\label{sd}

In this appendix we shall describe the normalization of the various hypermultiplet moduli
fields in our conventions and find their relations to those appearing in \S\ref{s2} and
\S\ref{s3} by comparing the tree level kinetic terms and the D-brane actions.

We begin by comparing the D-brane actions. We see from \refb{Dbraneact} and
\refb{ecentralcharge} that for $B=0$, the real part of the action
for an Euclidean D$p$-brane ($p=1,3,5$) wrapped on a cycle $L_\gamma$ is given by:
\be
\begin{array}{c|c|c|c}
&\ p=1 \ &  \ p=3 & \ p=5\
\\
\hline
-\cTR_\gamma=\ &\ -2\pi \tau_2 |q_a t^a|\ & \ -2\pi \tau_2 |(pt^2)/2|\ & \ -2\pi \tau_2 V\
\vphantom{\Bigl(}
\end{array}
\label{Tact}
\ee
On the other hand, $\cTR_\gamma$ should be given by the product of the tension of the brane
times the volume of the wrapped cycle, as in \eqref{Realact}.
Since the volume form on the holomorphic cycle is given by \eqref{defvn}
where the K\"ahler form is parametrized by $\bbom=\bbt^a\omega_a$,
the volume is found to be
\be
\begin{array}{c|c|c|c}
&\ p=1 \ &  \ p=3 & \ p=5\
\\
\hline
|\bbV_\gamma|=\ &\ |q_a \bbt^a|\ & \ |(p\bbt^2)/2|\ & \ \bbV\
\vphantom{\Bigl(}
\end{array}
\label{Vg}
\ee
where $\bbV=\frac16\, (\bbt^3)$ is the volume of $\CY$.
Multiplying by the tension $(2\pi)^{-p}\tau_2$ and comparing with \eqref{Tact},
we find that the two forms of the action coincide provided
\be\label{ecc9}
\bbt^a = (2\pi)^2 t^a.
\ee
The relation \refb{ecc9} also implies
\be \label{edetail}
\bbkap_{ab}=(2\pi)^2 \kappa_{ab}, \qquad \bbG_{ab}=(2\pi)^{-4} \, G_{ab}, \qquad
\bbG^{ab}=(2\pi)^4 \, G^{ab}, \qquad \bbV = (2\pi)^6\, V\, .
\ee

Let us now find the relative normalization of the moduli associated with the internal components of the
2-form field. In ten dimensions we begin with the action
\be \label{ebkinetic}
-{1\over 6} \int d^{10} x \, H_{MNP} H^{MNP}, \qquad H_{MNP}=\p_M \bbB_{NP}
+\p_N \bbB_{PM}
+\p_P \bbB_{MN}\, .
\ee
$\bbB_{MN}$ normalized this way has its vertex operator normalized as in
\refb{e2.1} \cite{Alexandrov:2021shf}.
Upon compactification we introduce the moduli $\bbb^a$ via \eqref{e626a}.
Using \eqref{Hodge-om} and \eqref{eb7}, it is easy to check that
the kinetic term of these moduli takes the form
\be
-\int d^4 x  \int_\CY \p_\mu \bbB \wedge \p^\mu\! \star\!\bbB=
-\int d^4 x\, {\bbV\over 4\kappa^2}\,\bbG_{ab}\,\p_\mu\bbb^a \p^\mu \bbb^b,
\ee
where $\bbG_{ab}$ has been defined in \refb{defbbG}.
However, this result is derived in the string frame when the Einstein-Hilbert action is multiplied by
$\bbV/(2\kappa^2)$. We need the result in the canonical frame in which the Einstein-Hilbert
action is multiplied by 1. This requires us to redefine the four-dimensional spacetime
metric by a multiplicative factor of
$2\kappa^2/\bbV$, which in turn multiplies the kinetic term of the scalar fields
by $2\kappa^2/\bbV$. This converts the kinetic term to
\be\label{ecc13}
-\hf\int d^4 x\, \bbG_{ab}\, \p_\mu\bbb^a \p^\mu \bbb^b.
\ee
On the other hand, according to \refb{escalaraction}, \refb{metricBtreelargeV}
the kinetic term for $b^a$ is given by
\be\label{ecc14}
-\hf\int d^4 x\,G_{ab}\,\p_\mu b^a \p^\mu b^b\, .
\ee
Comparing the two kinetic terms and using \refb{edetail}, we get
\be\label{ecc15}
\bbb^a = (2\pi)^2 b^a\, ,
\ee
up to a sign.

Next, we determine the normalization of the scalar field $\bbsigma$ dual to the NSNS 2-form $B$.
With the standard normalization of string theory, its kinetic term is given by (see \cite[Eq.(6.21)]{Alexandrov:2021shf})
\be\label{e720}
-\int d^4 x \, {\kappa^2\over 2 \bbV^2}\,  \p_\mu\bbsigma\, \p^\mu\bbsigma\, ,
\ee
where we have used the normalization in which the background
canonical metric is set to $\eta_{\mu\nu}$.
Comparing this with the kinetic term for $\sigma$ in \refb{escalaraction},
\refb{metricBtreelargeV} and using \eqref{edetail}, we get
\be\label{ebbpsipsi}
\bbsigma =- {(2\pi)^6\over 2\kappa \tau_2^2} \,\sigma ,
\ee
where the sign has been chosen to match the result \refb{eagammafin}.

Finally, we consider the moduli associated with the RR fields $\bbC^{(2k)}$ whose
kinetic term appears in \refb{ebb17}.
Upon compactification on a Calabi-Yau manifold and conversion to the canonical frame, it becomes
\be
-\int d^4 x \,\frac{32\kappa^2}{\bbV}\int_\CY \p_\mu\bbC^{(2k)} \wedge \star\, \p^\mu\bbC^{(2k)}.
\ee
On the other hand, due to \refb{RRiib}, \refb{basisforms}, \eqref{Hodge-om}, \eqref{Hodge-om2} and \eqref{edetail},
the kinetic terms for the RR scalars $c^0$, $c^a$, $\tc_a$ and $\tc_0$ in \refb{escalaraction},
\eqref{metricBtreelargeV} can be rewritten as
\be
-\int d^4 x \,\frac{(2\pi)^{4k}}{2\tau_2^2\bbV}\int_\CY \p_\mu C^{(2k)} \wedge \star\, \p^\mu C^{(2k)}.
\ee
This result is valid even when the background $b^a$ are non-zero as long as they are constants.
Thus, we conclude that
\be
\bbC^{(2k)}=s_k\, \frac{(2\pi)^{2k}}{8\kappa\tau_2}\,C^{(2k)}=s_k\, \frac{(2\pi)^{2k}}{2^6\pi^{7/2}}\,C^{(2k)},
\label{relCC}
\ee
where $s_k=\pm 1$ is a sign remaining undetermined at this stage.

In principle, the signs left undetermined by the above analysis
can be fixed, up to symmetry transformations,
by comparing the tree level S-matrix computed from the world-sheet theory to
those obtained from the action \refb{metricBtreelargeV}. Once this is done, we can
compute the disk one-point functions of various fields and check that they agree with the
ones obtained from the $e^{-\TT_\gamma}$ factor in the instanton amplitude. We shall
not do this. Instead we shall use the disk one-point functions to fix the signs.
The agreement between the magnitudes of these normalization constants computed
using the two approaches can be regarded as a check on our computation.

Let us compute the one-point function of $\bbC^{(2k)}$ on the disk,
and compare with the imaginary part $-2\pi\I\Theta_\gamma$ of
the D$(2k-1)$-instanton action. For this we need to use the vertex operator of
the $2k$-form field in the $(-1/2,-3/2)$
picture. Using \refb{efg3} with $A_{\alpha\beta}=0$ and \refb{eEBC},
we see that the relevant part of the vertex operator is given by
\be
\begin{split}
\wt V & = -{1\over 2\cdot (2k)!} \, \bbC^{(2k)}_{M_1\cdots M_{2k}}
\bigg[(\Gamma^{M_1\cdots M_{2k}})_{\alpha}^{~\beta}\,
c\, \bar c\, e^{-3\phi/2} S^\alpha e^{-\bar\phi/2} \bar S_\beta \\
& \hskip 2in - (\Gamma^{M_1\cdots M_{2k}})^{\alpha}_{~\beta}\, c\, \bar c\, \p\xi \, e^{-5\phi/2}
S_\alpha \,  \bar\eta \, e^{\bar\phi/2} \, \bar S^\beta\bigg] e^{\I p.X}\, ,
\end{split}
\ee
where we have dropped the term in the third line of \refb{efg3} since it does not satisfy the
$\xi$-$\eta$ charge conservation and therefore will have vanishing one-point function on the
disk.
Also for this computation we can drop the $e^{\I p.X}$ factor.
Due to \eqref{edisk1}, the disk one-point function of $\bbC^{(2k)}$ takes the form
\be \label{edd37a}
\begin{split}
& A=  -{\kappa\, \ts_{2k-1}\over 2 \cdot (2k)!} \, {1\over 2} \, {1\over 2}\,
 \bbC^{(2k)}_{M_1\cdots M_{2k}}\,
\Bigg[(\Gamma^{M_1\cdots M_{2k}})_{\alpha}^{~\beta}\, \Bigl\langle (\p c -\bar\p \bar c) \,
c\, \bar c\, e^{-3\phi/2} S^\alpha e^{-\bar\phi/2} \bar S_\beta (\I)
\\ & \hskip 1.7in - (\Gamma^{M_1\cdots M_{2k}})^{\alpha}_{~\beta}\,
\langle (\p c -\bar\p \bar c) c\, \bar c\, \p\xi \, e^{-5\phi/2}
S_\alpha \,  \bar\eta \, e^{\bar\phi/2} \, \bar S^\beta(\I)\,
\Bigr\rangle
\Bigg]\, .
\end{split}
\ee
Using the D$(2k-1)$-brane boundary condition \refb{SSbc},
the doubling trick and the operator product expansion \refb{eqb1}, the correlation function becomes
\be
A=-{\kappa\, \ts_{2k-1}\over 8\cdot  (2k)!}
\, \bbC^{(2k)}_{M_1\cdots M_{2k}} \times {1\over (2k)!} \, v^{M_1\cdots M_{2k}}
\times (-4 \I) \times(-1)^k  (2k)!
\times 16 \times (2\pi)^{2k} \, \delta^{(2k)}(0)\, .
\ee
Here the factor $(-4\I)$ represents the equal
contribution from the two terms inside the square bracket,
the 16 comes from the trace of the identity operator in the spinor representation
and the $(-1)^k (2k)!$ comes from contraction
of the gamma matrices:
\be
\Tr \(\Gamma_{M_1\cdots M_{2k}} \Gamma^{N_1\cdots N_{2k}}\)
=16\, {\delta_{M_1\cdots M_{2k}}}^{N_1\cdots N_{2k}} (-1)^{(2k-1)(2k)/2}\, .
\ee
Interpreting the $(2\pi)^{2k} \, \delta^{(2k)}(0)$ factor as the
integral of  over the $(2k)$-cycle $L_\gamma$, we get
\be\label{ecc37}
A = (-1)^k\,  8 \I  \kappa\, \ts_{2k-1} \int_{L_\gamma} \bbC^{(2k)}\, .
\ee

On the other hand, this one-point function can be extracted from the
$-2\pi i\Theta_\gamma$ term in the instanton action, where $\Theta_\gamma$ is specified in \eqref{Dbraneact}. Since the boundary condition \refb{SSbc} that we have used
to arrive at \refb{ecc37} is valid in the absence
of background $B$ field,
we need to consider
\be
-2\pi \I\int_{\CY}\gamma\wedge C^{\rm even}=2\pi \I\sum_{k=0}^3 (-1)^k \int_{L_\gamma} C^{(2k)},
\ee
where we used \eqref{mukaimap} and \eqref{defLg}.
Comparing with \eqref{ecc37}, one finds
\be
\bbC^{(2k)}=\frac{\pi}{4\kappa T_{2k-1}}\,C^{(2k)}= \frac{(2\pi)^{2k}}{2^6\pi^{7/2}}\,C^{(2k)},
\label{relCCsign}
\ee
which agrees with \refb{relCC} for $s_k=1$ for all $k$.

Note that our results \refb{ebkinetic}-\refb{relCC}, involving purely the closed string sector, are valid
whether or not a background NSNS 2-form field is present, provided
\refb{ecc9}, \refb{edetail} hold. However \refb{Tact}-\refb{edetail} and \refb{edd37a}-\refb{relCCsign}
involve the open string sector and therefore are valid only when the background NSNS 2-form field is absent.
Since \refb{ecc9}, \refb{edetail} were used in finding the relations between variables given in
\refb{ebkinetic}-\refb{relCC}, these relations should also be regarded as proven only in the
absence of background $B$ field. However, we have verified in
appendix \ref{sg} that \refb{ecc9}, \refb{ecc15},
\refb{relCCsign} correctly reproduce the disk one-point functions even when the
$B$ field is switched on.
Therefore, these relations, as well as \refb{ebbpsipsi}, remain valid
in the presence of the background $B$ field.

\section{D-instanton action in the presence of NSNS 2-form field} \label{sg}

In this appendix we shall compute the D-brane action in the presence
of a constant background $B$ field and compare the result with \refb{Dbraneact}.
First, we shall compute the imaginary part, which involves
the disk amplitude with just one RR vertex operator in the presence of a background $B$ field.
We take the RR vertex operator
in the $(-1/2,-3/2)$ picture as in appendix \ref{sd} and follow the procedure
described in \S\ref{s7.1x} to study the effect of the background $B$ field.
The net result is the same as in \S\ref{s7.1x}, namely that we replace the charges by the effective charges
given in \refb{def-qch}. Therefore, \refb{ecc37} is replaced by the sum:
\be\label{ecc37new}
\begin{split}
A =&\,  \sum_{k=0}^3 (-1)^k\,  8 \I  \kappa\, \ts_{2k-1} \int_{\chLgi{2k}} \bbC^{(2k)}
=  2\pi \I\sum_{k=0}^3 (-1)^k \int_{\CY} \omega_{\chgam}^{(6-2k)}\wedge C^{(2k)} =
-2\pi\I\Theta_\gamma\, ,
\end{split}
\ee
where we used the relation \eqref{relCCsign}, \eqref{hatgamTheta} and that
\be
\sum_k (-1)^{k}\omega_{\chgam}^{(6-2k)} = - \sum_k (-1)^{k}\omega_{\chgam}^{(2k)}
=-\iota(\omega_{\chgam}) =-\chgam
\ee
as remarked below \refb{list-volume}. The right hand side of \refb{ecc37new}
is precisely the imaginary part of $-\TT_\gamma$ given in \refb{Dbraneact}.

Next we shall compute the real part of the action of Euclidean D$p$-branes in the
presence of background $B$ field. Such action is given by
\be
\cTR_\gamma = T_{2n-1} \left| \int_{\Lgi{2n}} v^{(2n)}_\gamma \,
\sqrt{\mbox{det} (\bbg _\parallel+ 2\kappa\, \bbB_\parallel)}\,  \bigg/ \, \sqrt{\mbox{det}\bbg_\parallel} \right| .
\label{actionB}
\ee
To evaluate the integrand, let $M_{\bf ij}$ be a rank two tensor in $\CY$ with only mixed
components, i.e.\ $M_{st}=M_{\bs \bt}=0$, and suppose further that
$M_{s\bar t}$ may be expanded as $M^a (\omega_a)_{s\bar t}$.
Then its pullback $M_\parallel$ on $\Lgi{2n}$ will also be
a matrix with only mixed components in the holomorphic coordinates, i.e. $(M_\parallel)_{ab}=(M_\parallel)_{\ba \bb}=0$,
and its determinant factorizes into
$\bigl(\det (M_\parallel)_{\bb a}\bigr)\bigl(\det (M_\parallel)_{a\bb}\bigr)$,
where the indices $a,b$ now label holomorphic coordinates on $\Lgi{2n}$.
We shall use the notation $\det_{\rm h}(M_\parallel)$ to denote the determinant
of the matrix $(M_\parallel)_{a\bb}$.
Let $N$ be another rank two covariant tensor with similar properties.
Then we have on $\Lgi{2n}$
\be
{\det_{\rm h} (M_\parallel)\over \det_{\rm h} (N_\parallel)}
={\det_{\rm h} (M_P) \over \det_{\rm h} (N_P)}
={(M^n, v^{(2n)}_\gamma)\over (N^{n}, v^{(2n)}_\gamma)}
=\frac{M^n \wedge \omega_\gamma^{(6-2n)}}{N^n \wedge \omega_\gamma^{(6-2n)}}
=\left\{\begin{array}{ll}
1 & \qquad n=0,
\\
\frac{q_a M^a}{q_a N^a} &\qquad n=1,
\\
\frac{(pM^2)}{(pN^2)} &\qquad n=2,
\\
\frac{(M^3)}{(N^3)} &\qquad n=3,
\end{array} \right.
\label{evaldet}
\ee
where the first equality follows from the identities $M_\parallel = W M_P \bar W^T$,
$N_\parallel = W N_P \bar W^T$
as stated above \refb{e6.16}, the second equality follows from the definition of the determinant,
and we used \eqref{equal-scpr}, \eqref{innerwedge}, \eqref{list-volume}, \eqref{basisforms}
and \eqref{omaomb} for the rest of the equalities.
Note that while in the first two expressions we need to pullback / project
$M$ and $N$ onto $\Lgi{2n}$ before computing the determinant,
in the later expressions no such projection is needed.
Applying this result to the integrand in \eqref{actionB} where the matrices $M=\bbg + 2\kappa\, \bbB$ and  $N=\bbg$
satisfy the above assumptions, one finds
\be
\sqrt{\frac{\mbox{det} (\bbg_\parallel + 2\kappa\, \bbB_\parallel)}{\mbox{det} \bbg_\parallel}}
=\left\{\begin{array}{ll}
\left|\frac{q_a z^a}{q_a t^a}\right| &\qquad n=1,
\\
\left|\frac{(pz^2)}{(pt^2)}\right| &\qquad n=2,
\\
\left|\frac{(z^3)}{(t^3)}\right| &\qquad n=3.
\end{array} \right.
\label{evaldetB}
\ee
Since this quantity is constant on $\CY$, the integral in \eqref{actionB} gives rise just to the factor of
$|\bbV_\gamma|$. Therefore, the net effect of the background $B$ field on $\cT_\gamma^R$ is to
give an extra multiplicative factor  given by the r.h.s.\ of \refb{evaldetB}. Since we have already
verified in appendix \ref{sd} that in the absence of $B$ field the $\cT_\gamma^R$ computed from
the D-brane action agrees with the results given in \refb{Tact}, we can now use
\refb{edefzgamma} to conclude that in the presence of the background $B$ field,
the real part of the action is given by
\be \label{eg.7}
\cTR_\gamma=2\pi\tau_2 |Z_\gamma|
\ee
consistently with \refb{Dbraneact}.

\providecommand{\href}[2]{#2}\begingroup\raggedright\endgroup


\end{document}